\documentclass[twoside,letterpaper]{article}

\usepackage{mathtools}
\mathtoolsset{showonlyrefs=true}
\usepackage{cases}
\usepackage{microtype}
\usepackage{amsthm, amsmath, amssymb, amsfonts, graphicx, epsfig}
\usepackage{accents}
\usepackage[sorting=none,backend=biber]{biblatex}
\usepackage{bbm}
\usepackage{mathrsfs}
\usepackage{ulem}
\usepackage{cancel}
\usepackage{epstopdf}
\usepackage{graphicx}
\usepackage[font=sl,labelfont=bf]{caption}
\usepackage{subcaption}
\usepackage{float}
\restylefloat{table}
\usepackage{algorithmic}
\usepackage{algorithm}
\usepackage{enumerate}

\usepackage[dvipsnames]{xcolor}
\newif\ifshowkeys
\showkeysfalse %
\ifshowkeys

\usepackage[notref,notcite,color]{showkeys}
\definecolor{labelkey}{rgb}{0,0,1}
\fi

\usepackage[colorlinks=true, pdfstartview=FitV, linkcolor=blue,
            citecolor=blue, urlcolor=blue]{hyperref}

\usepackage{url}
\makeatletter\def\url@leostyle{%
 \@ifundefined{selectfont}{\def\UrlFont{\sf}}{\def\UrlFont{\scriptsize\ttfamily}}} \makeatother

\usepackage[framemethod=default]{mdframed}

\usepackage{array}

\usepackage[margin=2.5cm, letterpaper]{geometry}

\usepackage{multicol}
\usepackage{tikz,pgfplots}
\usetikzlibrary{shapes,arrows,trees,automata,positioning,matrix,decorations.pathreplacing,calc}
\pgfplotsset{compat=1.18, compat/show suggested version=false}

\newtheorem{theorem}{Theorem}[section]
\newtheorem{proposition}[theorem]{Proposition}
\newtheorem{lem}[theorem]{Lemma}
\newtheorem{cor}[theorem]{Corollary}

\theoremstyle{definition}

\newtheorem{hyp}{Assumption}

\theoremstyle{remark}
\newtheorem{remark}{Remark}

\definecolor{Red}{rgb}{1.0,0,0.0}
\definecolor{Blue}{rgb}{0,0.0,1.0}
\definecolor{Green}{rgb}{0.2,0.5,0.2}

\def\cA{\mathcal{A}}
\def\cB{\mathcal{B}}
\def\cC{\mathcal{C}}
\def\cD{\mathcal{D}}
\def\cE{\mathcal{E}}
\def\cF{\mathcal{F}}
\def\cG{\mathcal{G}}

\def\cL{\mathcal{L}}
\def\cM{\mathcal{M}}
\def\cN{\mathcal{N}}

\def\bF{\mathbb{F}}

\def\bH{\mathbb{H}}

\def\bN{\mathbb{N}}

\def\bP{\mathbb{P}}

\def\bR{\mathbb{R}}

\def\esp#1{\mathbb E\left[#1\right]}
\def\espc#1#2{\mathbb E_{#1}\left[#2\right]}
\def\pro#1{\mathbb P \left[#1\right]}
\def\ind{\mathbbm{1}}

\newcommand{\1}{\mathbbm{1}}            %

\renewcommand{\mid}{\;|\;}              %

\usepackage{todonotes}
\title{Error analysis for the numerical approximation of the quantile hedging problem}

\author{Cyril B\'en\'ezet\thanks{Université Paris-Saclay, CNRS, Univ Evry, ensIIE, Laboratoire de Mathématiques et Modélisation d'Evry, 91037, Evry-Courcouronnes, France ({\tt cyril.benezet@ensiie.fr}).}
\and
Cyril Nefzaoui-Blanchard\thanks{Université Paris-Saclay, CNRS, Univ Evry, Laboratoire de Mathématiques et Modélisation d'Evry, 91037, Evry-Courcouronnes, France ({\tt cyril.nefzaouiblanchard@univ-evry.fr}).}
\and
Sergio Pulido\thanks{\protect\raggedright Université Paris Cité, Laboratoire de Probabilités, Statistique et Modélisation (LPSM), CNRS UMR 8001 ({\tt pulido@lpsm.paris}).}\ \thanks{The work of Sergio Pulido benefited from the financial support of the research program Chaire Deep learning in finance and Statistics (Fondation de l'\'Ecole Polytechnique, \'Ecole Polytechnique and Qube R\&T).
}
}

\date{
\today
}

\begin{document}

\raggedbottom

\maketitle
\begin{abstract}
We study the numerical approximation of the quantile hedging problem \cite{FollmerLeukert1999QH} in a Markovian diffusion framework modeling a complete and linear market. In this setting, the quantile hedging problem, a stochastic target problem (see \cite{BET09}), writes as a stochastic control problem for an augmented state variable $(X,P)$, where $P$, controlled through its volatility coefficient in a degenerate and unbounded fashion, encodes the targeted success probability and, as such, is constrained to lie in $[0,1]$.

Our analysis targets approximation procedures that are both implementable and compatible with the state constraint $P \in [0,1]$: (i) truncation of the control set, and (ii) piecewise-constant-in-time controls (PCPT). As PCPT controls lead to conditionally Gaussian distributions, they do not satisfy the state constraint. One needs, in addition to a raw PCPT control, to consider the exit time from the interior of $[0,1]$, and nullify the control after this time. Then, following \cite{Krylov1999,Jakobsen2019}, we rely on one-step dynamic programming principles and derive a suitable generator inequality for a regularised PCPT value function.

Under standard regularity assumptions on the coefficients and a Lipschitz condition on the payoff, we obtain explicit error bounds on the interior interval and optimise the regularisation parameters. This yields an error exponent arbitrarily close to $1/8$, uniformly in the probability variable, for the combined effect of control truncation and PCPT time discretisation. In the situation where the control set is bounded from the outset, we obtain an error exponent arbitrarily close to $1/4$ from below, the exponent established by \cite{Jakobsen2019} in the absence of state constraints.

Building on the PCPT approximation, we implement a fully discrete scheme with neural-network controls. Projecting the probability process onto \([0,1]\) when an exit is detected on the time grid introduces a bias. We quantify this bias in a toy problem and mitigate it with a state-dependent control bound. Experiments in the Black–Scholes and Lipschitz local-volatility models illustrate the effect of this bound and the decrease in pricing errors under grid refinement.

\end{abstract}

{\small
\par
}

\clearpage

\tableofcontents

\normalem

\section{Introduction}\label{sec:introduction}

Quantile hedging addresses the problem of hedging a contingent claim with a prescribed probability of success. Unlike super-replication, it allows the terminal hedging error \(Y_T-g(X_T)\) to be negative on a set of positive probability, thereby reducing the required initial capital while controlling the probability of failure; see e.g. \cite{FollmerLeukert1999QH, FollmerLeukert2000EfficientHedging}.

Mathematically, quantile hedging is a stochastic target problem with a \emph{weak} terminal constraint, whereas super-replication corresponds to an almost-sure target. For $p \in [0,1]$ and a non-negative European payoff $g(X_T)$ at maturity $T$, we use the standard formulation
\begin{align}\label{eq:QH-intro}
    V(p) = \inf \Big\{y \in \bR &: \text{there exists a self-financing strategy $\nu$ such that } \\ \nonumber &Y^{y,\nu}_T \ge 0 \text{ a.s., and } \bP\left(Y^{y,\nu}_T \ge g(X_T)\right) \ge p \Big\},
\end{align}
where \(Y^{y,\nu}\) denotes the wealth process associated with initial wealth \(y\) and the self-financing strategy \(\nu\), trading in the risky assets with price process \(X\).

Under the Brownian filtration, the martingale representation theorem
allows one to reformulate the weak probabilistic constraint as an
almost-sure constraint on an additional controlled martingale; see
\cite{BET09}. More precisely, define
\[
P^{p,\alpha}_t
:=p+\int_0^t\alpha_s^\top dW_s,
\qquad
\cA_p
:=\left\{\alpha\in\bH^2:
P^{p,\alpha} \in[0,1]\ \text{ a.s. on } [0,\infty)\right\}.
\]
Then the value in \eqref{eq:QH-intro} can equivalently be written as
\begin{align*}
V(p)&=\inf_{\alpha\in\cA_p}
\inf\Big\{y\in\bR:\exists\nu,\;
Y_T^{y,\nu}\ge0\ \text{a.s.},\quad
\1_{\{Y_T^{y,\nu}\ge g(X_T)\}}\ge P_T^{p,\alpha}
\ \text{a.s.}\Big\}\\
&=\inf_{\alpha\in\cA_p}
\inf\Big\{y\in\bR:\exists\nu,\;
Y_T^{y,\nu}\ge
g(X_T)\1_{\{P_T^{p,\alpha}>0\}}
\ \text{a.s.}\Big\}.
\end{align*}

Thus quantile hedging is the infimum of super-replication prices of modified payoffs $g(X_T) \1_{P^{p,\alpha}_T>0}$, $\alpha \in \cA_p$. In a complete linear diffusion market, the super-replication price of a European option is given by the expectation of the discounted payoff under the unique equivalent martingale measure, and one can therefore obtain the equivalent formulation
\begin{align*}
    V(p)=\inf_{\alpha \in \cA_p} \esp{Z_T g(X_T) \1_{P^{p,\alpha}_T>0}}
\end{align*}
where $Z$ is the pricing kernel, whose dynamics are specified in Section~\ref{sec:problem}.

Finally, convexifying the discontinuous terminal condition \(g(x)\1_{\{p>0\}}\) with respect to \(p\) yields the equivalent representation
\begin{align}\label{eq:qh}
    V(p)=\inf_{\alpha \in \cA_p} \esp{Z_T g(X_T) P^{p,\alpha}_T},
\end{align}
see \cite{BET09} for details. This approach yields a nonlinear Hamilton--Jacobi--Bellman equation with a state constraint in the probability variable. The quantile hedging value is characterised as its constrained viscosity solution; see \cite{BET09} and \cite{BouveretChassagneux2018ComparisonApproxHedging}.

\medskip\noindent In related directions, \cite{BouchardElieReveillac2015WeakTerminal} study the non-Markovian case via BSDEs with weak terminal conditions, while Bermudan and American variants connect naturally with weak reflection and weak constraints at stopping times; see \cite{BouchardBouveretChassagneux2016BackwardDualBermudanQH, BriandElieHu2018MeanReflection, dumitrescu2023new}. These formulations should be contrasted with the classical reflected BSDE framework for superhedging American options; see \cite{ElKarouiKapoudjianPardouxPengQuenez1997ReflectedBSDE}.

When tackling the numerical approximation of the quantile hedging price, the main difficulties come from the fact that the diffusion $P$ is controlled through an unbounded volatility coefficient and must remain in $[0,1]$. Designing implementable numerical schemes in this constrained setting is therefore nontrivial; see \cite{BenezetChassagneuxReisinger2021NumericalQH} for a discretisation of the associated HJB equation. More generally, stochastic control has a rich numerical literature based on PDE schemes, BSDE methods, regression Monte Carlo, deep learning, and quantization; see for instance \cite{barles1991convergence, KushnerDupuis2001, BOUCHARD2004175, LongstaffSchwartz2001, gobet2005regression, E_2017, pages2004optimal, kharroubi2014numerical}.

Another approach consists in restricting admissible controls to policies that are held constant between consecutive decision dates on a grid with mesh size \(h\). This leads to the so-called Piecewise Constant Policy Timestepping (PCPT) approximation. Such methods are classical and effective in stochastic control; see \cite{Krylov1999} and the more recent refinements in \cite{Jakobsen2019}. We also refer to \cite{barles2007error} and \cite{reisinger2016piecewise}, where the standard control problem is approximated by a switching system with a small switching cost. PCPT methods were already considered in \cite{BenezetChassagneuxReisinger2021NumericalQH} to numerically approximate the quantile hedging price through the discretisation of the Hamilton-Jacobi-Bellman PDE, with no theoretical convergence rate.

\medskip\noindent
In this work, we design a PCPT scheme to approximate the quantile hedging problem in a complete linear diffusion market as in \eqref{eq:qh}. Compared with \cite{Krylov1999,Jakobsen2019}, we must also deal with unbounded controls, the state constraint in the probability variable, and the fact that raw PCPT controls are incompatible with $P\in[0,1]$. Our goal is to obtain an explicit error analysis for implementable approximation procedures in a complete Markovian diffusion framework. Our main contribution is an explicit error decomposition and convergence-rate
analysis for the piecewise-constant control approximation, accounting
simultaneously for control truncation, boundary effects, and time
discretisation.

Our approximation strategy combines three ingredients. First, we mollify the model coefficients and regularise the dual problem in order to construct \(\varepsilon\)-optimal smooth feedback controls (Theorem~\ref{thm:eps opti}). Second, these controls allow us to truncate the control set ($|\alpha| \le M$) and work on an interior interval ($P\in[\eta,1-\eta]$) to mitigate boundary effects (Theorem~\ref{cor:domain-truncation-vMeta-v}). Third, we analyse the PCPT scheme, where controls are restricted to be constant on a prescribed time grid.

A key technical point is that raw PCPT controls lead to Gaussian one-step increments for the controlled process $P$, which are not compatible with the state constraint $P\in[0,1]$. To bypass this difficulty, the PCPT controls we consider also encode the stopping time at which $P$ reaches the domain boundary; after this time, the control is nullified so that $P$ stays in $[0,1]$. We obtain an explicit error analysis by adapting the strategy of \cite{Krylov1999, Jakobsen2019} through shaking and mollification, while accounting for the boundary hitting time between two grid points, which directly impacts the final error bounds (Theorem~\ref{thm:decomposition-main-error}).

After optimising the regularisation and truncation parameters, we obtain a convergence rate arbitrarily close to \(h^{1/8}\) for the original problem with unbounded controls. When the control set is bounded from the outset, the PCPT time-discretisation error has an exponent arbitrarily close to $1/4$ from below. Thus, despite the state constraint, the limiting exponent is the same as the exponent proved by \cite{Jakobsen2019} for problems without state constraints.

We also implement a fully discrete scheme with neural-network controls in the spirit of \cite{han2016deep,E_2017}. We detect exits on the time grid and project the probability process onto \([0,1]\). This projection generally breaks the martingale property of \(P\). In a toy model, we show that the resulting bias vanishes under a suitable scaling of the control bound with the mesh size, while the experiments show that it can remain substantial on coarse grids. To mitigate the bias, we impose a state-dependent control bound based on the distance to the boundary. For put options in the Black–Scholes and Lipschitz local-volatility models, we compare the fixed and adaptive bounds and examine how pricing errors change under grid refinement. For the Black–Scholes put, we also compare the learned feedback with the explicit optimal feedback evaluated at the same states along the test paths.

\medskip\noindent Two directions for further work are suggested by this analysis. The first is to treat more general state constraints. This would require quantitative estimates of the error incurred by truncating nearly optimal controls, as well as piecewise-constant policies that preserve the state constraint between decision dates. In the present setting, regularising the dual problem yields explicit nearly optimal feedbacks with growth bounds that allow us to estimate the truncation error. The absorption of the probability martingale at $0$ and $1$ allows us to preserve the state constraint by stopping the control at the boundary.

The second is to obtain quantitative PCPT error estimates for quantile hedging in nonlinear markets, where the superhedging price is no longer given by a linear expectation under a fixed pricing measure. This would complement existing qualitative convergence results (see \cite{BenezetChassagneuxReisinger2021NumericalQH}) and require different arguments for the steps of our proof that rely on the linear pricing representation.

\medskip\noindent
The rest of the paper is organised as follows. Section~\ref{sec:problem} introduces the model and the control problem. We provide elementary properties for the value function, and we construct $\varepsilon$-optimal smooth feedback controls. Section~\ref{sec:trunc} addresses control and domain truncation. Section~\ref{sec:pcpt} is then devoted to the PCPT approximation and
its error bounds. Finally, Section~\ref{sec:numerics} investigates a fully discrete neural implementation. %

\section{The problem}\label{sec:problem}

We fix a finite horizon $T>0$ and work on a probability space $(\Omega,\cF,\bP)$ supporting a $q$-dimensional Brownian motion $W=(W_t)_{t \ge 0}$. We denote by $\bF$ its naturally augmented filtration.

For any $(t,x,p)\in[0,T]\times\bR^d\times\bR$ and for $\alpha \in \cA := \bH^2(\bF;\bR^q)$, we consider the controlled system $(X^{t,x},P^{p,\alpha})$ defined, for $s\ge0$, by
\begin{align} \label{eq:sde state}
    d X_s &= \mu(t+s,X_s)\, ds + \sigma(t+s,X_s)\, dW_s, \\
    d P_s &=  \alpha_s^\top dW_s, \\
    X_0 &= x, \quad P_0=p.
\end{align}
where $(\mu,\sigma):\bR\times\bR^d\to\bR^d\times\bR^{d\times q}$ is measurable.

\noindent Note that $P^{p,\alpha}$ is a square-integrable martingale for any $\alpha \in \cA$.

\noindent We also define the process $Z$ as follows:
\[
Z^{t,x}_s = \exp\Big(\int_0^s \left(b(t+u,X^{t,x}_u) - \frac12|\lambda(t+u,X^{t,x}_u)|^2 \right)\,du+\int_0^s \lambda(t+u,X^{t,x}_u)^\top dW_u\Big),\qquad s\ge0,
\]
where $(b,\lambda) : \bR\times\bR^d\to\bR\times\bR^q$ is measurable.

For later use, we also introduce the reciprocal process
\[
Q^{t,x}:=\frac{1}{Z^{t,x}}.
\]
An application of It\^o's formula  yields, for $s\ge0$,
\begin{align}
    dQ_s = \big(|\lambda(t+s,X_s)|^2-b(t+s,X_s)\big) Q_s\, ds - Q_s \lambda(t+s,X_s)^\top dW_s.
\end{align}
The following standard Lipschitz and growth conditions ensure well-posedness of the SDEs.
\begin{hyp} \label{h:coeffs}
    The map $(\mu,\sigma,b,\lambda):\bR\times\bR^d\to\bR^d\times\bR^{d\times q}\times\bR\times\bR^q$ is measurable. There exist constants $L\ge 0$, $\bar\lambda\ge 0$ and $\bar b\ge 0$ such that for all $t\in\bR$ and all $x,y\in\bR^d$,
    \begin{align}
        |\mu(t,x)-\mu(t,y)| + \Vert\sigma(t,x)-\sigma(t,y)\Vert + |\lambda(t,x)-\lambda(t,y)| + |b(t,x)-b(t,y)| \le L|x-y|,
    \end{align}
    and for all $s,t\in\bR$ and all $x\in\bR^d$,
    \begin{align}
        |\mu(t,x)-\mu(s,x)| + \Vert\sigma(t,x)-\sigma(s,x)\Vert + |\lambda(t,x)-\lambda(s,x)| + |b(t,x)-b(s,x)| \le L|t-s|^{1/2},
    \end{align}
    and
    \begin{align}
        |\mu(t,x)| + \Vert\sigma(t,x)\Vert \le L(1+|x|),\qquad |\lambda(t,x)|\le \bar\lambda,\qquad |b(t,x)|\le \bar b.
    \end{align}
\end{hyp}
Under Assumption~\ref{h:coeffs}, for any $(t,x,p) \in [0,T]\times\bR^d\times\bR$ and any $\alpha \in \cA$, the above SDE admits a unique strong solution $(X^{t,x}_s,Z^{t,x}_s,P^{p,\alpha}_s)_{s\ge0}$.

We take the control representation \eqref{eq:qh} in the Introduction as our starting point. In the complete-market interpretation, \(Z\) is the state-price density, \(b\) accounts for discounting, and \(\lambda\) is the corresponding Girsanov kernel. However, the analysis below only relies on
Assumption~\ref{h:coeffs} and therefore applies more generally.

\begin{remark}[Boundedness of $\lambda$ and $b$]\label{rem:boundedness-lambda-b}
    The boundedness assumption $|\lambda|\le\bar\lambda$, $|b|\le\bar b$ is stronger than the linear growth condition imposed on $\mu$ and $\sigma$, and is used throughout the paper. Since the proofs and error analysis are already rather technical, we retain these boundedness assumptions to keep the arguments transparent. However, it could in principle be relaxed to linear growth conditions, at the cost of replacing all uniform-in-$x$ moment bounds on~$Z$ by polynomial bounds of the form $C(1+|x|)^{km}$ and imposing additional integrability conditions (e.g.\ Kazamaki-type criteria). This would modify the polynomial dependence on the initial condition $x$ and possibly the convergence rates.
\end{remark}

\noindent For any $(t,x,p)\in[0,T]\times\bR^d\times\bR$, we consider the stochastic control problem
\begin{align} 
    v(t,x,p) %
    &= \inf_{\alpha \in \cA} \esp{Z^{t,x}_{T-t} \hat g(X^{t,x}_{T-t}, P^{p,\alpha}_{T-t})} \label{eq: control pb}
\end{align}
where $g \ge 0$ and 
\begin{align*}
\hat g(x,p):=
\begin{cases}
g(x)p, & p\in[0,1],\\
+\infty, & p\notin[0,1].
\end{cases}
\end{align*}
is straightforwardly seen to be a convex function of $p$. We work under the following Lipschitz assumption on $g$.

\begin{hyp} \label{h:g Lip}
For all $x,y \in \bR^d$, we have $|g(x)-g(y)| \le L|x-y|$.
\end{hyp}
\begin{remark} \label{rem:g-growth}
    In particular, under Assumption \ref{h:g Lip}, $g$ has linear growth: there exists a constant $C\ge 0$ such that for all $x\in\bR^d$, $|g(x)|\le C(1+|x|)$.
\end{remark}

\medskip\noindent We first record some estimates for the processes $X$, $Z$ and $Q$, which will be used throughout the paper. The classical proof of these results is postponed to Appendix~\ref{sec:proof-lem-props-XZQ}.
\begin{lem}\label{lem:props-XZQ}
    Let Assumptions~\ref{h:coeffs} and~\ref{h:g Lip} hold.
    \begin{enumerate}
        \item For every $m\in\bR$ and $\bar T > 0$, there exists a constant $C_{\bar T,m}\ge 0$ such that, for all $(t,x)\in[0,T]\times\bR^d$,
        \begin{align*}
                \esp{\sup_{s \in [0,\bar T]}\big(Z^{t,x}_s\big)^m}\le C_{\bar T,m}.
        \end{align*}
        
        \item For any $m \ge 2$ and $\bar T > 0$, there exists $C_{\bar T,m} \ge 0$ such that, for all $(t,x) \in [0,T]\times\bR^d$,
        \begin{align*}
            \esp{\sup_{s\in[0,\bar T]}|X^{t,x}_s|^m}\le C_{\bar T,m}\,(1+|x|)^{m}.
        \end{align*}
    \item For every $m\ge 2$ and $\bar T>0$, there exists a constant $C_{\bar T,m}\ge 0$ such that, for all $(t,x)\in[0,T]\times\bR^d$ and $\varepsilon\in(0,\bar T]$,
    \begin{align*}
        \esp{|X^{t,x}_{\bar T} - X^{t,x}_{\bar T-\varepsilon}|^m} \le C_{\bar T,m}\,(1+|x|)^m\,\varepsilon^{m/2}, \qquad \esp{|Z^{t,x}_{\bar T} - Z^{t,x}_{\bar T-\varepsilon}|^m} \le C_{\bar T,m}\,\varepsilon^{m/2}.
    \end{align*}
    \item For every $r\ge 1$ there exists a constant $C_r\ge 0$ such that, for all $(t,x)\in[0,T]\times\bR^d$,
    \[
        \esp{\Big(Z^{t,x}_{T-t}\,g(X^{t,x}_{T-t})\Big)^r}\le C_r\,(1+|x|)^{r}.
    \]
    \end{enumerate}
\end{lem}

\subsection{Elementary properties}

We now record a few elementary properties of $v$. First, we show that we can restrict our attention to controls making the controlled process $P$ stay in $[0,1]$. To this effect, we introduce the following classes of admissible controls:
\begin{align}
\cA_{t,p} &:= \{\alpha \in \cA \,|\, P^{p,\alpha} \in [0,1] \text{ a.s. on } [0,T-t]\}, \text{ and} \\ 
\label{eq:admissible controls}\cA_p &:= \{\alpha \in \cA \,|\, P^{p,\alpha} \in [0,1] \text{ a.s. on } [0, \infty)\}.
\end{align}
We further establish monotonicity, convexity and H\"older continuity of $v$ with respect to the probability variable $p$.
\begin{lem}\label{lem:props-v} Fix $(t,x) \in [0,T] \times \bR^d$. The following properties hold:
\begin{enumerate}
    \item For any $p \in \bR$, we have
    \begin{align}
        v(t,x,p) &= \inf_{\alpha \in \cA_{t,p}} \esp{
        Z^{t,x}_{T-t} g(X^{t,x}_{T-t}) P^{p,\alpha}_{T-t}} = \inf_{\alpha \in \cA_p} \esp{
        Z^{t,x}_{T-t} g(X^{t,x}_{T-t}) P^{p,\alpha}_{T-t}}.
    \end{align}

    In particular, $\cA_{t,p}=\cA_p=\emptyset$ and $v(t,x,p) = +\infty$ whenever $p \not\in [0,1]$.
    \item (Monotonicity in $p$) The map $p\mapsto v(t,x,p)$ is nondecreasing on $[0,1]$.
    \item (Convexity in $p$) The map $p\mapsto v(t,x,p)$ is convex on $[0,1]$.
    \item (H\"older continuity in $p$) For any $\kappa\in(0,1)$, the map $v(t,x,\cdot)$ is $(1-\kappa)$-H\"older continuous on $[0,1]$, i.e. there exists a constant $C_{\kappa}\ge 0$ (independent of $(t,x)$) such that for all $0 \le p < p' \le 1$,
        \[
            0\le v(t,x,p')-v(t,x,p)\le C_{\kappa}(1+|x|) (p'-p)^{1-\kappa}.
        \]
\end{enumerate}
\end{lem}
\begin{proof}
    \begin{enumerate}
        \item Let $\alpha\in\cA$.
    By definition of $\hat g$, $\{P^{p,\alpha}_{T-t}\not\in[0,1]\} \subset \{Z^{t,x}_{T-t}\hat g\big(X^{t,x}_{T-t},P^{p,\alpha}_{T-t}\big) = +\infty\}$. Hence, if $\pro{P^{p,\alpha}_{T-t}\not\in[0,1]}>0$, then
    \(
        \esp{Z^{t,x}_{T-t}\hat g\big(X^{t,x}_{T-t},P^{p,\alpha}_{T-t}\big)}=+\infty.
    \)
    Therefore, taking the infimum over $\cA$ amounts to restricting to those controls such that $P^{p,\alpha}_{T-t}\in[0,1]$, a.s.

    Finally, if $P^{p,\alpha}_{T-t}\in[0,1]$ a.s., then for each $0\le s\le T-t$ we have
    \(
        P^{p,\alpha}_s=\espc{s}{P^{p,\alpha}_{T-t}}\in[0,1]\ \text{a.s.}
    \)
    so such controls belong to $\cA_{t,p}$.

    Second, denote $w(t,x,p) := \inf_{\alpha \in \cA_p} \esp{Z^{t,x}_{T-t} g(X^{t,x}_{T-t}) P^{p,\alpha}_{T-t}}$. We have $v(t,x,p) \le w(t,x,p)$ as $\cA_p \subset \cA_{t,p}$. Conversely, for any $\alpha \in \cA_{t,p}$, define $\tilde\alpha := \alpha \1_{[0,T-t]}$, so that $\tilde\alpha \in \cA_p$. We then have
    \begin{align*}
        w(t,x,p) \le \esp{Z^{t,x}_{T-t} g(X^{t,x}_{T-t}) P^{p,\tilde\alpha}_{T-t}} = \esp{Z^{t,x}_{T-t} g(X^{t,x}_{T-t}) P^{p,\alpha}_{T-t}},
    \end{align*}
    and taking the infimum over $\alpha \in \cA_{t,p}$ yields $w(t,x,p) \le v(t,x,p)$, hence the result.

    \item Let $0 \le p \le p' \le 1$ and let $\alpha \in \cA_{p'}$. Since $P^{p',\alpha} \in [0,1]$, we have $P^{p,\alpha} \in [p-p', 1+p-p'] \subset (-\infty,1]$. Consider $\tau := \inf\{s \ge 0 \mid P^{p,\alpha}_s \le 0\}$, and define $\tilde\alpha := \alpha \1_{[0,\tau)}$, so that $P^{p,\tilde\alpha} \in [0,1]$, i.e. $\tilde\alpha \in \cA_p$. 
    
    Observe that $P^{p,\tilde\alpha} \le P^{p',\alpha}$ a.s.: for $s < \tau$, one has $P^{p',\alpha}_s - P^{p,\alpha}_s = p'-p+\int_0^s (\alpha-\tilde\alpha)_u dW_u = p'-p \ge 0$ as $\tilde\alpha_u = \alpha_u$ for $u \le s < \tau$. On $\{\tau < T-t\}$, one has $P^{p,\tilde\alpha}_s = P^{p,\tilde\alpha}_{\tau} = 0 \le P^{p',\alpha}$ for $\tau \le s \le T-t$. All in one, $P^{p,\tilde\alpha} \le P^{p',\alpha}$ a.s..

    We now compute
    \begin{align*}
        v(t,x,p) \le \inf_{\alpha \in \cA_{p'}} \esp{Z^{t,x}_{T-t} g(X^{t,x}_{T-t}) P^{p,\tilde\alpha}_{T-t}} \le \inf_{\alpha \in \cA_{p'}} \esp{Z^{t,x}_{T-t} g(X^{t,x}_{T-t}) P^{p',\alpha}_{T-t}} = v(t,x,p'),
    \end{align*}
    hence the result.
    \item Let $(t,x)\in[0,T]\times\bR^d$, let $p^1,p^2\in[0,1]$ and $\lambda\in[0,1]$. We set $p^3=\lambda p^1 + (1-\lambda) p^2$.
    For any $(\alpha^1,\alpha^2)\in\cA^2$, set $\alpha^3=\lambda \alpha^1 + (1-\lambda) \alpha^2$ and we have, by linearity of the dynamics of $P$,
    \(
        P^{p^3,\alpha^3} = \lambda P^{p^1,\alpha^1} + (1-\lambda) P^{p^2,\alpha^2}.
    \)
    One has, by convexity of $\hat g$,
    \begin{align}
        v(t,x,p^3) &\le \esp{Z^{t,x}_{T-t}\hat g(X^{t,x}_{T-t},P^{p^3,\alpha^3}_{T-t})} \\ 
    &=  \esp{Z^{t,x}_{T-t}
    \hat g(X^{t,x}_{T-t}, \lambda P^{p^1,\alpha^1}_{T-t} + (1-\lambda) P^{p^2,\alpha^2}_{T-t})} \\
    &\le \esp{Z^{t,x}_{T-t}\left(\lambda \hat g(X^{t,x}_{T-t}, P^{p^1,\alpha^1}_{T-t}) + (1-\lambda) \hat g(X^{t,x}_{T-t},P^{p^2,\alpha^2}_{T-t})\right)} \\ 
    &= \lambda \esp{Z^{t,x}_{T-t}\hat g(X^{t,x}_{T-t},P^{p^1,\alpha^1}_{T-t})} + (1-\lambda) \esp{Z^{t,x}_{T-t}\hat g(X^{t,x}_{T-t},P^{p^2,\alpha^2}_{T-t})}.
    \end{align}
    Taking the infimum over $(\alpha^1,\alpha^2) \in \cA^2$ proves $v(t,x,p^3) \le \lambda v(t,x,p^1)+(1-\lambda)v(t,x,p^2)$, hence the result.
    \item Let $\kappa\in(0,1)$. Choose $r>\frac{1}{\kappa}$ and set $\theta:=\frac{r-1}{r}$, so that $\theta\ge 1-\kappa$. Let $p, p' \in [0,1]$ such that $p \le p'$. We will first deal with the case $p'=1$, and then use convexity as above to propagate the H\"older continuity to $[0,1]$.
    
    \smallskip
    \noindent Assume first $p'=1$, and observe that  $\cA_{1}=\{0\}$ (a martingale in $[0,1]$ starting from $1$ must be identically equal to $1$), hence
    \[
        v(t,x,1)=\esp{Z^{t,x}_{T-t} g(X^{t,x}_{T-t})}.
    \]
    Then
    \begin{align*}
        0 \le v(t,x,1)-v(t,x,p)
        &= \esp{Z^{t,x}_{T-t} g(X^{t,x}_{T-t})}-\inf_{\alpha\in\cA_p}\esp{Z^{t,x}_{T-t} g(X^{t,x}_{T-t}) P^{p,\alpha}_{T-t}} \\
        &= \sup_{\alpha\in\cA_p}\esp{Z^{t,x}_{T-t} g(X^{t,x}_{T-t})\big(1-P^{p,\alpha}_{T-t}\big)}.
    \end{align*}
    Fix $\alpha\in\cA_p$. By H\"older's inequality and Lemma \ref{lem:props-XZQ}, 
    \begin{align*}
        \esp{Z^{t,x}_{T-t} g(X^{t,x}_{T-t})\big(1-P^{p,\alpha}_{T-t}\big)}
        &\le \esp{\big(Z^{t,x}_{T-t} g(X^{t,x}_{T-t})\big)^r}^{1/r}\,\esp{\big(1-P^{p,\alpha}_{T-t}\big)^{\frac{r}{r-1}}}^{\frac{r-1}{r}} \\
        &\le C(1+|x|)\,\esp{\big(1-P^{p,\alpha}_{T-t}\big)}^{\theta} \\
        &= C(1+|x|) (1-p)^\theta \le C_\kappa(1+|x|) (1-p)^{1-\kappa},
    \end{align*}
    where we used that since $\frac{r}{r-1}\ge 1$ and $0\le 1-P^{p,\alpha}_{T-t}\le 1$, we have $\big(1-P^{p,\alpha}_{T-t}\big)^{\frac{r}{r-1}}\le 1-P^{p,\alpha}_{T-t}$, and that $P^{p,\alpha}$ is a martingale.
    Therefore, taking the supremum over $\alpha\in\cA_p$ yields
    \[
        0\le v(t,x,1)-v(t,x,p)\le C_\kappa(1+|x|)(1-p)^{\theta}.
    \]

    \smallskip
    \noindent For the general case $0\le p<p'\le 1$, we have as for 1., with $\lambda:=\frac{p'-p}{1-p}\in[0,1]$,
    \[
        0\le v(t,x,p')-v(t,x,p)\le \lambda\big(v(t,x,1)-v(t,x,p)\big)
        \le \frac{p'-p}{1-p}\,C(1+|x|)(1-p)^{\theta} = C(1+|x|) (p'-p) (1-p)^{\theta-1}.
    \]
    Since $p'-p\le 1-p$ (because $p'\le 1$) and $\theta-1=-1/r<0$, we have $(1-p)^{\theta-1}\le (p'-p)^{\theta-1}$. Therefore
    \[
        0\le v(t,x,p')-v(t,x,p)\le C(1+|x|)|p'-p|^{\theta} \le C(1+|x|)|p'-p|^{1-\kappa},
    \]
    which concludes the proof.
\end{enumerate}
\end{proof}
\begin{remark}
    Note that $\cA_p$ is not empty if $p \in [0,1]$, as for instance, the null control belongs to $\cA_p$, (hence in particular $v(t,x,p) \le p \esp{Z^{t,x}_{T-t} g(X^{t,x}_{T-t})}$). In addition, given any $\alpha \in \cA$ and $p \in [0,1]$, defining $\tau := \inf\{s \ge 0 \mid P^{p,\alpha}_s \not\in [0,1]\}$ gives $\tilde\alpha := \alpha \1_{[0,\tau)} \in \cA_p$.
\end{remark}

\begin{remark}\label{rem:alpha-L2-bound}
    If $\alpha \in \cA_p$, we note, by It\^o's isometry, that, since $P^{p,\alpha}$ is a martingale in $[0,1]$, we have
    \begin{align}
        \esp{\int_0^{T-t} |\alpha_s|^2 ds} = \esp{\left(\int_0^{T-t}\alpha_s^\top dW_s\right)^2} = \esp{ \left(P^{p,\alpha}_{T-t} -p\right)^2} \le p(1-p) \le \frac14.
    \end{align}
\end{remark}

\begin{remark}
    The above argument yields global Hölder continuity under the available
moment assumptions. If
\(Z^{t,x}_{T-t}g(X^{t,x}_{T-t})\) is essentially bounded, the same
argument yields global Lipschitz continuity in \(p\).%

\end{remark}

\subsection{\texorpdfstring{$\varepsilon$}{epsilon}-optimal smooth feedback controls}\label{sec:eps-opt}

In order to perform differentiations with respect to the initial condition $x$ and justify feedback formulas, we will work locally with smooth space--time mollifications of the coefficients. We now introduce this regularisation explicitly.

Let $\rho\in\cC_c^\infty(\bR\times\bR^d)$ be a nonnegative mollifier supported in $(-1,0]\times\{x\in\bR^d\mid |x|\le 1\}$ such that $\int_{\bR\times\bR^d} \rho(r,y)\,drdy=1$. For $\varepsilon\in(0,1]$ we define
\[
    \rho_\varepsilon(r,y):=\varepsilon^{-(d+2)}\,\rho\Big(\frac{r}{\varepsilon^2},\frac{y}{\varepsilon}\Big),
\]

and we then define the regularisation of any coefficient $f\in\{\mu,\sigma,\lambda,b\}$ by
\[
    \tilde f(t,x) :=\int_{\bR}\!\int_{\bR^d} f(t-r,x-y)\,\rho_\varepsilon(r,y)\,dydr,\qquad (t,x)\in\bR\times\bR^d.
\]

\begin{lem}\label{lem:mollified-coeffs-props}
    Under Assumption~\ref{h:coeffs}, for each $\varepsilon\in(0,1]$, the coefficients $(\tilde\mu,\tilde\sigma,\tilde\lambda,\tilde b)$ are $\cC^\infty$ in $(t,x)$, satisfy the same global Lipschitz continuity in space, Hölder continuity in time and linear growth bounds as in Assumption~\ref{h:coeffs} (with constants independent of $\varepsilon$), and there exists a constant $C_0\ge0$ (depending only on the constant $L$ in Assumption~\ref{h:coeffs}, $\int_{\bR\times\bR^d} |y|\rho(r,y)\,drdy$ and $\int_{\bR\times\bR^d} |r|^{1/2}\rho(r,y)\,drdy$) such that
    \[
        |\mu(t,x)-\tilde\mu(t,x)| + \|\sigma(t,x)-\tilde\sigma(t,x)\| + |\lambda(t,x)-\tilde\lambda(t,x)| + |b(t,x)-\tilde b(t,x)| \le C_0\varepsilon
    \]
    for all $(t,x)\in\bR\times\bR^d$.

    Moreover, $|\tilde\lambda(t,x)|\le \bar\lambda$ and $|\tilde b(t,x)|\le \bar b$ for all $(t,x)$.
\end{lem}
\begin{proof}
    The $\cC^\infty$ regularity in $(t,x)$ follows from standard properties of convolution with smooth compactly supported kernels.

    Let $f$ be any of $\mu,\sigma,\lambda,b$. For any $t \in \bR$ and $x,x'\in\bR^d$,
    \[
        |\tilde f(t,x)-\tilde f(t,x')|
        \le \int_{\bR}\!\int_{\bR^d} |f(t-r,x-y)-f(t-r,x'-y)|\,\rho_\varepsilon(r,y)\,dydr
        \le L|x-x'|\int_{\bR}\!\int_{\bR^d}\rho_\varepsilon(r,y)\,dydr
        =L|x-x'|,
    \]
    hence the Lipschitz constant is preserved.

    Similarly, for $s,t\in\bR$ and $x\in\bR^d$,
    \[
        |\tilde f(t,x)-\tilde f(s,x)|
        \le \int_{\bR}\!\int_{\bR^d} |f(t-r,x-y)-f(s-r,x-y)|\,\rho_\varepsilon(r,y)\,dydr
        \le L|t-s|^{1/2},
    \]
    so the $1/2$-H\"older constant in time is preserved.

    For the linear growth bound, for $f \in \{\mu,\sigma\}$, using $\mathrm{supp}(\rho_\varepsilon)\subset (-\varepsilon^2,0]\times\{|y|\le \varepsilon\}$ and $|x-y|\le |x|+|y|$,
    \begin{align*}
        |\tilde f(t,x)|
        &\le \int_{\bR}\!\int_{\bR^d} |f(t-r,x-y)|\,\rho_\varepsilon(r,y)\,dydr \\
        &\le L\int_{\bR}\!\int_{\bR^d} (1+|x-y|)\,\rho_\varepsilon(r,y)\,dydr \\
        &\le L\Big(1+|x|+\varepsilon\int_{\bR\times\bR^d} |y|\,\rho(r,y)\,drdy\Big) \\
        &\le C(1+|x|),
    \end{align*}
    with a constant $C$ independent of $\varepsilon\in(0,1]$.

    Finally, by the Lipschitz property of $f$,
    \begin{align*}
        |f(t,x)-\tilde f(t,x)|
        &=\left|\int_{\bR}\!\int_{\bR^d} \big(f(t,x)-f(t-r,x-y)\big)\,\rho_\varepsilon(r,y)\,dydr\right| \\
        &\le \int_{\bR}\!\int_{\bR^d} \Big(|f(t,x)-f(t,x-y)| + |f(t,x-y)-f(t-r,x-y)|\Big)\,\rho_\varepsilon(r,y)\,dydr \\
        &\le L\int_{\bR\times\bR^d} |y|\,\rho_\varepsilon(r,y)\,drdy + L\int_{\bR\times\bR^d} |r|^{1/2}\,\rho_\varepsilon(r,y)\,drdy \\
        &= L\varepsilon\int_{\bR\times\bR^d}|y|\,\rho(r,y)\,drdy + L\varepsilon\int_{\bR\times\bR^d}|r|^{1/2}\,\rho(r,y)\,drdy,
    \end{align*}
    which yields the stated $O(\varepsilon)$ bound (and similarly for the matrix norm in the case of $\sigma$).

    Finally, since $|b|\le \bar b$ and $|\lambda|\le \bar\lambda$ by Assumption~\ref{h:coeffs}, the convolution representation yields
    \[
        |\tilde b(t,x)|\le \int_{\bR}\!\int_{\bR^d} |b(t-r,x-y)|\,\rho_\varepsilon(r,y)\,dydr \le \bar b,\qquad
        |\tilde\lambda(t,x)|\le \int_{\bR}\!\int_{\bR^d} |\lambda(t-r,x-y)|\,\rho_\varepsilon(r,y)\,dydr \le \bar\lambda.
    \]
    We are done.
\end{proof}
We denote by $(\tilde X^{t,x},\tilde Z^{t,x})$ the corresponding dynamics, we denote $\tilde Q^{t,x} := \frac{1}{\tilde Z^{t,x}}$ and we define the associated value function

\begin{align}\label{eq:control-pb-mollified}
    \tilde v(t,x,p) &:= \inf_{\alpha\in\cA}\;\esp{\tilde Z^{t,x}_{T-t}\,\hat g(\tilde X^{t,x}_{T-t}, P^{p,\alpha}_{T-t})} \\
    &= \inf_{\alpha\in\cA_p} \esp{\tilde Z^{t,x}_{T-t}\,g(\tilde X^{t,x}_{T-t})P^{p,\alpha}_{T-t}}, \quad (t,x,p) \in [0,T]\times\bR^d\times\bR.
\end{align}

\begin{remark}\label{rem:mollified-pb}
By the previous lemma, the mollified coefficients satisfy the same assumptions as the original ones, with constants independent of $\varepsilon$, so that all results of Section~\ref{sec:problem} apply to $\tilde v$ as well.
In particular, $\tilde v$ is convex in $p$ and satisfies the same regularity properties as $v$ (with constants independent of $\varepsilon$). One also has moment bounds (also independent of $\varepsilon$), mirroring those of Lemma~\ref{lem:props-XZQ} for the original dynamics.
\end{remark}

We have the following two results. The first is a stability result, which shows that the mollified dynamics are close to the original ones in $L^p$ sense. The second one shows that, for fixed $0 \le t \le T$ and $0 \le s \le T-t$, the map
$x\mapsto(\tilde X^{t,x}_s,\tilde Q^{t,x}_s)$ is almost surely differentiable and provides estimates for its spatial
derivatives. The proofs for these two results are standard and postponed to Appendix~\ref{sec:proof-lem-Lp-stability-mollified} and \ref{sec:proof-lem-bounds-JH}.

\begin{lem}\label{lem:Lp-stability-mollified}
    Under Assumption~\ref{h:coeffs}, for any $p \ge 1$, there exists a constant $C_p\ge 0$ (independent of $\varepsilon\in(0,1]$) such that for all $(t,x)\in[0,T]\times\bR^d$,
    \[
        \sup_{s\in[0,T-t]}\esp{\big|X^{t,x}_s-\tilde X^{t,x}_s\big|^{2p}}
        + \sup_{s\in[0,T-t]}\esp{\big|Z^{t,x}_s-\tilde Z^{t,x}_s\big|^{2p}}
        \le C_p\,\varepsilon^{2p}.
    \]
\end{lem}

\begin{lem}\label{lem:bounds-JH}
    Under Assumption~\ref{h:coeffs}, for $\varepsilon \in (0,1]$ and for any $t\in[0,T]$ and $s\in[0,T-t]$, the map $x\mapsto(\tilde X^{t,x}_s,\tilde Q^{t,x}_s)$ is almost surely differentiable, and 
\[
\tilde J^{t,x}_s := \partial_x \tilde X^{t,x}_s,\qquad \tilde H^{t,x}_s := \partial_x \tilde Q^{t,x}_s,
\]
solve the following SDEs:
\begin{align} \label{eq:sde deriv}
    d\tilde J^{t,x}_s &= \partial_x \tilde\mu(t+s,\tilde X^{t,x}_s)\, \tilde J^{t,x}_s\, ds + \sum_{k=1}^q \partial_x \tilde\sigma^{(k)}(t+s,\tilde X^{t,x}_s)\, \tilde J^{t,x}_s\, d(W_s)^k, \\
    d\tilde H^{t,x}_s &= \Bigl(\tilde H^{t,x}_s \big(|\tilde\lambda(t+s,\tilde X^{t,x}_s)|^2-\tilde b(t+s,\tilde X^{t,x}_s)\big) \\ &\qquad+ \tilde Q^{t,x}_s\, (\tilde J^{t,x}_s)^\top \big(2\, \partial_x \tilde\lambda(t+s,\tilde X^{t,x}_s)^\top \tilde\lambda(t+s,\tilde X^{t,x}_s)-\partial_x \tilde b(t+s,\tilde X^{t,x}_s)\big)\Bigr) ds \\
    &\quad -\Big(\tilde Q^{t,x}_s\, (\tilde J^{t,x}_s)^\top (\partial_x \tilde\lambda(t+s,\tilde X^{t,x}_s))^\top + \tilde H^{t,x}_s\, \tilde\lambda(t+s,\tilde X^{t,x}_s)^\top\Big) dW_s, \\
    \tilde J^{t,x}_0 &= I_d,\qquad \tilde H^{t,x}_0 = 0.
\end{align}
In addition, there exists a constant $C\ge 0$ such that for all $(t,x)\in[0,T]\times\bR^d$,
    \[
        \esp{\|\tilde J^{t,x}_{T-t}\|} \le C, \quad \text{and} \quad
        \esp{|\tilde H^{t,x}_{T-t}|} \le C.
    \]
\end{lem}

Using the stability result of Lemma~\ref{lem:Lp-stability-mollified}, we can now show that the mollified value function $\tilde v$ is close to the original value function $v$.

\begin{proposition}\label{prop:stability-v-mollified}
    Under Assumptions~\ref{h:coeffs} and \ref{h:g Lip}, there exists a constant $C\ge 0$ (independent of $\varepsilon$) such that for all $(t,x,p)\in[0,T]\times\bR^d\times[0,1]$,
    \[
        |v(t,x,p)-\tilde v(t,x,p)| \le C\,(1+|x|)\,\varepsilon.
    \]
\end{proposition}
\begin{proof}
    We have, since $P^{p,\alpha}_{T-t}\in[0,1]$ a.s. for any $\alpha\in\cA_p$,
    \begin{align*}
        |v(t,x,p)-\tilde v(t,x,p)| &= \left| \inf_{\alpha \in \cA_p} \esp{Z^{t,x}_{T-t} g(X^{t,x}_{T-t})P^{p,\alpha}_{T-t}} - \inf_{\alpha \in \cA_p} \esp{\tilde Z^{t,x}_{T-t} g(\tilde X^{t,x}_{T-t})P^{p,\alpha}_{T-t}} \right| \\ &\le \sup_{\alpha \in \cA_p} \left|\esp{Z^{t,x}_{T-t} g(X^{t,x}_{T-t})P^{p,\alpha}_{T-t}}-\esp{\tilde Z^{t,x}_{T-t} g(\tilde X^{t,x}_{T-t})P^{p,\alpha}_{T-t}}\right| \\
        &\le \sup_{\alpha \in \cA_p}\esp{\left|Z^{t,x}_{T-t} g(X^{t,x}_{T-t})-\tilde Z^{t,x}_{T-t} g(\tilde X^{t,x}_{T-t})\right|P^{p,\alpha}_{T-t}} \\
        &\le \esp{\left|Z^{t,x}_{T-t} g(X^{t,x}_{T-t})-\tilde Z^{t,x}_{T-t} g(\tilde X^{t,x}_{T-t})\right|}.
    \end{align*}
    Using the Lipschitz property of $g$ in Assumption~\ref{h:g Lip}, we have, by Cauchy--Schwarz inequality, Lemma \ref{lem:props-XZQ} and Lemma~\ref{lem:Lp-stability-mollified} (with $p=1$),
    \begin{align*}
        |v(t,x,p)-\tilde v(t,x,p)|
        &\le \esp{|Z_{T-t}-\tilde Z_{T-t}|\,|g(X_{T-t})| + \tilde Z_{T-t}\,|g(X_{T-t})-g(\tilde X_{T-t})|}
        \\ &\le C\esp{|Z_{T-t}-\tilde Z_{T-t}|(1+|X_{T-t}|) + \tilde Z_{T-t}\,|X_{T-t}-\tilde X_{T-t}|} \\
        &\le C\,\esp{\big|Z_{T-t}-\tilde Z_{T-t}\big|^2}^{\frac12}\,\esp{(1+|X_{T-t}|)^2}^{\frac12}
        + C\,\esp{\tilde Z_{T-t}^2}^{\frac12}\,\esp{\big|X_{T-t}-\tilde X_{T-t}\big|^2}^{\frac12} \\
        &\le C (1+|x|) \varepsilon,
    \end{align*}
    which concludes the proof.
\end{proof}

\medskip\noindent
In the next steps, we exploit the fact that the value function $\tilde v$ is convex in its last variable and, by Fenchel-Legendre duality, we construct an explicit feedback control that is $C\varepsilon$-optimal for the mollified value function $\tilde v$.
Proposition~\ref{prop:stability-v-mollified} then implies that the same control is $C'\varepsilon$-optimal for the original value function $v$, for some other constant $C' \ge 0$.

\medskip\noindent We start by computing the Fenchel transform of $v$ and $\tilde v$ with respect to the last variable.
\begin{lem}\label{lem:calcul v sharp}
    For $(t,x) \in [0,T]\times\bR^d$, consider the Fenchel transform of $v$, denoted $v^\sharp(t,x,\cdot)$ and defined by
    \begin{align}
        v^\sharp(t,x,q) &:= \sup_{p \in \bR} pq - v(t,x,p) \\
        &= \sup_{p \in [0,1]} pq - v(t,x,p).
    \end{align}
    We then have
    \begin{align}
        v^\sharp(t,x,q) = \esp{Z^{t,x}_{T-t} \hat g^\sharp(X^{t,x}_{T-t},qQ^{t,x}_{T-t})},
    \end{align}
    with, for any $(x,q) \in \bR^d \times \bR$,
    \begin{align}\label{eq: g^sharp}
        \hat g^\sharp(x,q) = (q-g(x))^+.
    \end{align}
    One similarly has $\tilde v^\sharp(t,x,q) = \esp{\tilde Z^{t,x}_{T-t} \hat g^\sharp(\tilde X^{t,x}_{T-t},q\tilde Q^{t,x}_{T-t})}$ for all $(t,x,q) \in [0,T]\times\bR^d\times\bR$.
\end{lem}
\begin{proof}
    Let $(t,x,q) \in [0,T] \times \bR^d \times \bR$. First, the defining equality of $v^\sharp$ is justified by the fact that $v(t,x,p) = +\infty$ whenever $p \not\in [0,1]$.

    Second, observing that $p = \esp{P^{p,\alpha}_{T-t}} = \esp{Z^{t,x}_{T-t} Q^{t,x}_{T-t} P^{p,\alpha}_{T-t}}$ for any $p \in [0,1]$ and $\alpha \in \cA$, we compute
    \begin{align}
        v^\sharp(t,x,q) &= \sup_{p \in [0,1]} pq - \inf_{\alpha \in \cA} \esp{Z^{t,x}_{T-t}  \hat g(X^{t,x}_{T-t},P^{p,\alpha}_{T-t})} \\
        &= \sup_{p \in [0,1],\alpha \in \cA} pq - \esp{Z^{t,x}_{T-t} \hat g(X^{t,x}_{T-t},P^{p,\alpha}_{T-t})} \\
        &= \sup_{p \in [0,1], \alpha \in \cA} \esp{Z^{t,x}_{T-t} \left(q Q^{t,x}_{T-t} P^{p,\alpha}_{T-t} - \hat g(X^{t,x}_{T-t},P^{p,\alpha}_{T-t})\right)} \\
        &\le \esp{Z^{t,x}_{T-t} \hat g^\sharp(X^{t,x}_{T-t},qQ^{t,x}_{T-t})}.
    \end{align}
    Finally, conversely, since
    \begin{align}
        \hat g^\sharp(x,q) = \sup_{p \in \bR} pq - \hat g(x,p) = \sup_{p \in [0,1]} pq - g(x)p = (q-g(x))^+,
    \end{align}
    we consider $\xi = \1_{\{qQ^{t,x}_{T-t} > g(X^{t,x}_{T-t})\}}$. In particular,
    \begin{align*}
        \esp{Z^{t,x}_{T-t} \hat g^\sharp(X^{t,x}_{T-t},qQ^{t,x}_{T-t})}
        &= \esp{Z^{t,x}_{T-t} \left(q Q^{t,x}_{T-t} - g(X^{t,x}_{T-t}) \right)^+} \\
        &= \esp{Z^{t,x}_{T-t} \left(q Q^{t,x}_{T-t} \xi - g(X^{t,x}_{T-t}) \xi\right)}.
    \end{align*}
    Since $\xi \in L^2(\cF_{T-t})$, by the martingale representation Theorem, there exists a predictable process $\alpha$ on $[0,T-t]$ such that $\espc{s}{\xi} = \esp{\xi} + \int_0^s \alpha_u^\top dW_u$ for all $s\in[0,T-t]$. Extending $\alpha$ by $0$ after $T-t$, we obtain $\alpha \in \cA_p$ and $\xi = P^{p,\alpha}_{T-t}$ with $p := \esp{\xi}$. We thus have
    \begin{align}
        v^\sharp(t,x,q) &\ge \esp{Z^{t,x}_{T-t} \left(q Q^{t,x}_{T-t} P^{p,\alpha}_{T-t} - \hat g(X^{t,x}_{T-t},P^{p,\alpha}_{T-t})\right)} \\ &= \esp{Z^{t,x}_{T-t} \left(q Q^{t,x}_{T-t} \xi - \hat g(X^{t,x}_{T-t},\xi)\right)} \\ &= \esp{Z^{t,x}_{T-t} \left(q Q^{t,x}_{T-t} \xi - g(X^{t,x}_{T-t})\xi\right)} \\ &= \esp{Z^{t,x}_{T-t} \hat g^\sharp(X^{t,x}_{T-t},qQ^{t,x}_{T-t})},
    \end{align}
    which concludes the computation for $v$, while the one for $\tilde v$ is identical.
\end{proof}

Now, in order to construct an $\varepsilon$-optimal feedback control, we need to differentiate, after regularization, the Fenchel transform $v^\sharp$ with respect to its last variable.

\medskip\noindent We let $\varpi \in \cC^\infty_c(\bR^d \times \bR)$ be a nonnegative mollifier with unit mass, supported in $\{x \in \bR^d \,|\, |x| \le 1\} \times (-1,1)$, and for any $\varepsilon \in (0,1]$, we define $\varpi_\varepsilon(x,q) := \frac{1}{\varepsilon^{d+1}}\varpi(\frac{x}{\varepsilon},\frac{q}{\varepsilon})$.

We now fix $\varepsilon\in(0,1]$, and we set $g^\sharp_\varepsilon := \hat g^\sharp \star \varpi_\varepsilon$, and define, for $(t,x,q) \in [0,T] \times \bR^d \times \bR$, 
\[
    w_\varepsilon(t,x,q) := \esp{\tilde Z^{t,x}_{T-t}g^\sharp_\varepsilon(\tilde X^{t,x}_{T-t},q\tilde Q^{t,x}_{T-t})}.
\]

For fixed \(\varepsilon>0\), the smoothness of the mollified stochastic flow and differentiation under the expectation show that \(w_\varepsilon(t,\cdot,\cdot)\) is smooth for every \(t\in[0,T]\).

We first show that $w_\varepsilon$ is close to $\tilde v^\sharp$. We then collect elementary results about $\partial_q w_\varepsilon$ that will turn out useful in the sequel.

\begin{lem}\label{lem:dist fenchel}
    Under Assumptions \ref{h:coeffs} and \ref{h:g Lip}, for all $(t,x,q) \in [0,T]\times\bR^d\times\bR$, we have
    \[
        |(\tilde v^\sharp-w_\varepsilon)(t,x,q)| \le (1+L) e^{\bar b T}\,\varepsilon.
    \]
\end{lem}
\begin{proof}
    First, we easily compute that $\hat g^\sharp(x,q)=(q-g(x))^+$ is $1$-Lipschitz continuous in $q$ and $L$-Lipschitz continuous in $x$ (Assumption~\ref{h:g Lip}), hence $(1+L)$-Lipschitz on $\bR^d\times\bR$.
    Since $\varpi_\varepsilon$ is supported on $\{x \in \bR^d \,|\, |x| \le \varepsilon\}\times(-\varepsilon,\varepsilon)$, for $(x,q) \in \bR^d\times\bR$,
    \begin{align}
        |(\hat g^\sharp-g^\sharp_\varepsilon)(x,q)|
        &= \left|\hat g^\sharp(x,q) - \int_{\bR^d\times\bR} \hat g^\sharp(x-y,q-r) \varpi_\varepsilon(y,r)\,dydr\right| \\
        &\le \int_{\bR^d\times\bR} |\hat g^\sharp(x,q)-\hat g^\sharp(x-y,q-r)|\,\varpi_\varepsilon(y,r)\,dydr
        \le (1+L)\varepsilon.
    \end{align}
    Let $(t,x,q) \in [0,T]\times\bR^d\times\bR$. By Lemma \ref{lem:calcul v sharp} (applied to the mollified coefficients) and Assumption~\ref{h:coeffs}, we have $\esp{\tilde Z^{t,x}_{T-t}}\le e^{\bar b (T-t)}\le e^{\bar b T}$, hence
    \begin{align}
        |(\tilde v^\sharp-w_\varepsilon)(t,x,q)| \le \esp{\tilde Z^{t,x}_{T-t}|(\hat g^\sharp-g^\sharp_\varepsilon)(\tilde X^{t,x}_{T-t},q\tilde Q^{t,x}_{T-t})|} \le (1+L) e^{\bar b T}\,\varepsilon.
    \end{align}
\end{proof}

\begin{lem} \label{lem:w_e infos}
    Under Assumptions \ref{h:coeffs} and \ref{h:g Lip}, one has the following properties for $w_\varepsilon$.
    \begin{enumerate}
    \item For any $(t,x,q) \in [0,T] \times \bR^d \times \bR$, we have 
    \begin{align}\label{eq :partial_q w_e}
        \partial_q w_\varepsilon(t,x,q) = \esp{\partial_q g^\sharp_\varepsilon(\tilde X^{t,x}_{T-t},q\tilde Q^{t,x}_{T-t})}.
    \end{align}
    In addition, the map \(q\mapsto\partial_q w_\varepsilon(t,x,q)\) is smooth and nondecreasing, with limits \(0\) at \(-\infty\) and \(1\) at \(+\infty\).%
    
    In particular, for any $(t,x,p) \in [0,T)\times\bR^d\times(0,1)$, there exists $q^\star_\varepsilon=q^\star_\varepsilon(t,x,p)$ such that $p=\partial_q w_\varepsilon(t,x,q^\star_\varepsilon)$.
    
    We further have $\partial_q w_\varepsilon \in \cC^{1,\infty,\infty}([0,T)\times\bR^d\times\bR)$ and, for all $(t,x,q) \in [0,T]\times\bR^d\times\bR$,
    \begin{align}
        \partial^2_{qq} w_\varepsilon(t,x,q) &= \esp{\tilde Q^{t,x}_{T-t} \partial^2_{qq}g^\sharp_\varepsilon(\tilde X^{t,x}_{T-t},q\tilde Q^{t,x}_{T-t})}, \text{ and}\\
        \partial^2_{xq} w_\varepsilon(t,x,q) &= \esp{(\partial_x \tilde X^{t,x}_{T-t})^\top \partial^2_{xq} g^\sharp_\varepsilon(\tilde X^{t,x}_{T-t},q\tilde Q^{t,x}_{T-t}) + q \partial_x \tilde Q^{t,x}_{T-t} \partial^2_{qq} g^\sharp_\varepsilon(\tilde X^{t,x}_{T-t},q\tilde Q^{t,x}_{T-t})},
    \end{align}
    Moreover, there exists $C\ge0$ such that, for all $(t,x,q) \in [0,T]\times\bR^d\times\bR$,
    \begin{align}
        |\partial^2_{qq}w_\varepsilon(t,x,q)| &\le C \,\varepsilon^{-1}, \text{ and } \\
        |\partial^2_{xq}w_\varepsilon(t,x,q)| &\le C\,(1+|q|)\,\varepsilon^{-1}.
    \end{align}
    \item Let $(t,x,p)\in[0,T)\times\bR^d\times(0,1)$. Then, for every $m\ge 1$, there exists a constant $C_m\ge 0$ (independent of $\varepsilon$) such that
    \begin{align*}
        -C_m^{\frac 1 m} \varepsilon p^{-\frac 1 m} \le q^\star_\varepsilon \le C_m^{\frac 1 m}(1+|x|+\varepsilon)(1-p)^{-\frac 1 m}.
    \end{align*}
    In particular,
    \begin{align}
        \label{eq:q_star_bound}&|q^\star_\varepsilon|\le C_m\,(1+|x|)\,(p\wedge(1-p))^{-\frac 1 m}, \\
        \label{eq:partial_xq_w_bound} &|\partial^2_{xq} w_\varepsilon(t,x,q^\star_\varepsilon)| \le C_m\,(1+|x|)\,(p\wedge(1-p))^{-\frac{1}{m}}\varepsilon^{-1}.
    \end{align}
\end{enumerate}
\end{lem}
\begin{proof}
\begin{enumerate}
\item The announced identities follow from the dominated convergence theorem and the chain rule. The stated estimates are classical bounds for derivatives of convolutions of Lipschitz-continuous functions with mollifiers, combined with the moment estimates for $\tilde X^{t,x}_{T-t}$, $\tilde Q^{t,x}_{T-t}$ and their spatial derivatives (see Lemma~\ref{lem:bounds-JH}). Note that from \eqref{eq :partial_q w_e} and the flow property, one deduces the time-regularity of $\partial_q w_\varepsilon$ by applying It\^o's formula over $[0,h]$ to the time-independent function $\partial_q w_\varepsilon(t+h,\cdot,\cdot)$, dividing by $h$ and letting $h\downarrow0$.
\item  %
We prove the bounds on $q^\star_\varepsilon$, the bound on $\partial^2_{xq} w_\varepsilon(t,x,q^\star_\varepsilon)$ then follows from the bound on $q^\star_\varepsilon$ together with the general bound on $\partial^2_{xq} w_\varepsilon$ stated in the first point.

First, using the explicit formula
\[
    \partial_q g^\sharp_\varepsilon(x,q)=\int_{\bR^d\times\bR}\1_{\{q-r>g(x-y)\}}\,\varpi_\varepsilon(y,r)\,dydr,
\]
one easily observes that
\begin{align*}
    \1_{q \ge g(x)+(L+1)\varepsilon} \le \partial_q g^\sharp_\varepsilon(x,q) \le \1_{q > g(x)-(L+1)\varepsilon}.
\end{align*}
Hence, applying the identity with $x$ replaced by $\tilde X^{t,x}_{T-t}$ and $q$ by $q\tilde Q^{t,x}_{T-t} = q (\tilde Z^{t,x}_{T-t})^{-1}$, and taking expectation, one gets
\begin{align}\label{eq:bounds dq w}
    \pro{q \ge \tilde Z^{t,x}_{T-t}g(\tilde X^{t,x}_{T-t})+\tilde Z^{t,x}_{T-t}(L+1)\varepsilon} &\le \partial_q w_\varepsilon(t,x,q) \\
    &\le \pro{q > \tilde Z^{t,x}_{T-t}g(\tilde X^{t,x}_{T-t})-\tilde Z^{t,x}_{T-t}(L+1)\varepsilon}.
\end{align}
Since $\tilde Z^{t,x}_{T-t}g(\tilde X^{t,x}_{T-t})\ge 0$, the right-hand side inequality implies, for $q<0$, 
\begin{align*}
\partial_q w_\varepsilon(t,x,q) &\le \pro{q > -\tilde Z^{t,x}_{T-t}(L+1)\varepsilon} \\ &= \pro{\tilde Z^{t,x}_{T-t} > \frac{-q}{(L+1)\varepsilon}} \\ &\le \varepsilon^m (L+1)^m \esp{(\tilde Z^{t,x}_{T-t})^m} (-q)^{-m} \\ &\le C_m \varepsilon^m (-q)^{-m}.
\end{align*} 
We deduce $q^\star_\varepsilon \ge -C_m^{\frac 1 m} \varepsilon p^{-\frac 1 m}$: the inequality is obvious if $q^\star_\varepsilon \ge 0$. Otherwise, the above computation yields $p \le C_m \varepsilon^m (-q^\star_\varepsilon)^{-m}$, i.e. $q^\star_\varepsilon \ge -C_m^{\frac 1 m} \varepsilon p^{-\frac 1 m}$.%

Coming back to \eqref{eq:bounds dq w}, the left-hand side inequality writes, using Markov's inequality, for $q>0$,
\begin{align*}
\partial_q w_\varepsilon(t,x,q) &\ge 1-\pro{q<\tilde Z^{t,x}_{T-t}g(\tilde X^{t,x}_{T-t})+(\tilde Z^{t,x}_{T-t})(L+1)\varepsilon} \\ &\ge  1 - \pro{\frac q 2 < \tilde Z^{t,x}_{T-t}g(\tilde X^{t,x}_{T-t})} - \pro{\frac q 2 < (L+1)\varepsilon \tilde Z^{t,x}_{T-t}} \\ &\ge 1 - \frac{2^m \esp{(\tilde Z^{t,x}_{T-t}g(\tilde X^{t,x}_{T-t}))^m}}{q^m} - \frac{2^m (L+1)^m \varepsilon^m \esp{(\tilde Z^{t,x}_{T-t})^m}}{q^m} \\ &\ge 1 - \frac{C_m((1+|x|)^m+\varepsilon^m)}{q^m}.
\end{align*}
We deduce $q^\star_\varepsilon \le C_m^{\frac 1 m}((1+|x|)+\varepsilon)(1-p)^{-1/m}$: the inequality is obvious if $q^\star_\varepsilon \le 0$. Otherwise, the above computation yields $1-p \le C_m ((1+|x|)^m+\varepsilon^m) (q^\star_\varepsilon)^{-m}$, i.e. $q^\star_\varepsilon \le C_m^{\frac 1 m}((1+|x|)+\varepsilon)(1-p)^{-1/m}$.%

\end{enumerate}
\end{proof}
We are now ready to prove the following result.
\begin{theorem}\label{thm:eps opti}
Under Assumptions \ref{h:coeffs} and \ref{h:g Lip}, let $(t,x,p) \in [0,T)\times\bR^d\times(0,1)$. We define
\[
    P^{t,x,p}_s := \partial_q w_\varepsilon(t+s,\tilde X^{t,x}_s,q^\star_\varepsilon \tilde Q^{t,x}_s),\qquad 0\le s\le T-t,
\]
where $q^\star_\varepsilon=q^\star_\varepsilon(t,x,p)$ is such that $\partial_q w_\varepsilon(t,x,q^\star_\varepsilon)=p$ (see Lemma~\ref{lem:w_e infos}).
Then $P^{t,x,p} = P^{p,\alpha}$ for $\alpha \in \cA_p$ defined, for $0 \le s \le T-t$, by
\begin{align}\label{eq: alpha epsilon optimal}
\alpha_s
:={}&
\tilde\sigma(t+s,\tilde X^{t,x}_s)^\top
\partial^2_{xq}w_\varepsilon
\bigl(t+s,\tilde X^{t,x}_s,
q^\star_\varepsilon\tilde Q^{t,x}_s\bigr)\\
&-
q^\star_\varepsilon\tilde Q^{t,x}_s
\partial^2_{qq}w_\varepsilon
\bigl(t+s,\tilde X^{t,x}_s,
q^\star_\varepsilon\tilde Q^{t,x}_s\bigr)
\tilde\lambda(t+s,\tilde X^{t,x}_s),
\end{align}
and by $\alpha_s=0$ for $s > T-t$. The control $\alpha$ is a $2 (1+L) e^{\bar b T}\,\varepsilon$-optimal control for the mollified problem $\tilde v(t,x,p)$, i.e.
\[
    0\le \esp{\tilde Z^{t,x}_{T-t} g(\tilde X^{t,x}_{T-t})P^{p,\alpha}_{T-t}}-\tilde v(t,x,p)\le 2 (1+L) e^{\bar b T}\,\varepsilon.
\]
In addition, $\alpha$ is a $(2(1+L) e^{\bar b T}+2C(1+|x|))\varepsilon$-optimal control for the original problem $v(t,x,p)$, where $C$ is the constant from Proposition~\ref{prop:stability-v-mollified}, i.e.
\[
    0\le \esp{Z^{t,x}_{T-t} g(X^{t,x}_{T-t})P^{p,\alpha}_{T-t}}-v(t,x,p)\le \big(2(1+L) e^{\bar b T}+2C(1+|x|)\big)\varepsilon.
\]
\end{theorem}
\begin{proof}
    Let $(t,x,p) \in [0,T)\times\bR^d\times(0,1)$.

    Observe that, for any $0 \le s \le T-t$, using Lemma \ref{lem:w_e infos}-1., we have
    \[
        P^{t,x,p}_s = \espc{s}{\partial_q g^\sharp_\varepsilon(\tilde X^{t,x}_{T-t},q^\star_\varepsilon \tilde Q^{t,x}_{T-t})},
    \]
    hence $P^{t,x,p}$ is a $[0,1]$-valued martingale. Then $P^{t,x,p}=P^{p,\alpha}$ by the martingale representation theorem, and the explicit formula for $\alpha$ in \eqref{eq: alpha epsilon optimal} straightforwardly comes from It\^o's formula applied to the process $P^{t,x,p} = \partial_q w_\varepsilon(\cdot,\tilde X^{t,x},q^\star_\varepsilon \tilde Q^{t,x})$.

    We now show that $\alpha$ is $2 (1+L) e^{\bar b T}\,\varepsilon$-optimal for $\tilde v(t,x,p)$.

    We define $g_\varepsilon(x,p) = (g^\sharp_\varepsilon)^\sharp(x,p)$ and we easily have, since $|\hat g^\sharp-g^\sharp_\varepsilon|\le (1+L)\varepsilon$ (see Lemma~\ref{lem:dist fenchel}), that
    \begin{align}
        g_\varepsilon(x,p) = \sup_{q\in\bR} pq - g^\sharp_\varepsilon(x,q) \ge \sup_{q\in\bR} pq - \hat g^\sharp(x,q) - (1+L)\varepsilon = g(x)p - (1+L)\varepsilon,
    \end{align}
    Since $P^{p,\alpha}_{T-t} = \partial_q g^\sharp_\varepsilon(\tilde X^{t,x}_{T-t},q^\star_\varepsilon \tilde Q^{t,x}_{T-t})$, we obtain, from Lemma \ref{lem:dist fenchel} and from the first order conditions of the smooth concave maximisation problem, that 
    \begin{align}
        g(\tilde X^{t,x}_{T-t})P^{p,\alpha}_{T-t} &= \sup_{q \in \bR} P^{p,\alpha}_{T-t} q - \hat g^\sharp(\tilde X^{t,x}_{T-t},q) 
        \\ &\le \sup_{q \in \bR} P^{p,\alpha}_{T-t} q - g^\sharp_\varepsilon(\tilde X^{t,x}_{T-t},q) + (1+L)\varepsilon 
        \\ &= P^{p,\alpha}_{T-t} q^\star_\varepsilon \tilde Q^{t,x}_{T-t} - g^\sharp_\varepsilon(\tilde X^{t,x}_{T-t},q^\star_\varepsilon \tilde Q^{t,x}_{T-t}) + (1+L)\varepsilon.
    \end{align}
    This allows us to compute, using Assumption~\ref{h:coeffs} (so that $\esp{\tilde Z^{t,x}_{T-t}}\le e^{\bar b T}$),
    \begin{align}
        \esp{\tilde Z^{t,x}_{T-t} g(\tilde X^{t,x}_{T-t})P^{p,\alpha}_{T-t}}
        &\le \esp{\tilde Z^{t,x}_{T-t} \left( P^{p,\alpha}_{T-t} q^\star_\varepsilon \tilde Q^{t,x}_{T-t} - g^\sharp_\varepsilon(\tilde X^{t,x}_{T-t},q^\star_\varepsilon \tilde Q^{t,x}_{T-t})\right)} + (1+L)\varepsilon\,\esp{\tilde Z^{t,x}_{T-t}} \\
        &= pq^\star_\varepsilon - w_\varepsilon(t,x,q^\star_\varepsilon) + (1+L)\varepsilon\,\esp{\tilde Z^{t,x}_{T-t}} \\
        &\le pq^\star_\varepsilon - \tilde v^\sharp(t,x,q^\star_\varepsilon) + |(\tilde v^\sharp-w_\varepsilon)(t,x,q^\star_\varepsilon)| + (1+L)\varepsilon\,\esp{\tilde Z^{t,x}_{T-t}} \\
        &\le \sup_r pr - \tilde v^\sharp(t,x,r) + 2 (1+L) e^{\bar b T}\,\varepsilon \\
        &= \tilde v(t,x,p) + 2 (1+L) e^{\bar b T}\,\varepsilon.
    \end{align}

    To transfer this estimate to the original value function $v$, note that for any admissible control $\tilde\alpha\in\cA_p$,
    \begin{align*}
        \left|\esp{Z^{t,x}_{T-t} g(X^{t,x}_{T-t})P^{p,\tilde\alpha}_{T-t}}-\esp{\tilde Z^{t,x}_{T-t} g(\tilde X^{t,x}_{T-t})P^{p,\tilde\alpha}_{T-t}}\right|
        &\le \esp{\left|Z^{t,x}_{T-t} g(X^{t,x}_{T-t})-\tilde Z^{t,x}_{T-t} g(\tilde X^{t,x}_{T-t})\right|} \\ &\le C(1+|x|)\varepsilon,
    \end{align*}
    where the last inequality comes from the proof of Proposition~\ref{prop:stability-v-mollified}.

    Combining this with $|v(t,x,p)-\tilde v(t,x,p)|\le C(1+|x|)\varepsilon$ and the above $\varepsilon$-optimality for $\tilde v$ gives
    \[
        0\le \esp{Z^{t,x}_{T-t} g(X^{t,x}_{T-t})P^{p,\alpha}_{T-t}}-v(t,x,p)\le \big(2(1+L) e^{\bar b T}+2C(1+|x|)\big)\varepsilon,
    \]
    which concludes the proof.
\end{proof}

\begin{remark}
    For \(p\in\{0,1\}\), the null control is optimal, so the corresponding optimality statements are immediate.
\end{remark}

\section{Approximations of the control problem}\label{sec:approx-bounded-control}

\subsection{Control space truncation}\label{sec:trunc}

We are now interested in approximating the stochastic control problem \eqref{eq: control pb} by a problem where admissible controls have to stay within a compact subset of $\bR^q$. More generally, as needed in the sequel (see Section \ref{sec:pcpt} below) in order to consider Piecewise Constant Policy Timestepping (PCPT for short) controls, we will work both with bounded controls and away from the boundary in the $p$ variable. As such, for $M > 0$ and $\eta \in [0,\frac12)$, we consider the following optimal control problem: For all $(t,x,p) \in [0,T]\times\bR^d\times\bR$,
\begin{align} \label{eq: control compact}
    v^{M,\eta}(t,x,p)
    &= \inf_{\alpha \in \cA^{M,\eta}_{p}} \esp{Z^{t,x}_{T-t} g(X^{t,x}_{T-t})P^{p,\alpha}_{T-t}},
\end{align}
with
\begin{align}
    \cA^{M} &:= \{\alpha \in \cA \,|\, |\alpha| \le M, ds \otimes d\bP\text{-a.s.}\}, \text{ and} \\
    \cA^{M,\eta}_{p} &:= \{\alpha \in \cA^M \,|\, P^{p,\alpha} \in [\eta,1-\eta] \text{ a.s. on } [0,\infty)\}.
\end{align}
Again, note that $\cA^{M,\eta}_{p} = \emptyset$, and hence $v^{M,\eta}(t,x,p) = +\infty$, as soon as $p \not\in [\eta,1-\eta]$, while $\alpha \equiv 0 \in \cA^{M,\eta}_{p}$ for $p \in [\eta,1-\eta]$. The case $\eta=0$ corresponds to controls bounded by $M$ but without restriction on the martingale $P^{p,\alpha}$ (except for the fact that it has to remain in $[0,1]$), and is interesting on its own.

As for $v$ in the previous section, for $M>0$ and $\eta \in [0,\frac12)$, we define the analogue with mollified coefficients:
\begin{align}\label{eq:control-compact-mollified}
    \tilde v^{M,\eta}(t,x,p):=\inf_{\alpha\in\cA^{M,\eta}_{p}}\esp{\tilde Z^{t,x}_{T-t} g(\tilde X^{t,x}_{T-t}) P^{p,\alpha}_{T-t}},\qquad (t,x,p)\in[0,T]\times\bR^d\times[\eta,1-\eta].
\end{align}

We have the following analogue of Proposition~\ref{prop:stability-v-mollified}, with identical, hence omitted, proof.

\begin{lem}[Stability for the $(M,\eta)$-truncated problem]\label{lem:stability-vMeta-mollified}
    Under Assumptions~\ref{h:coeffs} and \ref{h:g Lip}, there exists a constant $C\ge 0$ (independent of $\varepsilon$, $M$ and $\eta$) such that for all $M>0$, all $\eta\in[0,\frac12)$ and all $(t,x,p)\in[0,T]\times\bR^d\times[\eta,1-\eta]$,
    \begin{equation}\label{eq:stability-vMeta-mollified}
        \big|v^{M,\eta}(t,x,p)-\tilde v^{M,\eta}(t,x,p)\big|\le C\,(1+|x|)\,\varepsilon.
    \end{equation}
\end{lem}

\medskip\noindent We now estimate the difference $v^{M,\eta}(t,x,p)-v(t,x,p)$. 
As in the previous section, we will work with mollified coefficients and use the $\varepsilon$-optimal feedback control $\alpha$ for $\tilde v$ constructed in Section~\ref{sec:eps-opt}.
The derivative estimates obtained above provide a pathwise bound on \(\alpha\) in terms of the continuous processes \((\tilde X,\tilde Q)\), which allows us to introduce a stopping time $\tau$ such that $|\alpha_s|\le M$ on $[0,\tau]$.
In addition, for $\eta>0$ we will stop the control whenever $P^{p,\alpha}$ reaches $\eta$ or $1-\eta$.

Recall that, for any $(t,x,p) \in [0,T)\times\bR^d\times(0,1)$, formula \eqref{eq: alpha epsilon optimal} defines a $2(1+L) e^{\bar b T}\varepsilon$-optimal control $\alpha^{t,x,p}$. Setting the control equal to zero after some well-chosen stopping time allows us to construct $\alpha^{M,\eta} \in \cA^{M,\eta}_{p}$. The construction, together with estimates to control the error made, is provided in the following lemmas. 

\begin{lem} \label{lem:prob tau}
    Fix $(t,x,p) \in [0,T)\times\bR^d\times(0,1)$. 
    \begin{enumerate}
        \item There exists $C>0$ (independent of $t,x,p$ and $\varepsilon$) such that, for any $M \ge 1$, the stopping time
\begin{align}\label{eq: def stopping time M}
    \tau^{t,x,p} := \inf \left\{ s \ge 0 \,\middle|\, (1+|\tilde X^{t,x}_s|)(1+\,|q^\star_\varepsilon\tilde Q^{t,x}_s|) \ge C^{-1} M \varepsilon \right\}
\end{align}
induces a control $\alpha^{t,x,p,M}  := \alpha^{t,x,p} \1_{[0,\tau^{t,x,p}]} \in \cA^{M,0}_{p}$, i.e. $|\alpha^{t,x,p,M}| \le M$ and $P^{p,\alpha^{t,x,p,M}} \in [0,1]$.
        \item For every $m\ge 1$, there exists $C_m>0$ (independent of $t,x,p$ and $\varepsilon$) such that, for every $M \ge 1$,
    \begin{align*}
        \pro{\tau^{t,x,p} \le T-t} %
        &\le \left(\frac{C_m}{M\varepsilon}\right)^m\,(1+|x|)^{2m}\,(p \wedge (1-p))^{-1}.
    \end{align*}
\end{enumerate}
\end{lem}
\begin{proof}
    \begin{enumerate}
        \item Using the explicit expression \eqref{eq: alpha epsilon optimal}, we have, using Lemma~\ref{lem:w_e infos}-1. and Assumption~\ref{h:coeffs},
        \begin{align*}
        |\alpha^{t,x,p}_s| &\le \|\partial^2_{xq} w_\varepsilon(t+s,\tilde X^{t,x}_s,q^\star_\varepsilon \tilde Q^{t,x}_s)\| \|\tilde\sigma(t+s,\tilde X^{t,x}_s)\| \\ &\quad + |q^\star_\varepsilon|\,|\tilde Q^{t,x}_s|\,|\partial^2_{q^2} w_\varepsilon(t+s,\tilde X^{t,x}_s,q^\star_\varepsilon \tilde Q^{t,x}_s)| \|\tilde\lambda(t+s,\tilde X^{t,x}_s)\| \\
        &\le C(1+|q^\star_\varepsilon \tilde Q^{t,x}_s|)\varepsilon^{-1} (1+|\tilde X^{t,x}_s|) + C|q^\star_\varepsilon \tilde Q^{t,x}_s|\,\varepsilon^{-1} \\
        &\le C(1+|\tilde X^{t,x}_s|)(1+|q^\star_\varepsilon\tilde Q^{t,x}_s|)\varepsilon^{-1},
        \end{align*} 
        where the constants $C$ can change from line to line. In particular, for such $C$, the stopping time \eqref{eq: def stopping time M} defines $\alpha^{t,x,p,M} = \alpha^{t,x,p}\,\1_{[0,\tau^{t,x,p}]} \in \cA^{M,0}_{p}$. %
        \item We have, for all $(t,x,p) \in [0,T)\times\bR^d\times(0,1)$ and all $m\ge 1$,
    \begin{align*}
        \pro{\tau^{t,x,p} \le T-t} &= \pro{\sup_{s \in [0,T-t]} (1+|\tilde X^{t,x}_s|)(1+|q^\star_\varepsilon \tilde Q^{t,x}_s|) \ge C^{-1} M \varepsilon} \\
        &\le \Big(\frac{C}{M\varepsilon}\Big)^m\esp{\sup_{s \in [0,T-t]} (1+|\tilde X^{t,x}_s|)^m(1+|q^\star_\varepsilon \tilde Q^{t,x}_s|)^m} \\
        &\le \Big(\frac{C}{M\varepsilon}\Big)^m\esp{\sup_{s \in [0,T-t]} (1+|\tilde X^{t,x}_s|)^{2m}}^{\frac12}\esp{\sup_{s \in [0,T-t]}(1+|q^\star_\varepsilon \tilde Q^{t,x}_s|)^{2m}}^{\frac12} \\
        &\le \left(\frac{C_m}{M\varepsilon}\right)^m(1+|x|)^m(1+|q^\star_\varepsilon|)^m.
    \end{align*}
    Combining this estimate with Lemma~\ref{lem:w_e infos}-2. yields the result.
\end{enumerate}
\end{proof}

\begin{lem}\label{lem:absorption-improves-eta}
    Fix $\eta \in [0,\frac12)$, $(t,p)\in[0,T]\times[\eta,1-\eta]$ and let $\alpha\in\cA_p$. We consider the stopping time
    \begin{align}\label{eq: def stopping time eta}
        \tau^\alpha_\eta:=\inf\{s\ge 0:\ P^{p,\alpha}_s\notin(\eta,1-\eta)\} \wedge (T-t).
    \end{align}
    Set $\alpha^{\eta}:=\alpha\,\1_{[0,\tau^\alpha_\eta]}$. 
    For any $\kappa \in (0,1)$, there exists a constant $C \ge 0$ (independent of $t,p,\eta,\alpha$ and $\varepsilon$) such that, for any $x \in \bR^d$,
    \begin{align*}
                &\Big|\esp{Z^{t,x}_{T-t} g(X^{t,x}_{T-t})\,P^{p,\alpha}_{T-t}}-\esp{Z^{t,x}_{T-t} g(X^{t,x}_{T-t})\,P^{p,\alpha^{\eta}}_{T-t}}\Big|
        \le C(1+|x|)\big(\eta(1-\eta)\big)^{1-\kappa}, \quad \mbox{and}\\
        &\Big|\esp{\tilde Z^{t,x}_{T-t} g(\tilde X^{t,x}_{T-t})\,P^{p,\alpha}_{T-t}}-\esp{\tilde Z^{t,x}_{T-t} g(\tilde X^{t,x}_{T-t})\,P^{p,\alpha^{\eta}}_{T-t}}\Big| \le C(1+|x|)\big(\eta(1-\eta)\big)^{1-\kappa}.
    \end{align*}
\end{lem}
\begin{proof}
    Let $\eta\in[0,\frac12)$, $(t,p) \in [0,T]\times[\eta,1-\eta]$ and $\alpha\in\cA_p$. Let $\kappa \in (0,1)$ and consider $r=\frac 1 \kappa$.

    By H\"older's inequality and Lemma~\ref{lem:props-XZQ}, we have, since $|P^{p,\alpha}_{T-t}-P^{p,\alpha^{\eta}}_{T-t}| \le 1$,
    \begin{align*}
        \Big|\esp{Z^{t,x}_{T-t} g(X^{t,x}_{T-t})\,(P^{p,\alpha}_{T-t}-P^{p,\alpha^{\eta}}_{T-t})}\Big|&\le \|Z^{t,x}_{T-t} g(X^{t,x}_{T-t})\|_{L^r}\,\|P^{p,\alpha}_{T-t}-P^{p,\alpha^{\eta}}_{T-t}\|_{L^{\frac{r}{r-1}}} \\
        &\le C(1+|x|)\,\esp{|P^{p,\alpha}_{T-t}-P^{p,\alpha^{\eta}}_{T-t}|^{\frac{r}{r-1}}}^{1-\frac 1 r} \\
        &\le C(1+|x|)\,\esp{|P^{p,\alpha}_{T-t}-P^{p,\alpha^{\eta}}_{T-t}|}^{1-\frac 1 r}.
    \end{align*}

    Now, since $P^{p,\alpha}_{T-t}=P^{p,\alpha^\eta}_{T-t}$ on $\{\tau^\alpha_\eta\ge T-t\}$, we compute
    \begin{align*}
        \esp{|P^{p,\alpha}_{T-t}-P^{p,\alpha^{\eta}}_{T-t}|} &= \esp{|P^{p,\alpha}_{T-t}-\eta|\1_{\{\tau^\alpha_\eta< T-t\}} \1_{\{P^{p,\alpha^{\eta}}_{\tau^\alpha_\eta}=\eta\}}} \\ &\quad + \esp{|P^{p,\alpha}_{T-t}-(1-\eta)| \1_{\{\tau^\alpha_\eta< T-t\}} \1_{\{P^{p,\alpha^{\eta}}_{\tau^\alpha_\eta}=1-\eta\}}} \\
        &=  \esp{\esp{|P^{p,\alpha}_{T-t}-\eta| \middle| \cF_{\tau^\alpha_\eta}}\1_{\{\tau^\alpha_\eta< T-t\}} \1_{\{P^{p,\alpha^{\eta}}_{\tau^\alpha_\eta}=\eta\}}} \\ &\quad + \esp{\esp{|P^{p,\alpha}_{T-t}-(1-\eta)| \middle| \cF_{\tau^\alpha_\eta}} \1_{\{\tau^\alpha_\eta< T-t\}} \1_{\{P^{p,\alpha^{\eta}}_{\tau^\alpha_\eta}=1-\eta\}}}.
    \end{align*}
    Since $[0,1] \ni y \to |y-\eta|$ is convex, one has $|y-\eta| \le y|1-\eta| + (1-y)|0-\eta| = y(1-\eta)+(1-y)\eta=\eta+(1-2\eta)y$, hence, by optional stopping for the bounded martingale $P^{p,\alpha}$,
    \begin{align*}
        \esp{\esp{|P^{p,\alpha}_{T-t}-\eta|\middle| \cF_{\tau^\alpha_\eta}}\1_{\{\tau^\alpha_\eta< T-t\}} \1_{\{P^{p,\alpha^{\eta}}_{\tau^\alpha_\eta}=\eta\}}} &\le \esp{\left(\eta + (1-2\eta)\esp{P^{p,\alpha}_{T-t}\middle| \cF_{\tau^\alpha_\eta}}\right)\1_{\{\tau^\alpha_\eta< T-t\}} \1_{\{P^{p,\alpha^{\eta}}_{\tau^\alpha_\eta}=\eta\}}} \\ &\le \esp{\left(\eta + (1-2\eta)\eta\right)\1_{\{\tau^\alpha_\eta< T-t\}} \1_{\{P^{p,\alpha^{\eta}}_{\tau^\alpha_\eta}=\eta\}}} \\ &\le \esp{2\eta(1-\eta)\1_{\{\tau^\alpha_\eta< T-t\}}\1_{\{P^{p,\alpha^{\eta}}_{\tau^\alpha_\eta}=\eta\}}}.
    \end{align*}
    The second expectation is treated similarly, hence
    \begin{align*}
        \esp{|P^{p,\alpha}_{T-t}-P^{p,\alpha^{\eta}}_{T-t}|} \le 2\eta(1-\eta),
    \end{align*}
    and the result follows.
    
    The same computation holds with $\tilde Z^{t,x}_{T-t} g(\tilde X^{t,x}_{T-t})$ in place of $Z^{t,x}_{T-t} g(X^{t,x}_{T-t})$ (see Remark~\ref{rem:mollified-pb}), which concludes the proof.
\end{proof}

\begin{theorem}[Control and domain truncation]\label{cor:domain-truncation-vMeta-v}
    Under Assumptions~\ref{h:coeffs} and~\ref{h:g Lip}, for any $\kappa_1,\kappa_2\in(0,1)$, there exist constants $C_{\kappa_1}\ge 0$, $k_{\kappa_1}\ge 0$ and $C_{\kappa_2}\ge 0$ such that for all $\eta\in[0,\frac12)$, all $(t,x,p)\in[0,T]\times\bR^d\times([\eta,1-\eta]\cap(0,1))$, and all $M\ge 1$,
    \[
        0\le v^{M,\eta}(t,x,p)-v(t,x,p)
        \le C_{\kappa_1}\,(1+|x|)^{k_{\kappa_1}}%
        \, M^{-1+\kappa_1}
        + C_{\kappa_2}\,(1+|x|)\,(\eta(1-\eta))^{1-\kappa_2}.
    \]
    Moreover, when $\eta=0$ and $p\in\{0,1\}$, one has $v^{M,0}(t,x,p)= v(t,x,p)$.
\end{theorem}
\begin{proof}
    First, if $\eta=0$ and $p \in \{0,1\}$, then both values coincide, since the null control is the only admissible control. Hence the boundary cases are immediate. Similarly, the case $(T,x,p)$ for $(x,p)\in\bR^d\times[\eta,1-\eta]$ also yields $v(T,x,p)=v^{M,\eta}(T,x,p)$, since the terminal condition is independent of the control. We henceforth assume $(t,x,p)\in[0,T)\times\bR^d\times([\eta,1-\eta]\cap(0,1))$ and we fix $\varepsilon\in(0,1]$.

    Since $\cA^{M,\eta}_{p}\subset\cA_p$, one has $v(t,x,p)\le v^{M,\eta}(t,x,p)$.
    Using Proposition~\ref{prop:stability-v-mollified} and Lemma~\ref{lem:stability-vMeta-mollified}, we get
    \begin{align}
        \nonumber 0\le v^{M,\eta}(t,x,p)-v(t,x,p)
        &\le |v(t,x,p)-\tilde v(t,x,p)| + \big(\tilde v^{M,\eta}(t,x,p)-\tilde v(t,x,p)\big) \\ &\quad+ |\tilde v^{M,\eta}(t,x,p)-v^{M,\eta}(t,x,p)| \\
        &\le 2C\,(1+|x|)\varepsilon + \big(\tilde v^{M,\eta}(t,x,p)-\tilde v(t,x,p)\big). \label{eq: truncation error decomp}
    \end{align}

    We now bound the mollified truncation error.
    Let $\alpha:=\alpha^{t,x,p}$ be the $2(1+L)e^{\bar bT}\varepsilon$-optimal control for $\tilde v(t,x,p)$ given by \eqref{eq: alpha epsilon optimal} in Theorem \ref{thm:eps opti}, and consider the controls $\alpha^M := \alpha\1_{[0,\tau^{t,x,p}]} \in \cA^{M,0}_{p}$ and $\alpha^{M,\eta} = (\alpha^M)^\eta = \alpha^M\1_{[0,\tau_\eta^{\alpha^M}]} \in \cA^{M,\eta}_{p}$, where $\tau^{t,x,p}$ and $\tau_\eta^{\alpha^M}$ are respectively defined by \eqref{eq: def stopping time M} and \eqref{eq: def stopping time eta}.
    By definition of $\tilde v^{M,\eta}$ and $\varepsilon$-optimality of $\alpha$ for $\tilde v$,
    \begin{align*}
        0\le \tilde v^{M,\eta}(t,x,p)-\tilde v(t,x,p)
        &\le \esp{\tilde Z^{t,x}_{T-t} g(\tilde X^{t,x}_{T-t})\,\big(P^{p,\alpha^{M,\eta}}_{T-t}-P^{p,\alpha}_{T-t}\big)} + 2(1+L)e^{\bar bT}\varepsilon \\
        &= \esp{\tilde Z^{t,x}_{T-t} g(\tilde X^{t,x}_{T-t})\,\big(P^{p,\alpha^{M,\eta}}_{T-t}-P^{p,\alpha^{M}}_{T-t}\big)}\\ &\quad +\esp{\tilde Z^{t,x}_{T-t} g(\tilde X^{t,x}_{T-t})\,\big(P^{p,\alpha^{M}}_{T-t}-P^{p,\alpha}_{T-t}\big)}+ 2(1+L)e^{\bar bT}\varepsilon \\
        &\le C_{\kappa_2}\,(1+|x|)\,\left(\eta(1-\eta)\right)^{1-\kappa_2} \\ &\quad+ \esp{\tilde Z^{t,x}_{T-t} g(\tilde X^{t,x}_{T-t})\,\big(P^{p,\alpha^M}_{T-t}-P^{p,\alpha}_{T-t}\big)} + 2(1+L)e^{\bar bT}\varepsilon,
    \end{align*}
where we used Lemma~\ref{lem:absorption-improves-eta} for the first expectation.

    Observe that $0 \le |P^{p,\alpha^M}_{T-t}-P^{p,\alpha}_{T-t}| \le \1_{\{\tau^{t,x,p}\le T-t\}}$. Additionally, observe that $$\esp{|P^{p,\alpha^M}_{T-t}-P^{p,\alpha}_{T-t}|} \le \esp{P^{p,\alpha^M}_{T-t} + P^{p,\alpha}_{T-t}} = 2p,$$ and that $$\esp{|P^{p,\alpha^M}_{T-t}-P^{p,\alpha}_{T-t}|} \le 2 - \esp{P^{p,\alpha^M}_{T-t} + P^{p,\alpha}_{T-t}} = 2(1-p),$$ so that $$\esp{|P^{p,\alpha^M}_{T-t}-P^{p,\alpha}_{T-t}|} \le 2 (p \wedge (1-p)).$$  We then compute, using H\"older's inequality and Lemma \ref{lem:prob tau}, for any $m \ge 1$,
    \begin{align*}
    \esp{\tilde Z^{t,x}_{T-t} g(\tilde X^{t,x}_{T-t})\,\big(P^{p,\alpha^M}_{T-t}-P^{p,\alpha}_{T-t}\big)} =& \esp{\tilde Z^{t,x}_{T-t} g(\tilde X^{t,x}_{T-t})\,\big(P^{p,\alpha^M}_{T-t}-P^{p,\alpha}_{T-t}\big) \1_{\{\tau^{t,x,p}\le T-t\}}} \\ \le& \esp{|\tilde Z^{t,x}_{T-t} g(\tilde X^{t,x}_{T-t})|^4}^{\frac14}\,\esp{|P^{p,\alpha^M}_{T-t}-P^{p,\alpha}_{T-t}|^2}^{\frac12}\,\pro{\tau^{t,x,p}\le T-t}^{\frac14} \\ \le& \esp{|\tilde Z^{t,x}_{T-t} g(\tilde X^{t,x}_{T-t})|^4}^{\frac14}\,\esp{|P^{p,\alpha^M}_{T-t}-P^{p,\alpha}_{T-t}|}^{\frac12}\,\pro{\tau^{t,x,p}\le T-t}^{\frac14} \\ \le& C_m (1+|x|)^{1+\frac{m}{2}} (p \wedge (1-p))^{\frac14} (M\varepsilon)^{-\frac{m}{4}} \\ \le& C_m (1+|x|)^{1+\frac{m}{2}}  (M\varepsilon)^{-\frac{m}{4}}.
    \end{align*}
    Combining this estimate with the preceding bounds and choosing $\varepsilon = M^{-\frac{m}{m+4}}$ with $m$ such that $\frac{m}{m+4}>1-\kappa_1$, we obtain

    \[
        v^{M,\eta}(t,x,p)- v(t,x,p)
        \le C_{\kappa_1}\,(1+|x|)^{k_{\kappa_1}}%
        M^{-1+\kappa_1}
        + C_{\kappa_2}\,(1+|x|)\,(\eta(1-\eta))^{1-\kappa_2},
    \]
    for suitable constants $C_{\kappa_1},k_{\kappa_1}$.
\end{proof} 

\subsection{PCPT controls}\label{sec:pcpt}

We now fix $h,M>0$, and we introduce the class of $h$-piecewise constant on $[0,T-t]$, $M$-bounded controls. Since the controlled martingale must remain in $[0,1]$ almost surely, one cannot consider ``raw'' piecewise constant controls, as they lead to conditionally Gaussian distributions, which are not supported in $[0,1]$. Instead of considering truly constant controls on each interval of size $h$, we consider constant controls augmented with one stopping time, smaller than the exit time of the martingale from $[0,1]$. Hence, instead of taking exactly one value on each interval of size $h$, the control can take at most two values: the chosen constant control and potentially $0$ between the stopping time and the end of the interval.

Mathematically, for $k \ge 0$, $P \in L^0(\cF_{kh};\bR)$ and $\alpha \in L^0(\cF_{kh};\bR^q)$ such that $|\alpha| \le M$ a.s., we define 
\begin{align}\label{eq: pcpt elem}
\tau^{P,\alpha}_{h,k} := \inf\{s \ge kh : P + \alpha^\top (W_s-W_{kh}) \notin (0,1)\} \wedge (k+1)h,
\end{align}
and the building blocks for PCPT controls are of the form $\alpha \1_{[0,\tau]}$, where $\tau \le \tau^{P,\alpha}_{h,k}$ is a stopping time. We thus set, for $k \ge 0$ and $P \in L^0(\cF_{kh};\bR)$,
\begin{align}\label{eq: pcpt admissible}
\cA^{\text{elem}}_{k,P,M,h} := \left\{ (\alpha,\tau) : \alpha \in L^0(\cF_{kh};\bR^q), |\alpha| \le M, \tau \text{ stopping time}, \tau \le \tau^{P,\alpha}_{h,k}\right\},
\end{align}
and we define the class of admissible controls as follows:
\begin{align}\label{eq: pcpt admissible global}
\cA^{M,h}_{p} := \left\{ \alpha = \sum_{k=0}^\infty \alpha^k \1_{[kh,(k+1)h \wedge \tau^k]}, (\alpha^k,\tau^k) \in \cA^{\text{elem}}_{k,P^{p,\alpha}_{kh},M,h} \text{ a.s. for all } k \ge 0 \right\}.
\end{align}
\begin{remark}\label{rem:stop before boundary}
We allow stopping before reaching the boundary of $[0,1]$ for technical reasons that will become clear in the proofs below: due to state constraints, in order to regularize the PCPT value function, one will be forced to consider controls stopped when reaching the boundary of $[\varepsilon,1-\varepsilon]$ for some $\varepsilon>0$.
\end{remark}
Although the above definition is recursive, it obviously can be disentangled into an iterative construction: starting from $p \in \bR$, one first chooses $(\alpha^0,\tau^0)$, then from the $P+\alpha^0 W_\tau^0$, one chooses $(\alpha^1,\tau^1)$ in a measurable manner, and so on.

We easily observe, by construction of $\cA^{M,h}_p$, that if $p \in [0,1]$ and $\alpha \in \cA^{M,h}_p$, then $P^{p,\alpha}$ remains in $[0,1]$ on $[0,\infty)$, and thus $\cA^{M,h}_p \subset \cA^M_p \subset \cA_p$.

We now define the control problem with PCPT controls:
\begin{align}
    v^{M,h}(t,x,p) = \inf_{\alpha \in \cA^{M,h}_{p}} \esp{Z^{t,x}_{T-t} \hat g(X^{t,x}_{T-t}, P^{p,\alpha}_{T-t})}, \quad (t,x,p) \in [0,T]\times\bR^d\times\bR,
\end{align}
where $\hat g$ is defined below \eqref{eq: control pb}. One easily observes that if $p \not\in [0,1]$, then $\cA^{M,h}_p = \{0\}$, hence $v^{M,h}(t,x,p) = +\infty$ for all $(t,x) \in [0,T]\times\bR^d$. One can thus restrict its attention to $p \in [0,1]$ in the sequel, and we have
\begin{align}
    \label{eq: control pb pcpt} v^{M,h}(t,x,p) = \inf_{\alpha \in \cA^{M,h}_{p}} \esp{Z^{t,x}_{T-t} g(X^{t,x}_{T-t})P^{p,\alpha}_{T-t}}, \quad (t,x,p) \in [0,T]\times\bR^d\times[0,1].
\end{align}
In addition, since $\cA^{M,h}_p \subset \cA^M_p \subset \cA_p$ for $p \in [0,1]$, we have $v(t,x,p) \le v^{M,h}(t,x,p)$ for all $(t,x,p) \in [0,T]\times\bR^d\times[0,1]$.

\medskip\noindent In order to regularise the value function $v^{M,h}$, following \cite{Krylov1999, Jakobsen2019}, we consider the approximate value function with shaken and regularized coefficients. For fixed $\varepsilon\in(0,1]$,
we consider additional controls $\beta=(\beta^1,\beta^2) \in \cB^{\varepsilon,h}$, where
\begin{align}\label{eq:beta-set}
    \cB^{\varepsilon,h}
    :=\Big\{\beta=(\beta^1,\beta^2) \ \bR^{1+d}\mbox{-valued progressive process} \Big| &\beta_s = \sum_{k=0}^\infty \beta_k \1_{[kh,(k+1)h)}, \beta_k \in L^2(\cF_{kh};\bR^{1+d}), \\ &\beta_k \in(-\varepsilon^2,0]\times\{y\in\bR^d\mid |y|\le \varepsilon\} \text{ for all }k, \bP-\text{a.s.}\Big\}.
\end{align}

For $(t,x,p)\in[0,T+\varepsilon^2]\times\bR^d\times[0,1]$ and $\beta\in\cB^{\varepsilon,h}$, we define the \emph{shaken and regularised} processes $(\hat X^{t,x,\beta},\hat Z^{t,x,\beta}%
)$ on $[0,T+\varepsilon^2-t]$ by
\begin{align}\label{eq:shaken-sde}
    &d\hat X_s
    = \tilde\mu\big(t+s+\beta^1_s,\hat X_s+\beta^2_s\big)\,ds + \tilde\sigma\big(t+s+\beta^1_s,\hat X_s+\beta^2_s\big)\,dW_s,\qquad \hat X_0=x,\\
    &\hat Z_s = \exp\left(\int_0^s \left(\tilde b(t+u+\beta^1_u,\hat X_u+\beta^2_u) - \frac12 |\tilde \lambda(t+u+\beta^1_u,\hat X_u+\beta^2_u)|^2\right) du \right. \\&\qquad\qquad \left. + \int_0^s \tilde\lambda(t+u+\beta^1_u,\hat X_u+\beta^2_u)^\top dW_u\right),%
\end{align}
Note that under Assumption~\ref{h:coeffs}, the SDE for $\hat X^{t,x,\beta}$ is well posed on $[0,\infty)$, for any $\beta \in \cB^{\varepsilon,h}$. In addition, we have the following moment bounds and stability result with respect to the mollified dynamics.
\begin{lem}\label{lem:stability-shaken-mollified}
    Let Assumption~\ref{h:coeffs} hold.
    \begin{enumerate}
    \item For all $m \in \bR$ and $r \ge 1$, there exist constants $C_m,C_r\ge 0$ (independent of $\varepsilon$, $M$, $h$) such that, for all $(t,x)\in[0,T+\varepsilon^2]\times\bR^d$ and for all $\beta\in\cB^{\varepsilon,h}$,
    \begin{align*}
        &\sup_{0\le s \le T+\varepsilon^2-t} \esp{|\hat X^{t,x,\beta}_s|^{2r}} \le C_r(1+|x|)^{2r}, \qquad \sup_{0\le s \le T+\varepsilon^2-t} \esp{|\hat Z^{t,x,\beta}_s|^{m}} \le C_m \\
        & \esp{|\hat Z^{t,x,\beta}_{T+\varepsilon^2-t} g(\hat X^{t,x,\beta}_{T+\varepsilon^2-t})|^{2r}} \le C_r(1+|x|)^{2r}.
    \end{align*}
    \item There exists a constant $C\ge0$ (independent of $\varepsilon$, $M$ and $h$) such that, for all $(t,s,u,x,y) \in [0,T+\varepsilon^2]\times[0,T+\varepsilon^2]\times[0,T+\varepsilon^2]\times\bR^d\times\bR^d$ and for all $\beta\in\cB^{\varepsilon,h}$, we have
    \begin{align*}
        \esp{|\hat X^{t,x,\beta}_u - \hat X^{s,y,\beta}_u|^2} + \esp{|\hat Z^{t,x,\beta}_u - \hat Z^{s,y,\beta}_u|^2} \le C(|t-s| + |x-y|^2).
    \end{align*}
    \item For all $r \ge 1$, there exists a constant $C_r\ge 0$ (independent of $\varepsilon$, $M$, $h$) such that, for all $(t,x)\in[0,T]\times\bR^d$ and for all $\beta\in\cB^{\varepsilon,h}$,
    \begin{align}
        \sup_{0\le s \le T+\varepsilon^2-t} \esp{|\hat X^{t,x,\beta}_s - \tilde X^{t,x}_s|^{2r}} + \sup_{0\le s \le T+\varepsilon^2-t} \esp{|\hat Z^{t,x,\beta}_s - \tilde Z^{t,x}_s|^{2r}} \le C_r\varepsilon^{2r}.
    \end{align}
    \end{enumerate}
\end{lem}
\begin{proof}
    \begin{enumerate}
        \item The proof follows the computations of Lemma~\ref{lem:props-XZQ} with mollified coefficients, using the Lipschitz continuity and linear growth of $\tilde\mu$, $\tilde\sigma$, $\tilde b$ and $\tilde\lambda$ (see Lemma~\ref{lem:mollified-coeffs-props}) and the fact that $|\beta^1| \le \varepsilon^2 \le 1$ and $|\beta^2| \le \varepsilon \le 1$.
        \item We have, with $\delta X = \hat X^{t,x,\beta} - \hat X^{s,y,\beta}$ and $\delta Z = \hat Z^{t,x,\beta} - \hat Z^{s,y,\beta}$, using the Lipschitz continuity and linear growth of $\tilde\mu$ and $\tilde\sigma$, and the Burkholder--Davis--Gundy inequality,
        \begin{align*}
            \esp{|\delta X_u|^2} &\le 3|x-y|^2 + 3\esp{\left|\int_0^u \Big(\tilde\mu(t+v+\beta^1_v,\hat X^{t,x,\beta}_v+\beta^2_v) - \tilde\mu(s+v+\beta^1_v,\hat X^{s,y,\beta}_v+\beta^2_v)\Big) dv\right|^2} \\
            &\quad + 3 \esp{\sup_{0 \le v \le u}\left|\int_0^v \Big(\tilde\sigma(t+w+\beta^1_w,\hat X^{t,x,\beta}_w+\beta^2_w) - \tilde\sigma(s+w+\beta^1_w,\hat X^{s,y,\beta}_w+\beta^2_w)\Big) dW_w\right|^2} \\
            &\le C(|x-y|^2 + |t-s| + \int_0^u \esp{|\delta X_v|^2} dv),
        \end{align*}
        and we conclude by Gronwall's lemma. Applying the same argument with fourth moments first gives
        \[
            \esp{|\delta X_u|^4}\le C\big(|t-s|^2+|x-y|^4\big).
        \]
        The estimate for $\delta Z$ then follows from the Burkholder--Davis--Gundy and Cauchy--Schwarz inequalities, the Lipschitz continuity and boundedness of $\tilde b$ and $\tilde\lambda$, and the moment estimates from item 1.
        \item Computations similar to those in the proof of Lemma~\ref{lem:Lp-stability-mollified} allow us to obtain, for $0 \le s \le T+\varepsilon^2-t$,
    \begin{align*}
        \esp{|\delta X_s|^{2r}} \le& C_r \esp{\int_0^s |\delta X_u|^{2r} + |\delta\mu_u|^{2r} + \big\|\delta\sigma_u\big\|^{2r}\,du},
    \end{align*}
    where $\delta X = \hat X^{t,x,\beta} - \tilde X^{t,x}$, $\delta\mu_u = \tilde\mu(t+u+\beta^1_u,\hat X_u+\beta^2_u) - \tilde\mu(t+u,\tilde X_u)$ and $\delta\sigma_u = \tilde\sigma(t+u+\beta^1_u,\hat X_u+\beta^2_u) - \tilde\sigma(t+u,\tilde X_u)$.
    
    Using the Lipschitz continuity of $\tilde\mu$ in space and its $\frac12$-H\"older regularity in time (Lemma~\ref{lem:mollified-coeffs-props}), we have
    \begin{align*}
        |\delta\mu_u| = |\tilde\mu(t+u+\beta^1_u,\hat X_u+\beta^2_u) - \tilde\mu(t+u,\tilde X_u)| \le L(|\beta^1_u|^{\frac12} + |\delta X_u + \beta^2_u|)\le L(|\delta X_u| + \varepsilon),
    \end{align*}
    and similarly $\|\delta\sigma_u\| \le L(|\delta X_u| + \varepsilon)$. Hence
    \begin{align*}
        \esp{|\delta X_s|^{2r}} &\le C_r \esp{\int_0^s |\delta X_u|^{2r} + \varepsilon^{2r}\,du} \\
        &\le C_r \int_0^s \esp{|\delta X_u|^{2r}}\,du + C_r (T+1) \varepsilon^{2r}.
    \end{align*}
    Gronwall's lemma allows us to conclude that $\sup_{0 \le s \le T+\varepsilon^2-t} \esp{|\delta X_s|^{2r}} \le C_r \varepsilon^{2r}$.

    \medskip\noindent We now turn to the stability of $\hat Z^{t,x,\beta}$ with respect to $\tilde Z^{t,x}$. We have, for $0 \le s \le T+\varepsilon^2-t$,
    \begin{align*}
        \esp{|\delta Z_s|^{2r}} 
        \le& C_r \esp{\int_0^s \Big(|\delta Z_u|^{2r} \right.\\ &\qquad \left.+ \big|\tilde b(t+u,\tilde X_u)\tilde Z_u - \tilde b(t+u+\beta^1_u,\hat X_u+\beta^2_u)\hat Z_u\big|^{2r} \right. \\ &\qquad \left. + \big\|\tilde Z_u\tilde\lambda(t+u,\tilde X_u)^\top - \hat Z_u\tilde\lambda(t+u+\beta^1_u,\hat X_u+\beta^2_u)^\top\big\|^{2r}\Big)du},
    \end{align*}
    with $\delta Z = \hat Z^{t,x,\beta} - \tilde Z^{t,x}$.

    We compute
    \begin{align*}
        &\big|\tilde b(t+u,\tilde X_u)\tilde Z_u - \tilde b(t+u+\beta^1_u,\hat X_u+\beta^2_u)\hat Z_u\big| \\
        \le& \big|\tilde b(t+u,\tilde X_u)\tilde Z_u - \tilde b(t+u+\beta^1_u,\hat X_u+\beta^2_u)\tilde Z_u\big| \\ &\quad + \big|\tilde b(t+u+\beta^1_u,\hat X_u+\beta^2_u)\tilde Z_u- \tilde b(t+u+\beta^1_u,\hat X_u+\beta^2_u)\hat Z_u\big| \\
        \le& |\tilde Z_u| L(|\beta^1_u|^{\frac12} + |\delta X_u+\beta^2_u|) + \bar b |\delta Z_u| \\
        \le& \bar b |\delta Z_u| + C\varepsilon|\tilde Z_u| + C|\delta X_u||\tilde Z_u|,
    \end{align*}
    and similarly, $\big\|\tilde Z_u\tilde\lambda(t+u,\tilde X_u)^\top - \hat Z_u\tilde\lambda(t+u+\beta^1_u,\hat X_u+\beta^2_u)^\top\big\| \le \bar\lambda |\delta Z_u| + C\varepsilon|\tilde Z_u| + C|\delta X_u||\tilde Z_u|$. Therefore, using the Cauchy--Schwarz inequality, the fact that $\sup_u \esp{|\tilde Z_u|^m} \le C_m < \infty$ for all $m \ge 1$, and the previous estimates, we get
    \begin{align*}
        \esp{|\delta Z_s|^{2r}} \le& C_r \esp{\int_0^s |\delta Z_u|^{2r} + \big(|\delta Z_u| + \varepsilon|\tilde Z_u| + |\delta X_u||\tilde Z_u|\big)^{2r}\,du} \\
        \le& C_r \esp{\int_0^s |\delta Z_u|^{2r} + |\tilde Z_u|^{2r} \varepsilon^{2r} + |\tilde Z_u|^{2r} |\delta X_u|^{2r}\,du} \\
        \le& C_r \int_0^s \esp{|\delta Z_u|^{2r}}\,du + C_r\int_0^s \esp{|\tilde Z_u|^{4r}}^{\frac12}\esp{|\delta X_u|^{4r}}^{\frac12} du + C_r\,T\,\varepsilon^{2r} \\
        \le& C_r \int_0^s \esp{|\delta Z_u|^{2r}}\,du + C_r\,(T+1)\,\varepsilon^{2r},
    \end{align*}
    and we conclude by Gr\"onwall's inequality.
\end{enumerate}
\end{proof}

We now observe the following link between deterministically shaken processes and the mollified ones.
\begin{lem}\label{lem:link-shaken-mollified}
    For any $(t,x)\in [0,T+\varepsilon^2]\times\bR^d$, for any $(s,y)\in(-\varepsilon^2,0]\times B(0,\varepsilon)$ such that $t+s \in [0,T]$, we have, for $\beta \equiv (s,y)$ the constant control process,
    \begin{align*}
        \hat X^{t,x,\beta}_{u} + y = \tilde X^{t+s,x+y}_{u} ,\quad
        \hat Z^{t,x,\beta}_{u} = \tilde Z^{t+s,x+y}_{u},\quad u\ge 0.
    \end{align*}
    Equivalently, for $(t,x) \in [0,T]\times\bR^d$ and for all $(s,y)\in(-\varepsilon^2,0]\times B(0,\varepsilon)$,
    \begin{align} \label{eq:link-shaken-mollified}
        \tilde X^{t,x}_u = \hat{X}^{t-s,x-y,\beta}_{u} + y,\quad \tilde Z^{t,x}_u = \hat{Z}^{t-s,x-y,\beta}_{u},\quad u\ge 0,
    \end{align}
    with again $\beta \equiv (s,y)$.
\end{lem}
\begin{proof}
    Let $(t,x)\in [0,T+\varepsilon^2]\times\bR^d$ and $(s,y)\in(-\varepsilon^2,0]\times B(0,\varepsilon)$ be fixed such that $t+s \in [0,T]$, and consider $\beta\equiv(s,y)$.\\
    Let $Y_u = \hat X^{t,x,\beta}_{u} + y$ for $u\ge 0$. We have $Y_{0} = x+y = \tilde X^{t+s,x+y}_{0}$, and for $u\ge 0$,
    \begin{align*}
        dY_u
        &= d\hat X^{t,x,\beta}_{u}
        = \tilde\mu\big(t+u+s,\hat X^{t,x,\beta}_{u}+y\big)\,du + \tilde\sigma\big(t+u+s,\hat X^{t,x,\beta}_{u}+y\big)\,dW_u\\
        &= \tilde\mu\big(t+s+u,Y_u\big)\,du + \tilde\sigma\big(t+s+u,Y_u\big)\,dW_u.
    \end{align*}
    Hence $Y=\tilde X^{t+s,x+y}$ by uniqueness of the SDE solution.\\
    Similarly, setting $V_u = \hat Z^{t,x,\beta}_{u}$ for $u\ge 0$, we have $V_{0} = 1 = \tilde Z^{t+s,x+y}_{0}$, and for $u\ge 0$,
    \begin{align*}
        dV_u
        &= d\hat Z^{t,x,\beta}_{u}
        = \tilde b\big(t+u+s,\hat X^{t,x,\beta}_{u}+y\big)\,\hat Z^{t,x,\beta}_{u}\,du + \hat Z^{t,x,\beta}_{u}\,\tilde\lambda\big(t+u+s,\hat X^{t,x,\beta}_{u}+y\big)^\top dW_u\\
        &= \tilde b\big(t+s+u,\tilde X^{t+s,x+y}_u\big)\,V_u\,du + V_u\,\tilde\lambda\big(t+s+u,\tilde X^{t+s,x+y}_u\big)^\top dW_u.
    \end{align*}
    Hence $V=\tilde Z^{t+s,x+y}$ by uniqueness of the SDE solution.
\end{proof}

We then introduce the corresponding shaken PCPT value function
\begin{align}\label{eq:pcpt-shaken}
    \hat v^{M,h}(t,x,p)
    := \inf_{\alpha\in\cA^{M,h}_{p},\,\beta\in\cB^{\varepsilon,h}}\esp{\hat Z^{t,x,\beta}_{T+\varepsilon^2-t}\,g\big(\hat X^{t,x,\beta}_{T+\varepsilon^2-t}\big)\,P^{p,\alpha}_{T+\varepsilon^2-t}},
    \quad (t,x,p)\in[0,T+\varepsilon^2]\times\bR^d\times[0,1].
\end{align}

\begin{lem}\label{lem:stability-vMh-regularised}
    Under Assumptions~\ref{h:coeffs} and \ref{h:g Lip}, there exists a constant $C\ge 0$ (independent of $\varepsilon$, $M$ and $h$) such that for all $(t,x,p)\in[0,T]\times\bR^d\times[0,1]$,
    \[
        \big|v^{M,h}(t,x,p)-\hat v^{M,h}(t,x,p)\big| \le C\,(1+|x|)\,\varepsilon.
    \]
\end{lem}
\begin{proof}
    We introduce the auxiliary value function, with mollified coefficients but without shaking,
    \begin{align*}
        \tilde v^{M,h}(t,x,p)
        := \inf_{\alpha \in \cA^{M,h}_{p}} \esp{\tilde Z^{t,x}_{T-t} g(\tilde X^{t,x}_{T-t}) P^{p,\alpha}_{T-t}},
        \quad (t,x,p) \in [0,T]\times\bR^d\times[0,1].
    \end{align*}
    By the triangle inequality, we have
    \begin{align*}
        \big|v^{M,h}(t,x,p)-\hat v^{M,h}(t,x,p)\big| &\le \big|v^{M,h}(t,x,p)-\tilde v^{M,h}(t,x,p)\big| + \big|\tilde v^{M,h}(t,x,p)-\hat v^{M,h}(t,x,p)\big| \\
        &\le C(1+|x|)\varepsilon + \big|\tilde v^{M,h}(t,x,p)-\hat v^{M,h}(t,x,p)\big|,
    \end{align*}
    where the bound of the first term is obtained following the arguments of Proposition \ref{prop:stability-v-mollified}. Regarding the second term,
    we have, by the triangle inequality and since $P^{p,\alpha} \in [0,1]$,
    \begin{align*}
        \big|\tilde v^{M,h}(t,x,p)-\hat v^{M,h}(t,x,p)\big| &\le \sup_{\alpha\in\cA^{M,h}_{p},\,\beta\in\cB^{\varepsilon,h}} \Big|\esp{\tilde Z^{t,x}_{T-t} g\big(\tilde X^{t,x}_{T-t}\big) P^{p,\alpha}_{T-t}}-\esp{\hat Z^{t,x,\beta}_{T+\varepsilon^2-t} g\big(\hat X^{t,x,\beta}_{T+\varepsilon^2-t}\big) P^{p,\alpha}_{T+\varepsilon^2-t}}\Big| 
        \\ &\le \sup_{\alpha\in\cA^{M,h}_{p},\,\beta\in\cB^{\varepsilon,h}} \Big| \esp{\Big(\tilde Z_{T-t} g(\tilde X_{T-t}) - \tilde Z_{T+\varepsilon^2-t} g(\tilde X_{T+\varepsilon^2-t})\Big) P^{p,\alpha}_{T-t}} \\ &\qquad\qquad + \esp{\tilde Z_{T+\varepsilon^2-t} g(\tilde X_{T+\varepsilon^2-t})\Big(P^{p,\alpha}_{T-t} - P^{p,\alpha}_{T+\varepsilon^2-t}\Big)}\\ & \qquad\qquad + \esp{\Big( \tilde Z_{T+\varepsilon^2-t} g(\tilde X_{T+\varepsilon^2-t}) - \hat Z_{T+\varepsilon^2-t} g(\hat X_{T+\varepsilon^2-t})\Big) P^{p,\alpha}_{T+\varepsilon^2-t}}\Big| \\
        &\le \sup_{\alpha\in\cA^{M,h}_{p},\,\beta\in\cB^{\varepsilon,h}} \Big( \esp{\Big|\tilde Z_{T-t} g(\tilde X_{T-t}) - \tilde Z_{T+\varepsilon^2-t} g(\tilde X_{T+\varepsilon^2-t})\Big|} \\ &\qquad\qquad + \Big|\esp{\tilde Z_{T+\varepsilon^2-t} g(\tilde X_{T+\varepsilon^2-t})\Big(P^{p,\alpha}_{T-t} - P^{p,\alpha}_{T+\varepsilon^2-t}\Big)}\Big|\\ & \qquad\qquad + \esp{\Big| \tilde Z_{T+\varepsilon^2-t} g(\tilde X_{T+\varepsilon^2-t}) - \hat Z_{T+\varepsilon^2-t} g(\hat X_{T+\varepsilon^2-t})\Big|}\Big)
    \end{align*}
    Regarding the first term, we have, using the Lipschitz continuity and linear growth of $g$, together with the Cauchy--Schwarz inequality, and the estimates from Lemma \ref{lem:props-XZQ} (which also hold for the mollified processes, uniformly in $\varepsilon$ as per Lemma \ref{lem:mollified-coeffs-props}, see also Remark \ref{rem:mollified-pb}), that
    \begin{align*}
        &\esp{\Big|\tilde Z_{T-t} g(\tilde X_{T-t}) - \tilde Z_{T+\varepsilon^2-t} g(\tilde X_{T+\varepsilon^2-t})\Big|} \\
        \le& \esp{\Big|\tilde Z_{T-t} g(\tilde X_{T-t}) - \tilde Z_{T+\varepsilon^2-t} g(\tilde X_{T-t})\Big|} + \esp{\Big|\tilde Z_{T+\varepsilon^2-t} g(\tilde X_{T-t}) - \tilde Z_{T+\varepsilon^2-t} g(\tilde X_{T+\varepsilon^2-t})\Big|} \\
        \le& \esp{\Big|\tilde Z_{T-t} - \tilde Z_{T+\varepsilon^2-t}\Big| g(\tilde X_{T-t})} + \esp{\tilde Z_{T+\varepsilon^2-t} \Big|g(\tilde X_{T-t}) - g(\tilde X_{T+\varepsilon^2-t})\Big|} \\
        \le& C\esp{\Big|\tilde Z_{T-t} - \tilde Z_{T+\varepsilon^2-t}\Big| (1+|\tilde X_{T-t}|)} + C\esp{\tilde Z_{T+\varepsilon^2-t} |\tilde X_{T-t} - \tilde X_{T+\varepsilon^2-t}|} \\
        \le& C \esp{\Big|\tilde Z_{T-t} - \tilde Z_{T+\varepsilon^2-t}\Big|^2}^{\frac12} \esp{(1+|\tilde X_{T-t}|)^2}^{\frac12} + C\esp{\tilde Z_{T+\varepsilon^2-t}^2}^{\frac12} \esp{|\tilde X_{T-t} - \tilde X_{T+\varepsilon^2-t}|^2}^{\frac12} \\
        \le& C(1+|x|)\varepsilon.
    \end{align*}
    For the second term, we set $H:=\tilde Z^{t,x}_{T+\varepsilon^2-t} g\big(\tilde X^{t,x}_{T+\varepsilon^2-t}\big)\in L^2(\cF_{T+\varepsilon^2-t})$. From the Martingale representation theorem on $[T-t,T+\varepsilon^2-t]$, there exists a predictable process $\beta\in\bH^2$ such that
    \begin{align*}
        H=\espc{T-t}{H}+\int_{T-t}^{T+\varepsilon^2-t}\beta_s\,dW_s.
    \end{align*}
    We now compute, by the integration-by-parts formula, the Cauchy--Schwarz inequality, and It\^o's isometry,
    \begin{align*}
        \Big|\esp{\tilde Z_{T+\varepsilon^2-t} g(\tilde X_{T+\varepsilon^2-t})\Big(P^{p,\alpha}_{T-t} - P^{p,\alpha}_{T+\varepsilon^2-t}\Big)}\Big| &= \left|\esp{\int_{T-t}^{T+\varepsilon^2-t}\alpha_s^\top\beta_s\,ds}\right| \\ &\le \Big(\esp{\int_{T-t}^{T+\varepsilon^2-t}|\alpha_s|^2ds}\Big)^{\frac12}\,\Big(\esp{\int_{T-t}^{T+\varepsilon^2-t}|\beta_s|^2ds}\Big)^{\frac12} \\
        &\le C \esp{|H-\espc{T-t}{H}|^2}^{\frac12} \\
        &\le C \esp{|\tilde Z^{t,x}_{T+\varepsilon^2-t} g\big(\tilde X^{t,x}_{T+\varepsilon^2-t}\big)-\tilde Z^{t,x}_{T-t} g\big(\tilde X^{t,x}_{T-t}\big)|^2}^{\frac12},
    \end{align*}
    where we used Remark \ref{rem:alpha-L2-bound} to bound the $L^2$-norm of $\alpha$ (as for $\alpha \in \cA^{M,h}_{p}$, $P^{p,\alpha}_t \in [0,1]$ for all $t \in [0,\infty)$). Hence, a computation similar to that for the first term shows that the second term is also bounded by $C(1+|x|)\varepsilon$.
    
    Regarding the third term, we compute, using the Lipschitz continuity and linear growth of $g$, together with the Cauchy-Schwarz inequality, the estimates from Lemma \ref{lem:props-XZQ} (which also hold for the mollified and shaken processes, uniformly in $\varepsilon$ as per Lemma \ref{lem:mollified-coeffs-props}, see also Remark \ref{rem:mollified-pb}), and Lemma \ref{lem:stability-shaken-mollified}, that
    \begin{align*}
        &\esp{\Big| \tilde Z_{T+\varepsilon^2-t} g(\tilde X_{T+\varepsilon^2-t}) - \hat Z_{T+\varepsilon^2-t} g(\hat X_{T+\varepsilon^2-t})\Big|} \\ \le& \esp{g(\tilde X_{T+\varepsilon^2-t}) \Big|\tilde Z_{T+\varepsilon^2-t} - \hat Z_{T+\varepsilon^2-t}\Big|} + \esp{\hat Z_{T+\varepsilon^2-t} \Big|g(\tilde X_{T+\varepsilon^2-t}) - g(\hat X_{T+\varepsilon^2-t})\Big|} \\
        \le & C\esp{(1+|\tilde X_{T+\varepsilon^2-t}|) \Big|\tilde Z_{T+\varepsilon^2-t} - \hat Z_{T+\varepsilon^2-t}\Big|} + C\esp{\hat Z_{T+\varepsilon^2-t} |\tilde X_{T+\varepsilon^2-t} - \hat X_{T+\varepsilon^2-t}|} \\
        \le & C \esp{\Big|\tilde Z_{T+\varepsilon^2-t} - \hat Z_{T+\varepsilon^2-t}\Big|^2}^{\frac12} \esp{(1+|\tilde X_{T+\varepsilon^2-t}|)^2}^{\frac12} + C\esp{\hat Z_{T+\varepsilon^2-t}^2}^{\frac12} \esp{|\tilde X_{T+\varepsilon^2-t} - \hat X_{T+\varepsilon^2-t}|^2}^{\frac12} \\
        \le& C(1+|x|)\varepsilon.
    \end{align*}
    We thus conclude that $\big|\tilde v^{M,h}(t,x,p)-\hat v^{M,h}(t,x,p)\big| \le C(1+|x|)\varepsilon$, and the proof is complete.
\end{proof}

\medskip\noindent We finally establish analytical properties of $\hat v^{M,h}$ that will prove useful later.

\begin{lem}\label{lem:holder-p-hatvMh}
    Let Assumptions~\ref{h:coeffs} and \ref{h:g Lip} hold. For $\beta\in\cB^{\varepsilon,h}$, define
    \begin{align*}
        \hat v^{M,h}_\beta(t,x,p)
        := \inf_{\alpha\in\cA^{M,h}_{p}}\esp{H^\beta\,P^{p,\alpha}_{T+\varepsilon^2-t}},\qquad (t,x,p)\in[0,T+\varepsilon^2]\times\bR^d\times[0,1],
    \end{align*}
    where $H^\beta:=\hat Z^{t,x,\beta}_{T+\varepsilon^2-t}\,g(\hat X^{t,x,\beta}_{T+\varepsilon^2-t})\ge 0$.
    \begin{enumerate}
        \item Fix $(t,x)\in[0,T+\varepsilon^2]\times\bR^d$. For each $\beta\in\cB^{\varepsilon,h}$, the map $p\mapsto \hat v^{M,h}_\beta(t,x,p)$ is nondecreasing on $[0,1]$. The map $p\mapsto \hat v^{M,h}(t,x,p)$ is nondecreasing on $[0,1]$.
        \item For any $\kappa \in (0,1)$ and each $\beta\in\cB^{\varepsilon,h}$, there exists a constant $C\ge 0$ (independent of $\varepsilon$, $M$ and $h$) such that for all $(t,x)\in[0,T+\varepsilon^2]\times\bR^d$ and all $p,p'\in[0,1]$,
    \[
        \big|\hat v^{M,h}_\beta(t,x,p)-\hat v^{M,h}_\beta(t,x,p')\big|\le C\,(1+|x|)\,|p-p'|^{1-\kappa}.
    \]
        \item For any $\kappa\in(0,1)$ there exists a constant $C\ge 0$ (independent of $\varepsilon$, $M$ and $h$) such that for all $(t,x)\in[0,T+\varepsilon^2]\times\bR^d$ and all $p,p'\in[0,1]$, 
    \[
        \big|\hat v^{M,h}(t,x,p)-\hat v^{M,h}(t,x,p')\big|\le C\,(1+|x|)\,|p-p'|^{1-\kappa}.
    \]
        \item There exists a constant $C\ge0$ (independent of $\varepsilon$, $M$ and $h$) such that, for all $(t,s,x,y,p) \in [0,T+\varepsilon^2] \times [0,T+\varepsilon^2]\times\bR^d\times\bR^d\times[0,1]$, we have
    \begin{align*}
        |\hat v^{M,h}(t,x,p)-\hat v^{M,h}(s,y,p)| \le C\,(1+|x|)\,\big(|t-s|^{\frac12}+|x-y|\big),
    \end{align*}
    \end{enumerate}
\end{lem}
\begin{proof}
    \begin{enumerate}
    \item Fix $\beta\in\cB^{\varepsilon,h}$. Let $0 \le p < p' \le 1$ and let $\alpha' = \sum_{k=0}^\infty \alpha^k \1_{[0,\tau^k]} \1_{[kh,(k+1)h]} \in \cA^{M,h}_{p'}$. Define $\alpha = p/p' \alpha'$ and observe that $$P^{p,\alpha} = p+\int \alpha_s^\top dW_s = \frac{p}{p'} \left(p' + \int (\alpha'_s)^\top dW_s\right) = \frac{p}{p'} P^{p',\alpha'},$$ so $P^{p,\alpha} \in [0,p/p'] \subset [0,1]$, and hence $\alpha \in \cA^{M,h}_{p}$ (as the PCPT structure is obviously kept with the same stopping times if $p>0$ or with $\tau^k=kh$ if $p=0$).
    We thus have, as $p<p'$,
        \begin{align*}
            \hat v^{M,h}_\beta(t,x,p) \le \esp{H^\beta P^{p,\alpha}_{T+\varepsilon^2-t}} = \frac{p}{p'}\esp{H^\beta P^{p',\alpha'}_{T+\varepsilon^2-t}} \le \esp{H^\beta P^{p',\alpha'}_{T+\varepsilon^2-t}},
        \end{align*}
        hence taking the infimum over $\alpha' \in \cA^{M,h}_{p'}$ gives
        \begin{align*}
            \hat v^{M,h}_\beta(t,x,p) \le \hat v^{M,h}_\beta(t,x,p'),
        \end{align*}
        proving the monotonicity of $p\mapsto \hat v^{M,h}_\beta(t,x,p)$. Consequently, we have
    \begin{align*}
        \hat v^{M,h}(t,x,p) = \inf_{\beta\in\cB^{\varepsilon,h}} \hat v^{M,h}_\beta(t,x,p) \le \inf_{\beta\in\cB^{\varepsilon,h}} \hat v^{M,h}_\beta(t,x,p') = \hat v^{M,h}(t,x,p'),
    \end{align*}
    which proves monotonicity of $p \mapsto \hat v^{M,h}(t,x,p)$.
    \item For fixed $\beta \in \cB^{\varepsilon,h}$, let $0 \le p < p' \le 1$. For fixed $\eta>0$, let $\alpha \in \cA^{M,h}_p$ be an $\eta$-optimal control for $\hat v^{M,h}_\beta(t,x,p)$. Similarly as above, consider $\alpha' = (1-p')/(1-p) \alpha$, and observe that
    $$1-P^{p',\alpha'} = 1-p' - \int (\alpha'_s)^\top dW_s = \frac{1-p'}{1-p}\left(1-p - \int (\alpha_s)^\top dW_s\right)=\frac{1-p'}{1-p}(1-P^{p,\alpha}),$$
so that $P^{p',\alpha'} = 1-\frac{1-p'}{1-p}(1-P^{p,\alpha}) \in [1-\frac{1-p'}{1-p}, 1] \subset [0,1]$, and hence $\alpha' \in \cA^{M,h}_{p'}$ (the PCPT structure is again kept with the same stopping times if $p'<1$, or with $\tau^k=kh$ if $p'=1$). %
    We thus have, for $r>1$ such that $\theta := \frac{r-1}{r} \ge 1-\kappa$,
    \begin{align*}
        0 \le \hat v^{M,h}_\beta(t,x,p') - \hat v^{M,h}_\beta(t,x,p) &\le \esp{H^\beta (P^{p',\alpha'}_{T+\varepsilon^2-t} - P^{p,\alpha}_{T+\varepsilon^2-t})} + \eta. \\
        &\le \esp{(H^\beta)^r}^{\frac 1 r} \esp{(P^{p',\alpha'}_{T+\varepsilon^2-t} - P^{p,\alpha}_{T+\varepsilon^2-t})^{\frac r{r-1}}}^{\frac{r-1}{r}} + \eta \\
        &\le C(1+|x|) \esp{P^{p',\alpha'}_{T+\varepsilon^2-t} - P^{p,\alpha}_{T+\varepsilon^2-t}}^{\theta} + \eta \\
        &\le C(1+|x|) (p'-p)^\theta + \eta \le C(1+|x|) (p'-p)^{1-\kappa} + \eta,
    \end{align*}
    where $C$ is independent of $\beta$. As $\eta$ was arbitrary, this proves the desired result.

    \item Now, since $\hat v^{M,h}=\inf_{\beta\in\cB^{\varepsilon,h}}\hat v^{M,h}_\beta$, it follows that
    \begin{align*}
        \big|\hat v^{M,h}(t,x,p)-\hat v^{M,h}(t,x,p')\big| &\le \sup_{\beta\in\cB^{\varepsilon,h}}\big|\hat v^{M,h}_\beta(t,x,p)-\hat v^{M,h}_\beta(t,x,p')\big| \\
        &\le C(1+|x|)|p-p'|^{1-\kappa},
    \end{align*}
    as $C$ obtained above is independent of $\beta$.
    \item 
    By the triangle inequality,
    \begin{align}
        |\hat v^{M,h}(t,x,p)-\hat v^{M,h}(s,y,p)|
        &\le |\hat v^{M,h}(t,x,p)-\hat v^{M,h}(s,x,p)|\\
        &\quad+|\hat v^{M,h}(s,x,p)-\hat v^{M,h}(s,y,p)|. \label{eq triangle regul}
    \end{align}

    Regarding the second term, since $0\le P^{p,\alpha}\le1$ for any control $\alpha$,
    \begin{align*}
        |\hat v^{M,h}(s,x,p)-\hat v^{M,h}(s,y,p)|
        &\le\sup_{\beta\in\cB^{\varepsilon,h}}
        \esp{\left|\hat Z^{s,x,\beta}_{T+\varepsilon^2-s}\,g(\hat X^{s,x,\beta}_{T+\varepsilon^2-s})
        -\hat Z^{s,y,\beta}_{T+\varepsilon^2-s}\,g(\hat X^{s,y,\beta}_{T+\varepsilon^2-s})\right|}\\
        &\le C(1+|x|)|x-y|.
    \end{align*}
    Indeed, for $|x-y|\le1$, the Lipschitz continuity and linear growth of $g$, the Cauchy--Schwarz inequality, and items 1 and 2 of Lemma~\ref{lem:stability-shaken-mollified} give an upper bound $C(1+|y|)|x-y|$, and $1+|y|\le2(1+|x|)$. For $|x-y|>1$, the moment bounds instead give
    \[
        \esp{\left|\hat Z^{s,x,\beta}_{T+\varepsilon^2-s}\,g(\hat X^{s,x,\beta}_{T+\varepsilon^2-s})
        -\hat Z^{s,y,\beta}_{T+\varepsilon^2-s}\,g(\hat X^{s,y,\beta}_{T+\varepsilon^2-s})\right|}
        \le C(2+|x|+|y|)
        \le 3C(1+|x|)|x-y|.
    \]

    We now consider the first term in \eqref{eq triangle regul}. By symmetry in $t$ and $s$, assume $t\le s$.
    The usual increment estimates for SDEs yield, uniformly over $\beta\in\cB^{\varepsilon,h}$,
    \begin{align}\label{eq:H-shaken-time-increment}
        &\esp{\left|\hat Z^{t,x,\beta}_{T+\varepsilon^2-t}\,g(\hat X^{t,x,\beta}_{T+\varepsilon^2-t})-\hat Z^{t,x,\beta}_{T+\varepsilon^2-s}\,g(\hat X^{t,x,\beta}_{T+\varepsilon^2-s})\right|^2}
        \\ &\quad \le C(1+|x|)^2(T+\varepsilon^2-t-(T+\varepsilon^2-s))
        =C(1+|x|)^2(s-t).
    \end{align}
    This follows from the Lipschitz continuity and linear growth of $g$, the fourth-moment increment estimates for $\hat X$ and $\hat Z$, and the Cauchy--Schwarz inequality, as in Lemma~\ref{lem:props-XZQ}. At the common horizon $T+\varepsilon^2-s$, Lemma~\ref{lem:stability-shaken-mollified}1. and 2. also give
    \[
        \sup_{\beta\in\cB^{\varepsilon,h}}
        \esp{\left|\hat Z^{t,x,\beta}_{T+\varepsilon^2-s}\,g(\hat X^{t,x,\beta}_{T+\varepsilon^2-s})-\hat Z^{s,x,\beta}_{T+\varepsilon^2-s}\,g(\hat X^{s,x,\beta}_{T+\varepsilon^2-s})\right|}
        \le C(1+|x|)|t-s|^{1/2}.
    \]

    Fix admissible controls $\alpha,\beta$ 
    . Since $\hat Z^{t,x,\beta}_{T+\varepsilon^2-s}\,g(\hat X^{t,x,\beta}_{T+\varepsilon^2-s})$ is $\cF_{T+\varepsilon^2-s}$-measurable and $P^{p,\alpha}$ is a bounded martingale,
    \[
        \esp{\hat Z^{t,x,\beta}_{T+\varepsilon^2-s}\,g(\hat X^{t,x,\beta}_{T+\varepsilon^2-s})(P^{p,\alpha}_{T+\varepsilon^2-t}-P^{p,\alpha}_{T+\varepsilon^2-s})}=0.
    \]
    Therefore,
    \begin{align*}
        &\esp{\hat Z^{t,x,\beta}_{T+\varepsilon^2-t}\,g(\hat X^{t,x,\beta}_{T+\varepsilon^2-t})P^{p,\alpha}_{T+\varepsilon^2-t}-\hat Z^{t,x,\beta}_{T+\varepsilon^2-s}\,g(\hat X^{t,x,\beta}_{T+\varepsilon^2-s})P^{p,\alpha}_{T+\varepsilon^2-s}}
        \\ &\quad =\esp{(\hat Z^{t,x,\beta}_{T+\varepsilon^2-t}\,g(\hat X^{t,x,\beta}_{T+\varepsilon^2-t})-\hat Z^{t,x,\beta}_{T+\varepsilon^2-s}\,g(\hat X^{t,x,\beta}_{T+\varepsilon^2-s}))P^{p,\alpha}_{T+\varepsilon^2-t}}, \text { and }\\
        &\left|\esp{\hat Z^{t,x,\beta}_{T+\varepsilon^2-t}\,g(\hat X^{t,x,\beta}_{T+\varepsilon^2-t})P^{p,\alpha}_{T+\varepsilon^2-t}-\hat Z^{t,x,\beta}_{T+\varepsilon^2-s}\,g(\hat X^{t,x,\beta}_{T+\varepsilon^2-s})P^{p,\alpha}_{T+\varepsilon^2-s}}\right| \\
        &\quad \le\esp{\left|\hat Z^{t,x,\beta}_{T+\varepsilon^2-t}\,g(\hat X^{t,x,\beta}_{T+\varepsilon^2-t})-\hat Z^{t,x,\beta}_{T+\varepsilon^2-s}\,g(\hat X^{t,x,\beta}_{T+\varepsilon^2-s})\right|}
        \le C(1+|x|)|t-s|^{1/2},
    \end{align*}
    where we used $0\le P^{p,\alpha}_{T+\varepsilon^2-t}\le1$ and \eqref{eq:H-shaken-time-increment}. Since the same admissible sets are used at $t$ and $s$, we conclude that
    \begin{align*}
        |\hat v^{M,h}(t,x,p)-\hat v^{M,h}(s,x,p)|
        &\le\sup_{\alpha,\beta}
        \left|\esp{\hat Z^{t,x,\beta}_{T+\varepsilon^2-t}\,g(\hat X^{t,x,\beta}_{T+\varepsilon^2-t})P^{p,\alpha}_{T+\varepsilon^2-t}-\hat Z^{s,x,\beta}_{T+\varepsilon^2-s}\,g(\hat X^{s,x,\beta}_{T+\varepsilon^2-s})P^{p,\alpha}_{T+\varepsilon^2-s}}\right|\\
        &\le\sup_{\beta}\esp{\left|\hat Z^{t,x,\beta}_{T+\varepsilon^2-t}\,g(\hat X^{t,x,\beta}_{T+\varepsilon^2-t})-\hat Z^{t,x,\beta}_{T+\varepsilon^2-s}\,g(\hat X^{t,x,\beta}_{T+\varepsilon^2-s})\right|}\\
        &\quad +\sup_{\beta}\esp{\left|\hat Z^{t,x,\beta}_{T+\varepsilon^2-s}\,g(\hat X^{t,x,\beta}_{T+\varepsilon^2-s})-\hat Z^{s,x,\beta}_{T+\varepsilon^2-s}\,g(\hat X^{s,x,\beta}_{T+\varepsilon^2-s})\right|}\\
        &\le C(1+|x|)|t-s|^{1/2}.
    \end{align*}
    Combining the time and space estimates with the initial triangle inequality proves the result.
\end{enumerate}
\end{proof}

\medskip
\noindent%
We now regularise the shaken value function. Let $\bar\rho\in\cC_c^\infty(\bR\times\bR^d\times\bR)$ be a nonnegative kernel with unit mass and support contained in $(-1,0)\times\{|y|\le 1\}\times(-1,1)$.
For $\varepsilon\in(0,1]$, we define the rescaled kernel
\begin{align}\label{eq:kernel-txp}
    \bar\rho_{\varepsilon}(s,y,q):=\varepsilon^{-(d+3)}\,\bar\rho\Big(\frac{s}{\varepsilon^2},\frac{y}{\varepsilon},\frac{q}{\varepsilon}\Big),
\end{align}
and we define the regularised value function $\check v^{M,h}$ as the convolution of $\hat v^{M,h}$ with $\bar\rho_{\varepsilon}$ in all three variables:
\begin{align}\label{eq:checkv-def}
    \check v^{M,h}_{\varepsilon}(t,x,p)
    :=\int_{\bR}\!\int_{\bR^d}\!\int_{\bR}\hat v^{M,h}(t-s,x-y,p-q)\,\bar\rho_{\varepsilon}(s,y,q)\,dq\,dy\,ds.
\end{align}
Since $\hat v^{M,h}$ is finite on $[0,1]$ in the $p$-variable and the $q$-support of $\bar\rho_\varepsilon$ is contained in $(-\varepsilon,\varepsilon)$, the regularised value $\check v^{M,h}_\varepsilon$ is finite on $[\varepsilon,1-\varepsilon]$. We shall only use it on this interval.

In the next lemma, thanks to the H\"older regularity of $\hat v^{M,h}$, we provide an estimate for the convolution error. We also provide estimates for the derivatives of $\check v^{M,h}$, which will turn out useful later.

\begin{lem}\label{lem:convolution-error-checkv}
    Let Assumptions~\ref{h:coeffs} and \ref{h:g Lip} hold.
    \begin{enumerate}
        \item Fix $\kappa\in(0,1)$.
    Then there exists a constant $C\ge 0$ (independent of $\varepsilon$, $M$ and $h$) such that for all $\varepsilon\in(0,\frac12)$ and all $(t,x,p)\in[0,T]\times\bR^d\times[\varepsilon,1-\varepsilon]$,
    \[
        \big|\check v^{M,h}_{\varepsilon}(t,x,p)-\hat v^{M,h}(t,x,p)\big|
        \le C\,(1+|x|)\,\varepsilon^{1-\kappa}.
    \]
    \item Fix $\kappa\in(0,1)$. There exists a constant $C\ge 0$ (independent of $\varepsilon$, $M$ and $h$) such that for all $\varepsilon\in(0,\frac12)$ and all $(t,x,p)\in[0,T]\times\bR^d\times[\varepsilon,1-\varepsilon]$, for all $\alpha_1,\alpha_3\in\bN_0$ and all multi-indices $\alpha_2\in\bN_0^d$ such that $\alpha_1+|\alpha_2|+\alpha_3>0$,
    \begin{align*}
         |\partial^{\alpha_1}_t\partial^{\alpha_2}_x\partial^{\alpha_3}_p\check v^{M,h}_{\varepsilon}(t,x,p)|
        &\le C(1+|x|)\varepsilon^{1-\kappa-2\alpha_1-|\alpha_2|-\alpha_3}.
    \end{align*}
    \end{enumerate}
\end{lem}
\begin{proof}
    \begin{enumerate}
        \item Let $\varepsilon\in(0,\frac12)$ and $(t,x,p)\in[0,T]\times\bR^d\times[\varepsilon,1-\varepsilon]$.
    Since $\bar\rho_{\varepsilon}$ is nonnegative with unit mass and supported in $(-\varepsilon^2,0)\times\{|y|\le\varepsilon\}\times(-\varepsilon,\varepsilon)$, we have, using Lemma \ref{lem:holder-p-hatvMh}-3. and 4.,
    \begin{align*}
        &\big|\check v^{M,h}_{\varepsilon}(t,x,p)-\hat v^{M,h}(t,x,p)\big|\\
        &\le \int_{-\varepsilon^2}^0\!\int_{\{|y|\le\varepsilon\}}\!\int_{-\varepsilon}^{\varepsilon} \big|\hat v^{M,h}(t-s,x-y,p-q)-\hat v^{M,h}(t,x,p)\big|\,\bar\rho_{\varepsilon}(s,y,q)\,dq\,dy\,ds \\
        &\le \int_{-\varepsilon^2}^0\!\int_{\{|y|\le\varepsilon\}}\!\int_{-\varepsilon}^{\varepsilon} \Big(\big|\hat v^{M,h}(t-s,x-y,p-q)-\hat v^{M,h}(t,x,p-q)\big| \\ &\qquad\qquad+ \big|\hat v^{M,h}(t,x,p-q)-\hat v^{M,h}(t,x,p)\big|\Big)\bar\rho_{\varepsilon}(s,y,q)\,dq\,dy\,ds \\
        &\le C(1+|x|)\varepsilon^{1-\kappa}.
    \end{align*}
    \item Let $\alpha_1,\alpha_3\in\bN_0$ and $\alpha_2\in\bN_0^d$ satisfy $\alpha_1+|\alpha_2|+\alpha_3>0$, and write $\alpha=(\alpha_1,\alpha_2,\alpha_3)$. Since $\int\partial^\alpha\bar\rho_\varepsilon(s,y,q)\,ds\,dy\,dq=0$, Lemma~\ref{lem:holder-p-hatvMh}-3. and 4. yield
    \begin{align*}
        |\partial^{\alpha_1}_t\partial^{\alpha_2}_x\partial^{\alpha_3}_p\check v^{M,h}_{\varepsilon}(t,x,p)|
        &=|\int_{\bR}\!\int_{\bR^d}\!\int_{\bR}\hat v^{M,h}(t-s,x-y,p-q)\,\partial^\alpha \bar\rho_{\varepsilon}(s,y,q)\,dq\,dy\,ds| \\
        &\le\int_{\bR}\!\int_{\bR^d}\!\int_{\bR} \big|\hat v^{M,h}(t-s,x-y,p-q)-\hat v^{M,h}(t,x,p)\big|\,|\partial^\alpha \bar\rho_{\varepsilon}(s,y,q)|\,dq\,dy\,ds \\
        &\le C(1+|x|)\varepsilon^{1-\kappa}\int_{\bR}\!\int_{\bR^d}\!\int_{\bR} |\partial^\alpha \bar\rho_{\varepsilon}(s,y,q)|\,dq\,dy\,ds \\
        &\le C(1+|x|)\varepsilon^{1-\kappa-2\alpha_1-|\alpha_2|-\alpha_3}.
    \end{align*} 
    \end{enumerate}
\end{proof}

Fix $\varepsilon\in(0,\frac12)$. We are now going to see that the regularised shaken value function $\check v^{M,h}_{\varepsilon}$ satisfies a sub-DPP inequality. Since we only use $\check v^{M,h}_\varepsilon$ on $[\varepsilon,1-\varepsilon]$, we restrict our attention to controls that prevent $P$ from leaving this interval.

Importantly, notice that for $(t,x,p) \in [0,T]\times\bR^d\times[\varepsilon,1-\varepsilon]$ if $\alpha \in \cA^{M,h}_p$ is such that $P^{p,\alpha} \in [\varepsilon,1-\varepsilon]$ a.s., and if $q\in[-\varepsilon,\varepsilon]$, then $\alpha\in\cA^{M,h}_{p-q}$ since $P^{p-q,\alpha}=P^{p,\alpha}-q\in[0,1]$.

We first state a one-step sub-DPP for the shaken value function $\hat v^{M,h}$. The proof, given in Appendix A.4, uses a standard pasting argument, with a rescaling of the continuation controls to preserve the state constraint. We then transfer the inequality to the regularised value function $\check v^{M,h}_\varepsilon$.

\begin{proposition}[Sub-DPP inequality for $\hat v^{M,h}$ between $t$ and $t+h$]\label{prop:subdpp-hatvMh}
    Let $(t,x,p)\in[0,T+\varepsilon^2]\times\bR^d\times[0,1]$.
    Then
    \[
        \hat v^{M,h}(t,x,p)
        \le \inf_{\alpha\in\cA^{M,h}_{p}}\inf_{\beta\in\cB^{\varepsilon,h}}
        \esp{\hat Z^{t,x,\beta}_{(T+\varepsilon^2-t) \wedge h}\,\hat v^{M,h}\big((t+h) \wedge (T+\varepsilon^2),\hat X^{t,x,\beta}_{(T+\varepsilon^2-t) \wedge h},P^{p,\alpha}_{(T+\varepsilon^2-t) \wedge h}\big)}.
    \]
\end{proposition}

\begin{proposition}[Sub-DPP inequality for $\check v^{M,h}_{\varepsilon}$]\label{prop:subdpp-checkvMh-eps}
    Let $(t,x,p)\in[0,T-h]\times\bR^d\times[\varepsilon,1-\varepsilon]$.
    For any $a \in \bR^q$ with $|a|\le M$, set
    \[
        P^{p,a}_s:=p+\int_0^s a^\top dW_u = p + a^\top W_s,\qquad s\in[0,h].
    \]
    Define $\tau^{p,a} := \inf\{s \ge 0: P^{p,a}_s \notin (\varepsilon, 1-\varepsilon)\} \wedge h$.
    Then the stopped control $\alpha := a \mathbf{1}_{[0, \tau^{p,a}]}$ satisfies $P^{p,\alpha} \in [\varepsilon,1-\varepsilon]$ a.s. and
    \[
        \check v^{M,h}_{\varepsilon}(t,x,p)
        \le 
        \esp{\tilde Z^{t,x}_{h}\,\check v^{M,h}_{\varepsilon}\big(t+h,\tilde X^{t,x}_{h},P^{p,\alpha}_{h}\big)}.
    \]
\end{proposition}
\begin{proof}
    Fix $(t,x,p)\in[0,T-h]\times\bR^d\times[\varepsilon,1-\varepsilon]$ and let $a, \tau^{p,a}, \alpha$ be as in the statement.

    For $(s,y,q)$ in the support of $\bar\rho_{\varepsilon}$, consider the constant shaking control
    \[
        \beta \equiv(s,y)\in\cB^{\varepsilon,h}.
    \]

    Next, since $q\in(-\varepsilon,\varepsilon)$ and $-\varepsilon^2<s\le0$ on the support of $\bar\rho_{\varepsilon}$ and $P^{p,\alpha}_u\in[\varepsilon,1-\varepsilon]$ for all $u$ by construction of $\alpha$, we have $P^{p-q,\alpha}_u=P^{p,\alpha}_u-q\in[0,1]$, hence $\alpha \in \cA^{M,h}_{p-q}$.

    From Proposition~\ref{prop:subdpp-hatvMh} applied at $(t-s,x-y,p-q)$ between $t-s$ and $(t+h)-s$, and from Lemma~\ref{lem:link-shaken-mollified}, we get
    \begin{align*}
        \hat v^{M,h}(t-s,x-y,p-q)
        &\le \esp{\hat Z^{t-s,x-y,\beta}_{h}\,\hat v^{M,h}\big(t+h-s,\hat X^{t-s,x-y,\beta}_{h},P^{p-q,\alpha}_{h}\big)} \\
        &= \esp{\tilde Z^{t,x}_{h}\,\hat v^{M,h}\big(t+h-s,\tilde X^{t,x}_{h}-y,P^{p,\alpha}_{h}-q\big)}.
    \end{align*}
    Multiplying by $\bar\rho_{\varepsilon}(s,y,q)$, integrating in $(s,y,q)$ and using Fubini's theorem, we obtain
    \[
        \check v^{M,h}_{\varepsilon}(t,x,p)
        \le \esp{\tilde Z^{t,x}_{h}\,\check v^{M,h}_{\varepsilon}\big(t+h,\tilde X^{t,x}_{h},P^{p,\alpha}_{h}\big)}.
    \]
    This is the claimed inequality.
\end{proof}

We now consider, for any $a \in \bR^q$, the linear operators defined for any $\varphi \in \cC^{1,2,2}([0,T]\times\bR^d\times\bR)$ by \begin{align*}
    \tilde\cL_a\varphi(t,x,p)
        :=&\;\partial_t\varphi(t,x,p)
        + \tilde\mu(t,x)\cdot \partial_x\varphi(t,x,p)
        + \frac12\mathrm{Tr}\!\Big(\tilde\sigma\tilde\sigma^\top(t,x)\,\partial^2_{xx}\varphi(t,x,p)\Big) \\
        &\quad + \frac12 |a|^2\,\partial^2_{pp}\varphi(t,x,p)
        + a^\top \tilde\sigma^\top(t,x)\,\partial^2_{xp}\varphi(t,x,p) \\
        &\; + \tilde b(t,x)\,\varphi(t,x,p)
        + \tilde\lambda(t,x)^\top\tilde\sigma^\top(t,x)\,\partial_x\varphi(t,x,p)
        + \tilde\lambda(t,x)^\top a\,\partial_p\varphi(t,x,p), \\
        \tilde\cM_a\varphi(t,x,p) :=&\; \partial_x \varphi(t,x,p)^\top \tilde\sigma(t,x) + \partial_p \varphi(t,x,p) a^\top + \tilde\lambda(t,x)^\top\,\varphi(t,x,p).
\end{align*}
It\^o's formula gives
\begin{align}\label{eq:generator-tildeLa}
    d\Big(\tilde Z^{t,x}_s\,\varphi\big(t+s,\tilde X^{t,x}_s,P^{p,a}_s\big)\Big)
        = \tilde Z^{t,x}_s\,\big(\tilde\cL_a\varphi\big)\big(t+s,\tilde X^{t,x}_s,P^{p,a}_s\big)\,ds + \tilde Z^{t,x}_s (\tilde\cM_a\varphi)\big(t+s,\tilde X^{t,x}_s,P^{p,a}_s\big) dW_s.
\end{align}

We now prove, following \cite{Krylov1999,Jakobsen2019}, a generator inequality for the regularised shaken value function $\check v^{M,h}_{\varepsilon}$, which is a key ingredient in the proof of the main theorem.

We first collect some elementary estimates on the exit time $\tau^{p,a}$ defined in Proposition~\ref{prop:subdpp-checkvMh-eps}.

\begin{lem}[Exit-time bounds]\label{lem:exit-time-bounds}
    Fix $0<\varepsilon<\eta<\frac12$, $p\in[\eta,1-\eta]$, $a\in\bR^q$ with $|a|\le M$, and $h>0$.
    Recall $P^{p,a}_s=p+a^\top W_s$ and $\tau^{p,a}:=\inf\{s\ge0:P^{p,a}_s\notin(\varepsilon,1-\varepsilon)\}\wedge h$.
    Set $\delta(p):=\min(p-\varepsilon,1-\varepsilon-p)\ge \eta-\varepsilon>0$.
    \begin{enumerate}
        \item %
        We have $\mathbb P(\tau^{p,a}<h)
\le \min\left(\frac{M^2h}{(\eta-\varepsilon)^2},4\exp\left(-\frac{(\eta-\varepsilon)^2}{2M^2h}\right)\right)$.
        \item If $h\le \frac{(\eta-\varepsilon)^2}{2M^2}$, then $\bP(\tau^{p,a}<h)\le \frac12$ and $
            \esp{\tau^{p,a}}\ge \frac{h}{2}$.
    \end{enumerate}
\end{lem}
\begin{proof}
    \begin{enumerate}
        \item Since $\{\tau^{p,a}<h\} \subset \{\sup_{s\in[0,h]}|a^\top W_s|\ge\delta(p)\}$, the bound $\mathbb P(\tau^{p,a}<h)
\le 4\exp\left(-\frac{(\eta-\varepsilon)^2}{2M^2h}\right)$follows from the reflection principle for Brownian motion. The other bound follows from Doob's maximal inequality:
    \begin{align*}
        \pro{\tau^{p,a}<h} \le \pro{\sup_{s\in[0,h]}|a^\top W_s|\ge\delta(p)}
        \le \frac{\esp{|a^\top W_h|^2}}{\delta(p)^2}
        = \frac{|a|^2\,h}{\delta(p)^2}
        \le \frac{M^2\,h}{(\eta-\varepsilon)^2}.
    \end{align*}
        \item Under $h\le(\eta-\varepsilon)^2/(2M^2)$, item~1.\ gives $\bP(\tau^{p,a}<h)\le M^2h/(\eta-\varepsilon)^2\le 1/2$.
    Since $\tau^{p,a}\le h$ a.s.\ and $\tau^{p,a}=h$ on $\{\tau^{p,a}\ge h\}$,
    \[
        \esp{\tau^{p,a}}
        \ge h\,\bP(\tau^{p,a}=h)
        = h\,\big(1-\bP(\tau^{p,a}<h)\big)
        \ge \frac{h}{2},
    \]
    and we are done.
    \end{enumerate}
\end{proof}

\begin{proposition}\label{prop:generator-ineq-checkv}
    Under Assumptions \ref{h:coeffs} and \ref{h:g Lip}, fix $\kappa\in(0,1)$ and $0<\varepsilon
    <\frac12$.
    Let $a\in\bR^q$ satisfy $|a|\le M$ and define $P^{p,a}$ and $\tau^{p,a}$ as in Proposition~\ref{prop:subdpp-checkvMh-eps}.

    Assume that
    \begin{align*}
        h\le \frac{\varepsilon^2}{2M^2}.
    \end{align*}
        
    Then there exists a constant $C\ge 0$ (depending only on the bounds in Assumption~\ref{h:coeffs}, on $\kappa$ and on the kernel $\bar\rho$, but not on $M$, $h$ nor $\varepsilon$) such that, for all $(t,x,p)\in[0,T-h]\times\bR^d\times
    [2\varepsilon,1-2\varepsilon]$,
    \begin{align*}
        \tilde\cL_a\check v^{M,h}_{\varepsilon}(t,x,p) \ge -C (1+|x|)^5\left((1+M^4)\varepsilon^{-3-\kappa}h + \varepsilon^{-1-\kappa} \exp\left(-\frac{\varepsilon^2}{4M^2h}\right)\right).%
    \end{align*}
        
\end{proposition}
\begin{proof}
    Fix $(t,x,p)$ and $a$ as in the statement and define the stopped control $\alpha:=a\1_{[0,\tau^{p,a}]}$, so that $P^{p,\alpha}_s=P^{p,a}_s$ for $s\le\tau^{p,a}$ and $P^{p,\alpha}_s=P^{p,a}_{\tau^{p,a}}$ for $s\ge \tau^{p,a}$.

    By Proposition~\ref{prop:subdpp-checkvMh-eps}, we have
    \begin{align}
        0 \le& \esp{\tilde Z^{t,x}_{h}\,\check v^{M,h}_\varepsilon\big(t+h,\tilde X^{t,x}_{h},P^{p,\alpha}_{h}\big) - \check v^{M,h}_\varepsilon(t,x,p)} \\
        \label{eq:subdpp-one-step-for-generator}=& \esp{\tilde Z^{t,x}_{h}\,\check v^{M,h}_\varepsilon\big(t+h,\tilde X^{t,x}_{h},P^{p,\alpha}_{h}\big) - \tilde Z^{t,x}_{\tau^{p,a}}\,\check v^{M,h}_\varepsilon\big(t+\tau^{p,a},\tilde X^{t,x}_{\tau^{p,a}},P^{p,\alpha}_{\tau^{p,a}}\big)} \\ &+ \esp{\tilde Z^{t,x}_{\tau^{p,a}}\,\check v^{M,h}_\varepsilon\big(t+\tau^{p,a},\tilde X^{t,x}_{\tau^{p,a}},P^{p,\alpha}_{\tau^{p,a}}\big) - \check v^{M,h}_\varepsilon(t,x,p)}
    \end{align}

    Since $\check v^{M,h}_\varepsilon$ is smooth on $[0,T]\times\bR^d\times(\varepsilon,1-\varepsilon)$, we apply It\^o's formula on $[0,\tau^{p,a}]$ to the semimartingale $s\mapsto \tilde Z^{t,x}_s\,\check v^{M,h}_\varepsilon\big(t+s,\tilde X^{t,x}_s,P^{p,a}_s\big)$ (note that $P^{p,\alpha}=P^{p,a}$ on $[0,\tau^{p,a}]$, and $P^{p,\alpha} \in (\varepsilon,1-\varepsilon)$ on $[0,\tau^{p,a})$):
    \begin{align*}
        &\tilde Z^{t,x}_{\tau^{p,a}}\,\check v^{M,h}_\varepsilon\big(t+\tau^{p,a},\tilde X^{t,x}_{\tau^{p,a}},P^{p,a}_{\tau^{p,a}}\big) - \check v^{M,h}_\varepsilon(t,x,p) \\
        =& \int_0^{\tau^{p,a}} \tilde Z^{t,x}_s\,\big(\tilde\cL_a \check v^{M,h}_\varepsilon\big)\big(t+s,\tilde X^{t,x}_s,P^{p,a}_s\big)\,ds + \int_0^{\tau^{p,a}} \tilde Z^{t,x}_s\,(\tilde\cM_a\check v^{M,h}_\varepsilon)\big(t+s,\tilde X^{t,x}_s,P^{p,a}_s\big) dW_s.
    \end{align*}
    Note that, from Lemma \ref{lem:convolution-error-checkv}-2. and the growth bounds on the coefficients from Assumption~\ref{h:coeffs}, there exists a constant $C\ge 0$ such that
    \begin{align*}
        &\esp{\int_0^h \1_{s \le \tau^{p,a}}|\tilde Z^{t,x}_s\,\big(\tilde\cM_a \check v^{M,h}_\varepsilon\big)\big(t+s,\tilde X^{t,x}_s,P^{p,a}_s\big)|^2\,ds} \\ \le& C(1+M^2) \varepsilon^{-2\kappa} \esp{\int_0^h \big| \tilde Z^{t,x}_s (1+|\tilde X^{t,x}_s|)^2 \big|^2\,ds} \\ \le& C(1+M^2) h \varepsilon^{-2\kappa} \sup_{s\le h} \esp{(\tilde Z^{t,x}_s)^4}^{\frac12} \esp{(1+|\tilde X^{t,x}_s|)^8}^{\frac12} < \infty,
    \end{align*}
    hence the local martingale term is a true martingale and its expectation vanishes. Taking expectations,
    \begin{align*}
        \esp{\tilde Z^{t,x}_{\tau^{p,a}}\,\check v^{M,h}_\varepsilon\big(t+\tau^{p,a},\tilde X^{t,x}_{\tau^{p,a}},P^{p,a}_{\tau^{p,a}}\big)}
        = \check v^{M,h}_\varepsilon(t,x,p) + \esp{\int_0^{\tau^{p,a}} \tilde Z^{t,x}_s\,\big(\tilde\cL_a \check v^{M,h}_\varepsilon\big)\big(t+s,\tilde X^{t,x}_s,P^{p,a}_s\big)\,ds}.
    \end{align*}
    Next, for each $s \in [0,h]$, we apply It\^o's formula on $[0,s\wedge\tau^{p,a}]$ once more to the smooth function $(u,x,p)\mapsto (\tilde\cL_a \check v^{M,h}_\varepsilon)(u,x,p)$ along the same diffusion and use again the product form \eqref{eq:generator-tildeLa} (now with $\varphi=\tilde\cL_a \check v^{M,h}_\varepsilon$).
    Integrating between $0$ and $\tau^{p,a}$ and taking expectations yields the second-order expansion
    \begin{align}
        &\esp{\tilde Z^{t,x}_{\tau^{p,a}}\,\check v^{M,h}_\varepsilon\big(t+\tau^{p,a},\tilde X^{t,x}_{\tau^{p,a}},P^{p,a}_{\tau^{p,a}}\big)}
        - \check v^{M,h}_\varepsilon(t,x,p) \\ =& \esp{\tau^{p,a}}\,\tilde\cL_a \check v^{M,h}_\varepsilon(t,x,p) + \esp{\int_0^{\tau^{p,a}}\!\int_0^{s} \tilde Z^{t,x}_r\,\big(\tilde\cL_a\tilde\cL_a \check v^{M,h}_\varepsilon\big)\big(t+r,\tilde X^{t,x}_r,P^{p,a}_r\big)\,dr\,ds} \\
        &+ \esp{\int_0^{\tau^{p,a}}\!\int_0^{s} \tilde Z^{t,x}_r\,\big(\tilde\cM_a\tilde\cL_a \check v^{M,h}_\varepsilon\big)\big(t+r,\tilde X^{t,x}_r,P^{p,a}_r\big)\,dW_r\,ds} \\
        \le& \esp{\tau^{p,a}}\,\tilde\cL_a \check v^{M,h}_\varepsilon(t,x,p) + \esp{h\int_0^{h} \1_{r \le \tau^{p,a}} \tilde Z^{t,x}_r\,\left|\big(\tilde\cL_a\tilde\cL_a \check v^{M,h}_\varepsilon\big)\big(t+r,\tilde X^{t,x}_r,P^{p,a}_r\big)\right|\,dr} \\
        &+ \int_0^h \esp{\1_{s \le \tau^{p,a}}\int_0^{s\wedge\tau^{p,a}} \tilde Z^{t,x}_r\,\big(\tilde\cM_a\tilde\cL_a \check v^{M,h}_\varepsilon\big)\big(t+r,\tilde X^{t,x}_r,P^{p,a}_r\big)\,dW_r} ds \\
        \le& \esp{\tau^{p,a}}\,\tilde\cL_a \check v^{M,h}_\varepsilon(t,x,p) + C (1+M^4)\,\varepsilon^{-3-\kappa} h^2 \sup_{0 \le r \le h}\esp{\tilde Z^{t,x}_r (1+|\tilde X^{t,x}_r|)^5} \\ 
        &- \int_0^h \esp{\1_{s > \tau^{p,a}}\int_0^{\tau^{p,a}} \tilde Z^{t,x}_r\,\big(\tilde\cM_a\tilde\cL_a \check v^{M,h}_\varepsilon\big)\big(t+r,\tilde X^{t,x}_r,P^{p,a}_r\big)\,dW_r} ds \\
        \le& \esp{\tau^{p,a}}\,\tilde\cL_a \check v^{M,h}_\varepsilon(t,x,p) + C (1+M^4)(1+|x|)^5\,\varepsilon^{-3-\kappa}h^2 \\ 
        &+ \left|\esp{(h-\tau^{p,a})\int_0^{\tau^{p,a}} \tilde Z^{t,x}_r\,\big(\tilde\cM_a\tilde\cL_a \check v^{M,h}_\varepsilon\big)\big(t+r,\tilde X^{t,x}_r,P^{p,a}_r\big)\,dW_r}\right| \\
        \le& \esp{\tau^{p,a}}\,\tilde\cL_a \check v^{M,h}_\varepsilon(t,x,p) + C (1+M^4)(1+|x|)^5\,\varepsilon^{-3-\kappa}h^2 \\ 
        &+ \esp{(h-\tau^{p,a})^2}^{\frac12}\esp{\int_0^{\tau^{p,a}} (\tilde Z^{t,x}_r)^2\,\left|\big(\tilde\cM_a\tilde\cL_a \check v^{M,h}_\varepsilon\big)\big(t+r,\tilde X^{t,x}_r,P^{p,a}_r\big)\right|^2 dr }^{\frac12} \\
        \le& \esp{\tau^{p,a}}\,\tilde\cL_a \check v^{M,h}_\varepsilon(t,x,p) + C (1+M^4)(1+|x|)^5\,\varepsilon^{-3-\kappa}h^2 \\ 
        &+ C(1+M^3)\varepsilon^{-2-\kappa}h^{\frac32} \frac{M}{\varepsilon}\esp{\int_0^h (\tilde Z^{t,x}_r)^2\,(1+|\tilde X^{t,x}_r|)^8 dr }^{\frac12} \\
        \le& \esp{\tau^{p,a}}\,\tilde\cL_a \check v^{M,h}_\varepsilon(t,x,p) + C (1+M^4)(1+|x|)^5\,\varepsilon^{-3-\kappa}h^2,
        \label{eq:ito-2-expansion}
    \end{align}
    where the bounds $\|\partial_t^j\partial_x^\nu\tilde f\|_\infty\le C_{j,\nu}\varepsilon^{1-2j-|\nu|}$ for $2j+|\nu|>0$ and $f\in\{\mu,\sigma,b,\lambda\}$ come from classical bounds on derivatives of mollified coefficients. Combining these bounds with Lemma~\ref{lem:convolution-error-checkv} yields the bounds on $\tilde\cM_a\tilde\cL_a \check v^{M,h}_\varepsilon$ and $\tilde\cL_a\tilde\cL_a \check v^{M,h}_\varepsilon$.

    On $[\tau^{p,a},h]$, the control is null so that $P^{p,\alpha}_s=P^{p,a}_{\tau^{p,a}}$ is constant; hence we may apply It\^o's formula to $s\mapsto \tilde Z^{t,x}_s\,\check v^{M,h}_\varepsilon\big(t+s,\tilde X^{t,x}_s,P^{p,a}_{\tau^{p,a}}\big)$ (which only requires smoothness of $\check v^{M,h}_\varepsilon(\cdot,\cdot,p')$ in $(t,x)$ for fixed $p'\in\{\varepsilon,1-\varepsilon\}$).
    This gives, the local martingale term being again a true martingale,
    \begin{align}
        &\esp{\tilde Z^{t,x}_{h}\,\check v^{M,h}_\varepsilon\big(t+h,\tilde X^{t,x}_{h},P^{p,\alpha}_{h}\big)-\tilde Z^{t,x}_{\tau^{p,a}}\,\check v^{M,h}_\varepsilon\big(t+\tau^{p,a},\tilde X^{t,x}_{\tau^{p,a}},P^{p,a}_{\tau^{p,a}}\big)} \\ \le& \esp{\int_0^h \1_{s>\tau^{p,a}} \tilde Z^{t,x}_s\,\big|\big(\tilde\cL_0 \check v^{M,h}_\varepsilon\big)\big(t+s,\tilde X^{t,x}_s,P^{p,\alpha}_s\big)\big|\, ds} \\
        \le& C\varepsilon^{-1-\kappa}\int_0^h \pro{s>\tau^{p,a}}^{\frac12}\,\esp{(\tilde Z^{t,x}_s)^2\,\big(1+|\tilde X^{t,x}_s|\big)^6}^{\frac12} ds \\
        \le& C(1+|x|)^3\varepsilon^{-1-\kappa} h \exp\left(-\frac{\varepsilon^2}{4M^2h}\right).\label{eq:ito-1b-checkv} %
    \end{align}
    We hence obtain, combining \eqref{eq:subdpp-one-step-for-generator} with \eqref{eq:ito-2-expansion} and \eqref{eq:ito-1b-checkv}, that
    \begin{align*}
        0\le& \esp{\tau^{p,a}}\,\tilde\cL_a \check v^{M,h}_\varepsilon(t,x,p) + C (1+|x|)^5\left((1+M^4)\varepsilon^{-3-\kappa}h^2 + \varepsilon^{-1-\kappa} h \exp\left(-\frac{\varepsilon^2}{4M^2h}\right)\right),%
    \end{align*}
    hence, since $\esp{\tau^{p,a}}\ge h/2$,
    \begin{align*}
        \tilde\cL_a \check v^{M,h}_\varepsilon(t,x,p) &\ge -\frac{1}{\esp{\tau^{p,a}}}C (1+|x|)^5\left((1+M^4)\varepsilon^{-3-\kappa}h^2 + \varepsilon^{-1-\kappa} h \exp\left(-\frac{\varepsilon^2}{4M^2h}\right)\right) \\
        &\ge -C (1+|x|)^5\left((1+M^4)\varepsilon^{-3-\kappa}h + \varepsilon^{-1-\kappa} \exp\left(-\frac{\varepsilon^2}{4M^2h}\right)\right),%
    \end{align*}
    proving the result.

\end{proof}

\begin{theorem}\label{thm:decomposition-main-error}
    Under Assumptions \ref{h:coeffs} and \ref{h:g Lip}, fix $\kappa\in(0,\frac12)$, $\varepsilon\in(0,\frac14)$%
    , and let $M\ge 1$ and $h>0$.
    Assume in addition that $h\le 
    \varepsilon^2/(2M^2)$.
    Then for all $(t,x,p)\in[0,T]\times\bR^d\times%
    [2\varepsilon,1-2\varepsilon]$ one has
    \begin{align} \label{eq:decomposition-main-error}
        0\le v^{M,h}(t,x,p)-v(t,x,p)
        &\le C\,(1+|x|)^{k_{\kappa}}\,\Big(%
        M^{-1+\kappa}
        + \varepsilon^{1-\kappa} \\ &\qquad + (1+M^4)\varepsilon^{-3-\kappa}h + \varepsilon^{-1-\kappa} \exp\left(-\frac{\varepsilon^2}{4M^2h}\right)\Big).%
    \end{align}
\end{theorem}

\begin{proof}
    Since $v\le v^{M,h}$, it suffices to upper bound $v^{M,h}-v$.
    We decompose as follows:
    \begin{align}\label{eq:triangle-main-error}
        (v^{M,h}-v)(t,x,p)
        &\le \big|v^{M,h}-\hat v^{M,h}\big|
        + \big|\hat v^{M,h}-\check v^{M,h}_{\varepsilon}\big| + \big(\check v^{M,h}_{\varepsilon}-\tilde v^{M,2\varepsilon}\big)
        + \big|\tilde v^{M,2\varepsilon}-v^{M,2\varepsilon}\big|+\big|v^{M,2\varepsilon}-v\big| \\
        &\le
        C\,(1+|x|)^{k_{\kappa}}\,(%
        M^{-1+\kappa}
        + \varepsilon^{1-\kappa}) + \big(\check v^{M,h}_{\varepsilon}-\tilde v^{M,2\varepsilon}\big) , \label{eq:preliminary-error-decomposition}
    \end{align}
    where we used Lemma~\ref{lem:stability-vMh-regularised} for the first term, Lemma~\ref{lem:convolution-error-checkv} for the second term, Lemma~\ref{lem:stability-vMeta-mollified} for the fourth term and Theorem~\ref{cor:domain-truncation-vMeta-v} for the last term.

    \medskip
    \noindent We now focus on the term $\check v^{M,h}_{\varepsilon}-\tilde v^{M,2\varepsilon}$. First, for any $(x,p) \in \bR^d \times [2\varepsilon,1-2\varepsilon]$, as $\tilde v^{M,2\varepsilon}(T,x,p) = g(x)p = \hat v^{M,h}(T+\varepsilon^2,x,p)$, one has, using Lemma~\ref{lem:holder-p-hatvMh}-3. and 4., that
    \begin{align}
        \big|\check v^{M,h}_\varepsilon(T,x,p)-\tilde v^{M,2\varepsilon}(T,x,p)\big| &= \big|\check v^{M,h}_\varepsilon(T,x,p)-\hat v^{M,h}(T+\varepsilon^2,x,p)\big| \\ &\le \int_\bR \int_{\bR^d}\int_{\bR}\big|\hat v^{M,h}(T-s,x-y,p-q)-\hat v^{M,h}(T+\varepsilon^2,x,p)\big|\,\bar\rho_\varepsilon(s,y,q)\,ds\,dy\,dq \\
        &\le C(1+|x|)\,\varepsilon^{1-\kappa}.\label{eq:terminal-error}
    \end{align}

    Fix $(t,x,p)\in[0,T]\times\bR^d\times[2\varepsilon,1-2\varepsilon]$ and take any $\alpha\in\cA^{M,2\varepsilon}_p$. Write $S:=T-t$, $P_s:=P^{p,\alpha}_s$ and abbreviate $(\tilde X_s,\tilde Z_s):=(\tilde X^{t,x}_s,\tilde Z^{t,x}_s)$. Applying It\^o's formula up to $T-t$ gives
    \begin{align*}
        \check v^{M,h}_\varepsilon(t,x,p)
        &=\esp{\tilde Z^{t,x}_{T-t}\check v^{M,h}_\varepsilon(T,\tilde X^{t,x}_{T-t},P^{p,\alpha}_{T-t})}\\
        &\quad-\esp{\int_0^{T-t}\tilde Z^{t,x}_s
        \big(\tilde\cL_{\alpha_s}\check v^{M,h}_\varepsilon\big)
        (t+s,\tilde X^{t,x}_s,P^{p,\alpha}_s)\,ds} \\
        &=\esp{\tilde Z^{t,x}_{T-t}\check v^{M,h}_\varepsilon(T,\tilde X^{t,x}_{T-t},P^{p,\alpha}_{T-t})}\\
        &\quad-\esp{\int_0^{(T-t-h)^+}\tilde Z^{t,x}_s
        \big(\tilde\cL_{\alpha_s}\check v^{M,h}_\varepsilon\big)
        (t+s,\tilde X^{t,x}_s,P^{p,\alpha}_s)\,ds} \\
        &\quad-\esp{\int_{(T-t-h)^+}^{T-t}\tilde Z^{t,x}_s
        \big(\tilde\cL_{\alpha_s}\check v^{M,h}_\varepsilon\big)
        (t+s,\tilde X^{t,x}_s,P^{p,\alpha}_s)\,ds}.
    \end{align*}
    Here the stochastic integral has zero expectation by the derivative bounds of Lemma~\ref{lem:convolution-error-checkv}, the bound on $\alpha$, and the moment estimates for $(\tilde X^{t,x},\tilde Z^{t,x})$.
    
    Regarding the integral between $0$ and $(T-t-h)^+$ (of nonzero length only if $t<T-h$), since $0 \le s \le (T-t-h)^+$ gives $t+s\le T-h$, and since $P^{p,\alpha}_s\in[2\varepsilon,1-2\varepsilon]$, Proposition~\ref{prop:generator-ineq-checkv} gives
    \begin{align*}
        &\esp{\int_0^{(T-t-h)^+}\tilde Z^{t,x}_s
        \big(\tilde\cL_{\alpha_s}\check v^{M,h}_\varepsilon\big)
        (t+s,\tilde X^{t,x}_s,P^{p,\alpha}_s)\,ds} \\ &\ge -C\left((1+M^4)\varepsilon^{-3-\kappa}h
        +\varepsilon^{-1-\kappa}\exp\left(-\frac{\varepsilon^2}{4M^2h}\right)\right)\esp{\int_0^{(T-t-h)^+}\tilde Z^{t,x}_s(1+|\tilde X^{t,x}_s|)^5 ds} \\
        &\ge -C(1+|x|^5)\left((1+M^4)\varepsilon^{-3-\kappa}h
        +\varepsilon^{-1-\kappa}\exp\left(-\frac{\varepsilon^2}{4M^2h}\right)\right).
    \end{align*}
    Regarding the integral between $(T-t-h)^+$ and $T-t$, one has from Lemma~\ref{lem:convolution-error-checkv}, since $T-t - (T-t-h)^+ \le h$ and $h(1+M^2) \le \varepsilon^2$,
    \begin{align*}  \left|\esp{\int_{(T-t-h)^+}^{T-t}\tilde Z^{t,x}_s
        \big(\tilde\cL_{\alpha_s}\check v^{M,h}_\varepsilon\big)
        (t+s,\tilde X^{t,x}_s,P^{p,\alpha}_s)\,ds}\right| %
        &\le C(1+M^2)\varepsilon^{-1-\kappa}\esp{\int_{(T-t-h)^+}^{T-t} \tilde Z^{t,x}_s(1+|\tilde X^{t,x}_s|)^3 ds} \\ &\le C(1+|x|)^3h(1+M^2)\varepsilon^{-1-\kappa} \le C(1+|x|)^3 \varepsilon^{1-\kappa}.
\end{align*}
Finally, applying \eqref{eq:terminal-error} at $(\tilde X^{t,x}_{T-t},P^{p,\alpha}_{T-t})$
and using $\mathbb E[\tilde Z^{t,x}_{T-t}(1+|\tilde X^{t,x}_{T-t}|)]\le C(1+|x|)$,
we obtain
\begin{align*}
\esp{\tilde Z^{t,x}_{T-t}\check v^{M,h}_\varepsilon(T,\tilde X^{t,x}_{T-t},P^{p,\alpha}_{T-t})} - \esp{\tilde Z^{t,x}_{T-t} g(\tilde X^{t,x}_{T-t})P^{p,\alpha}_{T-t}} \le  C(1+|x|)\varepsilon^{1-\kappa} \le C(1+|x|)^3\varepsilon^{1-\kappa}.
\end{align*}
Hence, combining the three inequalities gives
\begin{align*}
\check v^{M,h}_\varepsilon(t,x,p) - \esp{\tilde Z^{t,x}_{T-t}g(\tilde X^{t,x}_{T-t})P^{p,\alpha}_{T-t}} \le& C(1+|x|)^5\left((1+M^4)\varepsilon^{-3-\kappa}h
        +\varepsilon^{-1-\kappa}\exp\left(-\frac{\varepsilon^2}{4M^2h}\right)\right) \\ &+ C(1+|x|)^3\varepsilon^{1-\kappa},
\end{align*}%
and taking the supremum over $\alpha\in\cA^{M,2\varepsilon}_p$ gives
    \begin{align*}
        (\check v^{M,h}_\varepsilon-\tilde v^{M,2\varepsilon})(t,x,p)
        &\le C(1+|x|)^3\varepsilon^{1-\kappa}+C(1+|x|)^5\left((1+M^4)\varepsilon^{-3-\kappa}h
        +\varepsilon^{-1-\kappa}\exp\left(-\frac{\varepsilon^2}{4M^2h}\right)\right).
    \end{align*}
    Altogether, combining the above estimate with \eqref{eq:preliminary-error-decomposition}, we get
    \begin{align}
        (v^{M,h}-v)(t,x,p)
        &\le 
        C\,(1+|x|)^{k_{\kappa}}\,(%
        M^{-1+\kappa}
        + \varepsilon^{1-\kappa} + (1+M^4)\varepsilon^{-3-\kappa}h + \varepsilon^{-1-\kappa} \exp\left(-\frac{\varepsilon^2}{4M^2h}\right)).%
    \end{align}
\end{proof}

The regularity of $v$ in the probability variable allows us to extend the interior estimate uniformly to $p\in[0,1]$.

\begin{cor}\label{cor:global-p-error}
    Under Assumptions~\ref{h:coeffs} and~\ref{h:g Lip}, fix $\kappa\in(0,\frac12)$. There exist constants $C_\kappa\ge0$ and $k_\kappa\ge1$, independent of $M$, $h$ and $\varepsilon$, such that, for every $M\ge1$, $\varepsilon\in(0,\frac14)$ and $h>0$ satisfying $h\le\varepsilon^2/(2M^2)$, and every $(t,x,p)\in[0,T]\times\bR^d\times[0,1]$,
    \begin{align*}
        0\le v^{M,h}(t,x,p)-v(t,x,p)
        &\le C_\kappa(1+|x|)^{k_\kappa}\Big(
        M^{-1+\kappa}+\varepsilon^{1-\kappa}    +(1+M^4)\varepsilon^{-3-\kappa}h
        +\varepsilon^{-1-\kappa}\exp\!\left(-\frac{\varepsilon^2}{4M^2h}\right)\Big).
    \end{align*}
\end{cor}
\begin{proof}
    Fix $(t,x,p)\in[0,T]\times\bR^d\times[0,1]$. From Theorem~\ref{thm:decomposition-main-error}, it is enough to consider $p \in [0,2\varepsilon] \cup [1-2\varepsilon,1]$.
    
    The null control is admissible for every initial probability, and $\cA^{M,h}_p\subset\cA_p$. Hence
    \[
        0\le v(t,x,p)\le v^{M,h}(t,x,p)\le p\esp{Z^{t,x}_{T-t}g(X^{t,x}_{T-t})},
        \qquad p\in[0,1].
    \]
    Moreover, both values equal $0$ at $p=0$ and $\esp{Z^{t,x}_{T-t}g(X^{t,x}_{T-t})}$ at $p=1$, since a $[0,1]$-valued martingale starting at an endpoint is constant.

    If $p\in[0,2\varepsilon]$, Lemma~\ref{lem:props-XZQ} gives
    \[
        0\le v^{M,h}(t,x,p)-v(t,x,p)
        \le 2\varepsilon\esp{Z^{t,x}_{T-t}g(X^{t,x}_{T-t})}
        \le C(1+|x|)\varepsilon
        \le C(1+|x|)\varepsilon^{1-\kappa}.
    \]
    If $p\in[1-2\varepsilon,1]$, the H\"older estimate in Lemma~\ref{lem:props-v} yields
    \begin{align*}
        0\le v^{M,h}(t,x,p)-v(t,x,p)
        &\le v(t,x,1)-v(t,x,p)\\
        &\le C_\kappa(1+|x|)(1-p)^{1-\kappa}
        \le C_\kappa(1+|x|)\varepsilon^{1-\kappa}.
    \end{align*}
\end{proof}

For fixed $M$, sending $h$ to zero does not in general remove the control-space truncation error. Conversely, at fixed $h$, sending $M$ to infinity violates the step-size condition in Theorem~\ref{thm:decomposition-main-error}. However, the following corollary gives a joint choice of $M$ and $h$ (and $\varepsilon$) to balance the error terms.

\begin{cor}\label{cor:rates}
    Under Assumptions~\ref{h:coeffs} and~\ref{h:g Lip}, fix $\kappa\in(0,\frac12)$.
    There exist constants $C_\kappa\ge0$ and $k_\kappa\ge1$ such that the following assertions hold.
    \begin{enumerate}[(i)]
        \item \emph{(Rate in $h$.)} Set
        \[
            M_h=h^{-1/8},\qquad \varepsilon_h=h^{1/8}.
        \]
        For $h$ small enough and every $(t,x,p)\in[0,T]\times\bR^d\times[0,1]$,
        \[
            0\le v^{M_h,h}(t,x,p)-v(t,x,p)
            \le C_\kappa(1+|x|)^{k_\kappa}h^{(1-\kappa)/8}.
        \]
        As $\kappa\downarrow0$, the exponent approaches $1/8$.
        \item \emph{(Equivalent rate in $M$.)} Set
        \[
            h_M=M^{-8},\qquad \varepsilon_M=M^{-1}.
        \]
        For $M$ large enough and every $(t,x,p)\in[0,T]\times\bR^d\times[0,1]$,
        \[
            0\le v^{M,h_M}(t,x,p)-v(t,x,p)
            \le C_\kappa(1+|x|)^{k_\kappa}M^{-1+\kappa}.
        \]
    \end{enumerate}
\end{cor}
\begin{proof}
    For $h$ small enough, $M_h\ge1$, $\varepsilon_h<1/4$, and
    $\varepsilon_h^2/M_h^2=h^{1/2}$, so that $h\le\varepsilon_h^2/(2M_h^2)$.
    Since $M_h=h^{-1/8}$ and $\varepsilon_h=h^{1/8}$, the error terms in \eqref{eq:decomposition-main-error} satisfy
    \[
        M_h^{-1+\kappa}=\varepsilon_h^{1-\kappa}
        =M_h^4\varepsilon_h^{-3-\kappa}h=h^{(1-\kappa)/8},
    \]
    and $1+M_h^4\le2M_h^4$. The remaining term satisfies
    \[
        \varepsilon_h^{-1-\kappa}
        \exp\!\left(-\frac{\varepsilon_h^2}{4M_h^2h}\right)
        = h^{-(1+\kappa)/8}\exp\!\left(-\frac{1}{4\sqrt h}\right)
        = o\!\left(h^{(1-\kappa)/8}\right).
    \]
    Assertion (i) follows from Corollary \ref{cor:global-p-error}.
    For (ii), substitute $h=h_M$ in (i); then $M_h=M$ and $\varepsilon_h=\varepsilon_M$.
\end{proof}

\begin{remark}[Fixed control bound]
Suppose that the target value is $v^{M,0}$ for a fixed $M\ge1$, i.e. that the control set is bounded from the outset.
In the proof of Theorem~\ref{thm:decomposition-main-error}, replace $v$ by $v^{M,0}$.
For $p\in[2\varepsilon,1-2\varepsilon]$, stopping a nearly optimal $M$-bounded control at the first hit of $\{2\varepsilon,1-2\varepsilon\}$ preserves the bound $M$.
By Lemma~\ref{lem:absorption-improves-eta}, the resulting error
$v^{M,2\varepsilon}-v^{M,0}$ is bounded by $C_\kappa(1+|x|)\varepsilon^{1-\kappa}$.
The term $M^{-1+\kappa}$ is therefore absent from the error bound on $[2\varepsilon,1-2\varepsilon]$, hence on $[0,1]$ by the boundary estimates in the proof of Corollary \ref{cor:global-p-error}.

For $\varepsilon_h=h^{1/4}$, the step-size condition holds for all sufficiently small $h$.
The two error terms are of order $h^{(1-\kappa)/4}$ and the exponential term is of higher order, since $\varepsilon_h^2/(M^2h)=1/(M^2\sqrt h)$.
Thus for every $\kappa\in(0,\frac12)$ and all sufficiently small $h$,
\[
    0\le v^{M,h}(t,x,p)-v^{M,0}(t,x,p)
    \le C_{\kappa,M}(1+|x|)^{k_\kappa}h^{(1-\kappa)/4},
    \qquad (t,x,p)\in[0,T]\times\bR^d\times[0,1].
\]
As $\kappa\downarrow0$, the exponent approaches $1/4$.
\end{remark}

\section{Numerical experiments}\label{sec:numerics}

In this section, we investigate the fully discrete implementation of the PCPT approximation. The PCPT value \(v^{M,h}\) is defined using controls that are piecewise constant on the decision grid, adapted to the continuous Brownian filtration, and stopped continuously upon hitting \(\{0,1\}\). The numerical implementation differs in three respects: controls are adapted to the grid filtration, \(X\) is approximated by an Euler scheme, and exits of \(P\) are detected only at grid times and followed by projection onto \([0,1]\). The resulting finite-dimensional control problem is parametrized by neural networks and optimized using the method of \cite{han2016deep} (see also \cite{han2018solving} for the solution of semilinear PDEs or BSDEs by deep neural networks). Neural approximation, Monte Carlo, and optimization errors are present but are not theoretically analyzed here.

We consider the uniform grid $t_n:=nh$, $0\le n\le N$, with $h:=T/N$, and the discrete completed filtration
\begin{align*}
    \cG_n^h:=\sigma\bigl(W_{t_{k+1}}-W_{t_k}:0\le k<n\bigr) \vee \cN^{\bP}
    =\sigma(W_{t_k}:0\le k\le n) \vee \cN^{\bP},
    \qquad 0\le n\le N,
\end{align*}
where $\cN^{\bP}$ denotes the collection of $\bP$-null sets. We also fix $M>0$.

Regarding the uncontrolled diffusion $X$, we consider its standard Euler scheme. Starting from $x\in\bR^d$ at time $t_n$, define $X^{n,x}_0:=x$ and, for $0\le k<N-n$,
\begin{align}
\label{eq:x disc redline}
X^{n,x}_{k+1}
&:=X^{n,x}_k
+\mu(t_{n+k},X^{n,x}_k)h
+\sigma(t_{n+k},X^{n,x}_k)
\bigl(W_{t_{k+1}}-W_{t_{k}}\bigr).
\end{align}
For $0\le k\le N-n$, define also
\begin{align}
\label{eq:z disc redline}
Z^{n,x}_k
&:=\exp\Bigg(
\sum_{j=0}^{k-1}
\left(b(t_{n+j},X^{n,x}_j)-\frac12|\lambda(t_{n+j},X^{n,x}_j)|^2\right)h\\
&\hspace{3.5cm}
+\sum_{j=0}^{k-1}\lambda(t_{n+j},X^{n,x}_j)^\top
\bigl(W_{t_{j+1}}-W_{t_{j}}\bigr)
\Bigg).
\end{align}

Regarding the controlled process, for $p\in[0,1]$ and a control
\[
\alpha=(\alpha_k)_{0\le k<N},
\qquad
\alpha_k\in L^0(\cG_k^h;\bR^q),
\qquad
|\alpha_k|\le M,
\]
define
\begin{align}
\label{eq:p disc raw redline}
\widetilde P^{p,\alpha}_0&:=p,
&
\widetilde P^{p,\alpha}_{n+1}
&:=\widetilde P^{p,\alpha}_n
+\alpha_n^\top(W_{t_{n+1}}-W_{t_n}),
\end{align}
and, with the convention $\inf\varnothing=+\infty$,
\begin{align}
\label{eq:p disc tau redline}
\tau^{p,\alpha}
&:=\inf\bigl\{1\le n\le N:\widetilde P^{p,\alpha}_n\notin(0,1)\bigr\}.
\end{align}
The projected and absorbed process is
\begin{align}
\label{eq:p disc redline}
P^{p,\alpha}_n
&:=\widetilde P^{p,\alpha}_n\mathbf 1_{\{n<\tau^{p,\alpha}\}}
+\mathbf 1_{\{n\ge\tau^{p,\alpha},\widetilde P^{p,\alpha}_{\tau^{p,\alpha}}\ge1\}},
\qquad 0\le n\le N.
\end{align}
If $p\in\{0,1\}$, we set $P^{p,\alpha}_n=p$ for every $n$.

\begin{remark}
The construction above resembles the PCPT processes of Section~\ref{sec:pcpt}, but the two admissible classes should not be confused. Here the controls and exit rule are adapted to the discrete filtration $(\cG_n^h)$, whereas the PCPT controls and their continuous hitting times are adapted to the full Brownian filtration. %
\end{remark}

\begin{remark}\label{rem:projection-bias-redline}
The projection in \eqref{eq:p disc redline} generally destroys the martingale property. The resulting bias is a genuine feature of the discrete control problem and can therefore be exploited by the optimizer. The toy example below quantifies this effect and shows that, although the bias vanishes under a suitable joint limit for $M$ and $h$, it can remain substantial for the values of $N$ and $M$ used in practice.
\end{remark}

\subsection{Neural-network approximation}

We now fix $(x,p)\in\bR^d\times[0,1]$ and approximate the fully discrete problem starting from $(0,x,p)$. Write $X_n:=X^{0,x}_n$, $Z_n:=Z^{0,x}_n$, and let $P_n$ denote the projected process at time $t_n$.

In the implementation, in contrast to \cite{han2016deep}, we represent the controls at all dates by a single time-shared neural network (where the normalized time $t_n/T$ is included among the inputs), which we call ``shared neural network'' in the sequel. More specifically, for a parameter set $\Theta^N$, let
\[
\cN^N:[0,1]\times\bR^d\times[0,1]\times\Theta^N
\longrightarrow\{a\in\bR^q:|a|\le M\}.
\]
The bound is enforced through
\begin{align}
\label{eq:NN tanh M redline}
\cN^N(s,x,p,\theta)
&:=\Psi^M\bigl(\widetilde\cN^N(s,x,p,\theta)\bigr),
\end{align}
where $\widetilde\cN^N$ is unconstrained and $\Psi^M$ maps $\bR^q$ into the Euclidean ball of radius $M$. We use $\Psi^M(z)=M\tanh(z)$ in dimension one; in higher dimension, one may use, for instance, the radial map $\Psi^M(z)=Mz/\sqrt{1+|z|^2}$.

We denote by $\widehat X_n$ the normalized state feature passed to the network. In the financial experiments below, in which $d=1$, we take
\[
\widehat X_n:=\frac{\log(\frac{X_n\vee 10^{-6}}{x})}{s_X},
\]
where $s_X>0$ is a fixed scale and the maximum with $10^{-6}$ is to avoid taking the logarithm of a nonpositive number (which could happen due to the Euler discretization). Thus, the network inputs are $(t_n/T,\widehat X_n,P_n)$.

Given $\theta\in\Theta^N$, set $\widetilde P^{p,\theta}_0=P^{p,\theta}_0=p$ and, recursively,
\begin{align}\label{eq:p disc theta redline}
\widetilde P^{p,\theta}_{n+1}
&:=\widetilde P^{p,\theta}_n
+\cN^N(t_n/T,\widehat X_n,P^{p,\theta}_n,\theta)^\top
(W_{t_{n+1}}-W_{t_n}),
\end{align}
with $\tau^{p,\theta}$ and $P^{p,\theta}$ defined as in
\eqref{eq:p disc tau redline}--\eqref{eq:p disc redline}. The ideal value associated with this neural class is
\begin{align}
\label{eq:deep-learning-problem-redline}
V_{\mathrm{NN}}^{M,h}(0,x,p)
:=\inf_{\theta\in\Theta^N}
\mathbb E\left[Z_Ng(X_N)P_N^{p,\theta}\right].
\end{align}
Thus, $V_{\mathrm{NN}}^{M,h}$ includes the Euler and projection errors, as well as the restriction to the chosen neural class, but not the Monte Carlo and optimization errors incurred when solving \eqref{eq:deep-learning-problem-redline}.

The training procedure is described in Algorithm~\ref{alg:deep-learning-redline}. The parameter vector is selected on a fixed validation sample, whereas the final test sample is independent and is evaluated only once after model selection. The reported Monte Carlo standard error is therefore conditional on the selected neural control.

Unless warm-started, the network is initialized with a zero last layer, so the initial candidate is the feasible zero control. In all the numerical experiments, the shared network has two hidden ReLU layers of width $32$ and a hyperbolic-tangent output (as $q=1$), and whenever $X$ is used as an input, we take $s_X=0.2$. The batch size is $L_{\text{batch}}=2048$ Brownian paths, the validation size is $L_{\text{val}}=2^{15}$ and the test size is $L_{\text{test}}=2^{17}$. Adam uses a learning rate of $5\times10^{-3}$ in the toy examples and $10^{-3}$ in the financial examples, with gradient clipping at norm one.

\begin{algorithm}[htbp]
    \caption{Neural approximation of the fully discrete projected problem}
    \label{alg:deep-learning-redline}
    \begin{algorithmic}[1]
        \STATE Fix initial parameters $\theta^{(0)}$, a training batch size $L_{\mathrm{batch}}$, a validation size $L_{\mathrm{val}}$, a final test size $L_{\mathrm{test}}$, a validation frequency, and a learning rate $\rho>0$.
        \STATE Generate independent fixed validation and test samples $\cD^{\mathrm{val}}$ and $\cD^{\mathrm{test}}$.
        \STATE Set $r\gets0$, $\cL_{\mathrm{best}}\gets+\infty$, and $\theta^{\mathrm{best}}\gets\theta^{(0)}$.
        \REPEAT
            \STATE Generate an independent training batch $\cD^{(r)}$ of $L_{\mathrm{batch}}$ Brownian paths.
            \STATE Simulate $(X_n,Z_n,P_n^{\theta^{(r)}})_{0\le n\le N}$ and compute
            \[
            \widehat\cL_{\mathrm{train}}(\theta^{(r)})
            :=\frac1{L_{\mathrm{batch}}}
            \sum_{\ell=1}^{L_{\mathrm{batch}}}
            Z_N^{(\ell)}g(X_N^{(\ell)})P_N^{\theta^{(r)},(\ell)}.
            \]
            \STATE Perform one Adam update with learning rate $\rho$ and clipped gradient.
            \STATE At the prescribed validation frequency, evaluate $\theta^{(r+1)}$ on $\cD^{\mathrm{val}}$ and update $(\cL_{\mathrm{best}},\theta^{\mathrm{best}})$ if the validation loss decreases.
            \STATE Set $r\gets r+1$.
        \UNTIL{the maximal number of iterations or a validation-based stopping criterion is reached.}
        \STATE Evaluate $\theta^{\mathrm{best}}$ once on $\cD^{\mathrm{test}}$ and report the test estimate and its Monte Carlo standard error.
    \end{algorithmic}
\end{algorithm}

The rest of the section is organized as follows. We first use a toy example to identify and quantify the projection bias. Motivated by the toy example, we then stabilize the scheme by imposing an additional state-dependent bound tied to the distance of $P_n$ from the boundary $\{0,1\}$. %
Finally, we apply the stabilized algorithm to quantile hedging in the Black--Scholes and local-volatility models.

\subsection{Toy example}

Assume $g\equiv1$, $b\equiv0$, and $\lambda\equiv0$, so that $Z\equiv1$, and $q=1$. The continuous-time value is
\[
v(t,x,p)=\inf_{\alpha\in\cA_p}\mathbb E[P_{T-t}^{p,\alpha}]=p,
\]
because every admissible $P^{p,\alpha}$ is a bounded martingale. Likewise, for every $M>0$ and $h>0$, the continuously stopped PCPT value satisfies
\[
v^{M,h}(t,x,p)=p.
\]

The fully discrete projected problem behaves differently. We nevertheless have the following convergence result. The diffusion $X$ plays no role. We therefore omit the corresponding input and, with a slight abuse of notation, still denote the reduced
network by $\cN^N(s,p,\theta)$ (compare with \eqref{eq:NN tanh M redline}). 

\begin{proposition}\label{prop:V-M-h-redline}
Assume that the neural-network class contains the zero control: there exists $\theta^0\in\Theta^N$ such that $\cN^N(\cdot,\cdot,\theta^0)\equiv0$. Then, for every $M>0$, $N\ge1$, and $p\in(0,1)$,
\begin{align}
p-M\sqrt{\frac{2T\log(2N)}{N}}
\le V_{\mathrm{NN}}^{M,h}(x,p)
\le p.
\end{align}
Consequently, if $M=M_N$ satisfies
\[
M_N\sqrt{\frac{\log N}{N}}\longrightarrow0,
\]
then $V_{\mathrm{NN}}^{M_N,h}(x,p)\to p$. This holds in particular for fixed $M$, and for $M=h^{-1/2+\kappa}$ with any $\kappa\in(0,1/2)$.
\end{proposition}

\begin{proof}
    Considering the parameters $\theta^0$, one has $P^{p,\theta^0}_N = p$ a.s., hence $V_{\mathrm{NN}}^{M,h}(x,p) \le p$.

    Conversely, fix $\theta\in\Theta^N$. Using \eqref{eq:p disc redline} and \eqref{eq:p disc theta redline}, together with optional stopping at the bounded stopping time $\tau^\theta\wedge N$ for the martingale $\widetilde P^\theta$, we obtain
    \begin{align*}
    \esp{P^{\theta}_N} &= \esp{\tilde P^{\theta}_N \ind_{\tau^\theta > N}} + \esp{\ind_{\tau^\theta \le N} \ind_{\tilde P^\theta_{\tau^\theta} \ge 1}} \\
    &= p + \esp{\ind_{\tau^\theta \le N} \ind_{\tilde P^\theta_{\tau^\theta} \ge 1}} - \esp{\tilde P^{\theta}_{\tau^\theta} \ind_{\tau^\theta \le N}} \\
    &= p + \esp{(-\tilde P^{\theta}_{\tau^\theta}) \ind_{\tau^\theta \le N} \ind_{\tilde P^\theta_{\tau^\theta} < 0}}- \esp{(\tilde P^\theta_{\tau^\theta}-1)\ind_{\tau^\theta \le N} \ind_{\tilde P^\theta_{\tau^\theta} \ge 1}} \\
    &\ge p - \esp{|\tilde P^{\theta}_{\tau^\theta}-\tilde P^\theta_{\tau^\theta-1}| \ind_{\tau^\theta \le N}} \\
    &\ge p - \esp{\max_{0 < n \le N} |\tilde P^{\theta}_n-\tilde P^\theta_{n-1}|} \\
    &= p - \esp{\max_{0 < n \le N} |\cN^N(t_{n-1}/T,P^\theta_{n-1},\theta)| \sqrt{h} |\xi^n|} \\
    &\ge p - M \sqrt{h} \esp{\max_{0 < n \le N} |\xi^n|},
    \end{align*}
    where the random variables $\xi^n := \frac{W_{t_n}-W_{t_{n-1}}}{\sqrt{h}} \sim \cN(0,1)$, $1 \le n \le N$, are i.i.d. standard normal. Using the classical bound $\esp{\max_{1 \le n \le N} |\xi^n|} \le \sqrt{2\ln(2N)}$, we obtain $\esp{P^\theta_N} \ge p - M\sqrt{\frac{2T\ln(2N)}{N}}$, and the result follows by taking the infimum over $\theta \in \Theta^N$.
\end{proof}

\begin{figure}[htbp]
    \centering
    \includegraphics[width=0.8\textwidth]{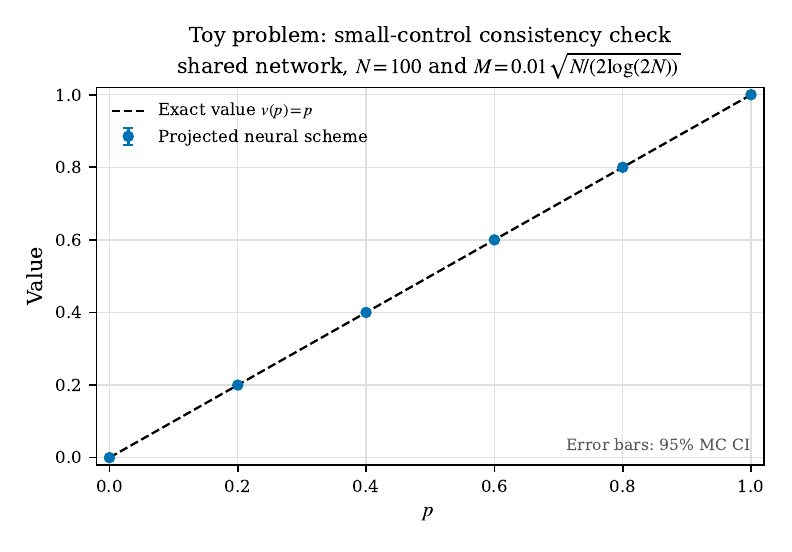}
    \caption{Toy example with $T=1$, $N=100$, and $M=0.01\sqrt{N/(2\log(2N))}$. The markers report the final out-of-sample performance of the selected neural control, used as an estimate of $V_{\mathrm{NN}}^{M,h}$, while the dashed line is the exact value $p\mapsto p$. Test sample size: $2^{17}$; number of training iterations: $150$.}
    \label{fig:toy-example-1-redline}
\end{figure}

Given a target tolerance $\varepsilon>0$, choosing
\[
M=\varepsilon\sqrt{\frac{N}{2T\log(2N)}}
\]
ensures
\[
p-\varepsilon\le V_{\mathrm{NN}}^{M,h}(x,p)\le p.
\]
Figure~\ref{fig:toy-example-1-redline} provides a numerical sanity check for this small-control regime.

However, this experiment does not establish the accuracy of the projected scheme for the larger truncation levels $M=M_h=h^{-1/8}$ used in Corollary~\ref{cor:rates}.

Since $M_h\sqrt h=h^{3/8}$, Proposition~\ref{prop:V-M-h-redline} gives asymptotic convergence of the toy fully discrete value. Its bound nevertheless allows for a substantial bias at the moderate values of $N$ used in practice. For $T=1$,
\[
M_h\sqrt h=N^{-3/8}\simeq0.279\quad(N=30),
\qquad
M_h\sqrt h\simeq0.178\quad(N=100).
\]
Thus, the proposition gives only approximately $p-0.799$ for $N=30$ and $p-0.579$ for $N=100$ as lower bounds.

The mechanism can be seen directly at one step. Defining $H(r):=\varphi(r)-r\Phi(-r)$ for any $r\ge0$, where $\varphi$ and $\Phi$ respectively denote the standard normal density and distribution function, a direct computation gives, for $a \neq 0$,
\begin{align}
\label{eq:local-projection-bias-redline}
\mathbb E\left[\Pi_{[0,1]}(p+|a| \sqrt h \xi)\right]-p
=|a| \sqrt h\left(H\left(\frac p {|a| \sqrt h}\right)
-H\left(\frac{1-p}{|a| \sqrt h}\right)\right),
\end{align}
while the left-hand side is zero for $a=0$. Since $H$ is decreasing, the bias is positive for $p<1/2$, zero for $p=1/2$, and negative for $p>1/2$. This asymmetry can be exploited by the optimizer, as suggested by the numerical results below, by selecting large controls when $p>\frac12$, where the bias is negative, and smaller controls when $p\le\frac12$.

Formula~\eqref{eq:local-projection-bias-redline} shows that the relevant quantities are $\frac{p}{|a|\sqrt h}$ and $\frac{1-p}{|a|\sqrt h}$.
This motivates imposing that the one-step standard deviation be no larger than a fraction of the current distance to the boundary. For a safety parameter $\eta\in(0,1]$, set
\[
d(p):=\operatorname{dist}(p,\{0,1\})=\min(p,1-p)
\]
and replace \eqref{eq:NN tanh M redline} by
\begin{align}
\!m^{\eta_{\mathrm{stab}}}_{h,M}(p)
&:=\min\left(M,\eta_{\mathrm{stab}}\frac{d(p)}{\sqrt h}\right),\\
\label{eq:NN tanh M adaptive redline}\cN^{N,\mathrm{ad}}(s,x,p,\theta)
&:=\Psi^{m^{\eta_{\mathrm{stab}}}_{h,M}(p)}
\bigl(\widetilde\cN^N(s,x,p,\theta)\bigr).
\end{align}
Then, if $|a|\le m^{\eta_{\mathrm{stab}}}_{h,M}(p)$, one computes
\begin{align}
\mathbb P\bigl(p+a \sqrt h \xi\notin[0,1]\bigr)
&\le2\Phi(-1/\eta_{\mathrm{stab}}),\\
\left|
\mathbb E[\Pi_{[0,1]}(p+a\sqrt h\,\xi)]-p
\right| &\le2 M \sqrt h\,H(1/\eta_{\mathrm{stab}}),
\end{align}
where we used \eqref{eq:local-projection-bias-redline} for the second inequality. Thus, both the probability of a projection and the one-step bias can be made arbitrarily small by choosing $\eta_{\mathrm{stab}}$ small enough. This reduction is obtained at the cost of restricting the set of admissible discrete controls, so \(\eta_{\mathrm{stab}}\) reflects a trade-off between bias and control flexibility.

Finally, note that for fixed \(\eta_{\mathrm{stab}}>0\) and \(M=M_h=h^{-1/8}\), the adaptive and fixed bounds differ only when
\[
d(p)<\frac{M_h\sqrt h}{\eta_{\mathrm{stab}}}
=\frac{h^{3/8}}{\eta_{\mathrm{stab}}}.
\]
This region shrinks to the boundary as \(h\to0\). In the toy problem, Proposition \ref{prop:V-M-h-redline} shows that both discretizations converge to \(p\).

\begin{figure}[htbp]
    \centering
    \begin{subfigure}{0.48\textwidth}
        \centering
        \includegraphics[width=\textwidth]{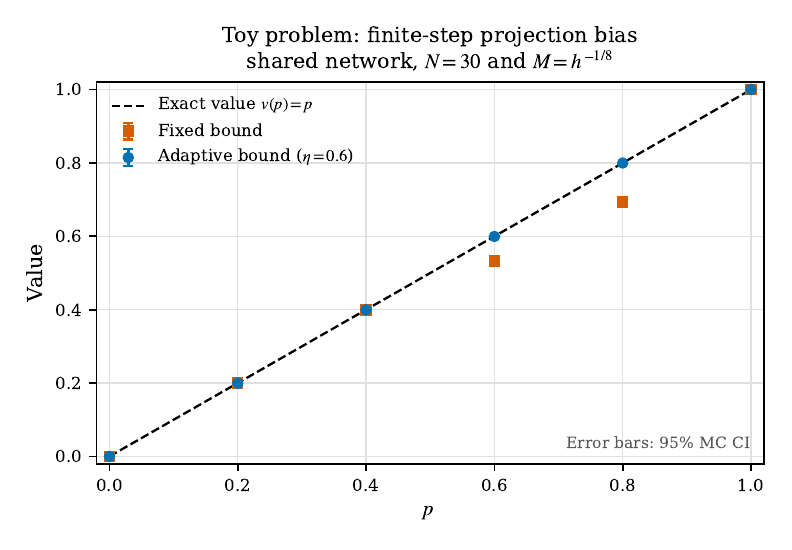}
        \caption{$N=30$.}
    \end{subfigure}
    \begin{subfigure}{0.48\textwidth}
        \centering
        \includegraphics[width=\textwidth]{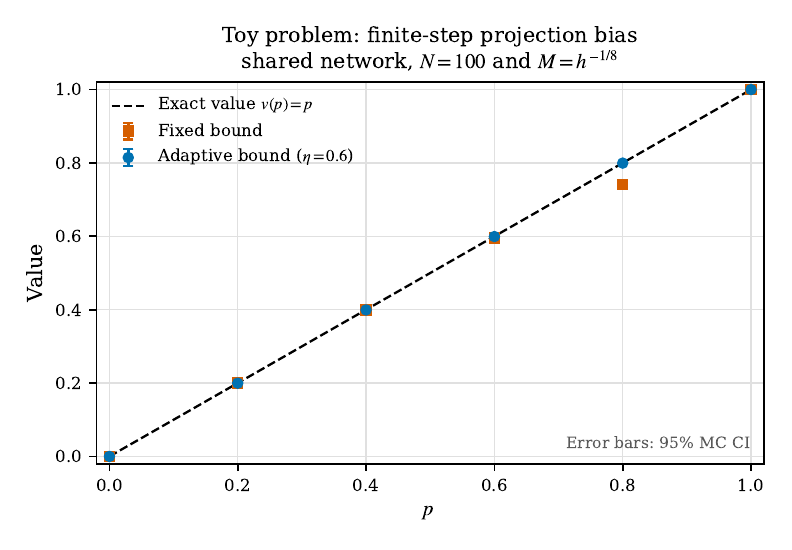}
        \caption{$N=100$.}
    \end{subfigure}
    \caption{Toy example with $T=1$, $M=N^{1/8}$, and $\eta_{\mathrm{stab}}=0.6$. Fixed bound $|\alpha|\le M$ versus adaptive bound $|\alpha|\le m^{\eta_{\mathrm{stab}}}_{h,M}(p)$. The dashed line is the exact value $p\mapsto p$. Test sample size: $2^{17}$; number of training iterations: $150$.}
    \label{fig:toy-example-2-redline}
\end{figure}

Figure~\ref{fig:toy-example-2-redline} compares the fixed and adaptive bounds. The fixed-bound results display the expected downward bias, especially for $p>1/2$ and $N=30$, whereas the adaptive results are much closer to the identity. For instance, at $p=0.8$ and $N=30$, the two test estimates are $0.6934$ and $0.7998$, respectively.

We next apply the stabilized parametrization \eqref{eq:NN tanh M adaptive redline} to quantile hedging in the Black--Scholes and local-volatility models.

\subsection{Quantile hedging in the Black--Scholes and local-volatility models}

We now consider the following more realistic examples
\begin{align}
dX_t&=\mu X_t\,dt+X_t\ell(X_t)\,dW_t,
&dZ_t&=Z_t\lambda(X_t)\,dW_t,
\end{align}
with $T=1$, $X_0=x=100$, $\mu=0.1$, and either
\begin{align*}
    &\ell(x) \equiv \sigma \equiv 0.2, \mbox{ and} \quad \lambda \equiv -\frac{\mu}{\sigma} = -0.5 \quad \mbox{(Black--Scholes model), or} \\
    &\ell(x) = 0.16+0.10\min\left(\left|\frac{x}{100}-1\right|,1\right), \mbox{ and} \quad  \lambda(x)=-\frac{\mu}{\ell(x)}. \quad \mbox{(local-volatility model).}
\end{align*}
In both situations, Assumption \ref{h:coeffs} is satisfied, but in the local-volatility model, $x\mapsto x\ell(x)$ and $\lambda$ are only piecewise differentiable (while still globally Lipschitz continuous), providing a simple nonsmooth test case. The nonsmoothness enters the PCPT error bound through the regularisation and boundary terms, hence we expect a more challenging and closer-to-reality numerical approximation in this case.

We consider the case of a put option:
\begin{align*}
g(x)=(K-x)^+ %
, \qquad K=100.
\end{align*}

It is well known (see e.g. \cite{FollmerLeukert1999QH,benezet2023optimal}) that the value function is given, for any $p \in (0,1)$, by
\begin{align}\label{eq:explicit value}
v(0,x,p)=\mathbb E\big[Z_T g(X_T)\mathbf 1_{\{Z_T g(X_T)<q_p\}}\big]
+q_p\big(p-\mathbb P(Z_T g(X_T)<q_p)\big),
\end{align}
where $q_p$ denotes the $p$-quantile of the random variable $Z_T g(X_T)$. It is also well known that the value function induced by the put option in the Black--Scholes model is not Lipschitz continuous at $p=1$. This lack of regularity is also important in the theoretical study as it also enters the PCPT error bound. Hence, the put option in the nonsmooth local-volatility model should be understood as our challenging test case.

In the Black--Scholes case, $Z_Tg(X_T)$ can be written as a deterministic function of a standard normal random variable. We therefore evaluate \eqref{eq:explicit value}, without time discretization, using a common sample of $2^{20}$ standard normal
variables for all values of $p$.

In the local-volatility case, since there is no closed-form distribution for $Z_T(K-X_T)^+$, we construct the reference curve by a Monte Carlo approximation of \eqref{eq:explicit value} with $2^{18}$ Euler paths with $1000$ time steps. Note that a coupled computation with $500$ and $1000$ time steps gives
reference curves differing by at most $1.79\times10^{-4}$ on the probability grid used to construct the curves.

For the neural approximation, for both models, we use $N=100$, i.e. $h=10^{-2}$, $M=h^{-1/8}\simeq1.7783$, and $\eta_{\mathrm{stab}}=0.6$ as the stabilization parameter. The same Brownian samples are used for the fixed/adaptive comparison, while the test sample remains independent of training and validation.

\begin{figure}[htbp]
    \centering
    \begin{subfigure}{0.49\textwidth}
        \centering
        \includegraphics[width=\textwidth]{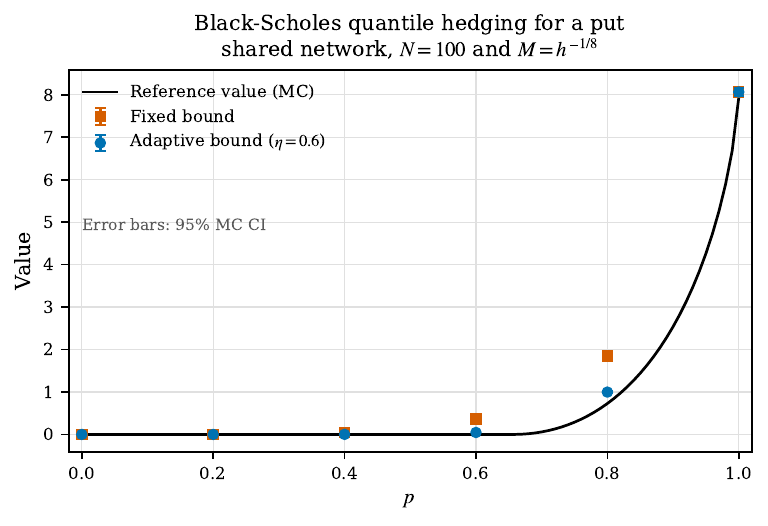}
        \caption{Black--Scholes case.}
    \end{subfigure}
    \begin{subfigure}{0.49\textwidth}
        \centering
        \includegraphics[width=\textwidth]{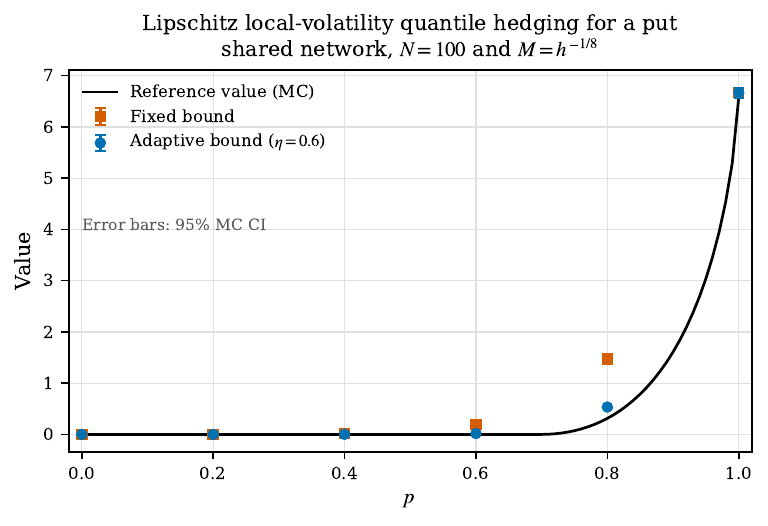}
        \caption{Local volatility case.}
    \end{subfigure}
    \caption{Quantile-hedging value for an ATM put in the Black--Scholes and Lipschitz local-volatility models, as a function of $p$, with $(t,x)=(0,100)$. Here $T=1$, $X_0=K=100$, $N=100$, $M=h^{-1/8}$, $\eta_{\mathrm{stab}}=0.6$, and $500$ training iterations. The error bars are pointwise $95\%$ test-sample Monte Carlo confidence intervals conditional on the selected controls.}
    \label{fig:qh-value}
\end{figure}

Figure~\ref{fig:qh-value} compares the reference curve with the neural estimates obtained under the fixed and adaptive bounds. In both models, the adaptive estimates are closer to the reference at $p=0.4,0.6,0.8$. For example, at $p=0.8$ the absolute difference decreases from $1.12$ to $0.27$ in Black--Scholes and from $1.16$ to $0.22$ in the local-volatility model.

We next examine convergence for $(x,p)=(100,0.7)$ and
\[
N\in\{10,20,30,50,75,100\},\qquad M=N^{1/8}.
\]
Only the adaptive bound \eqref{eq:NN tanh M adaptive redline} is used in this experiment. For both models, we run three independent warm-started training chains, with $500$ Adam iterations at each value of $N$. Since the time-shared architecture is independent of $N$, the network weights obtained at one value of $N$ are used to initialize the next, larger value of $N$ within each chain.

\begin{figure}[htbp]
    \centering
    \begin{subfigure}{0.49\textwidth}
        \centering
        \includegraphics[width=\textwidth]{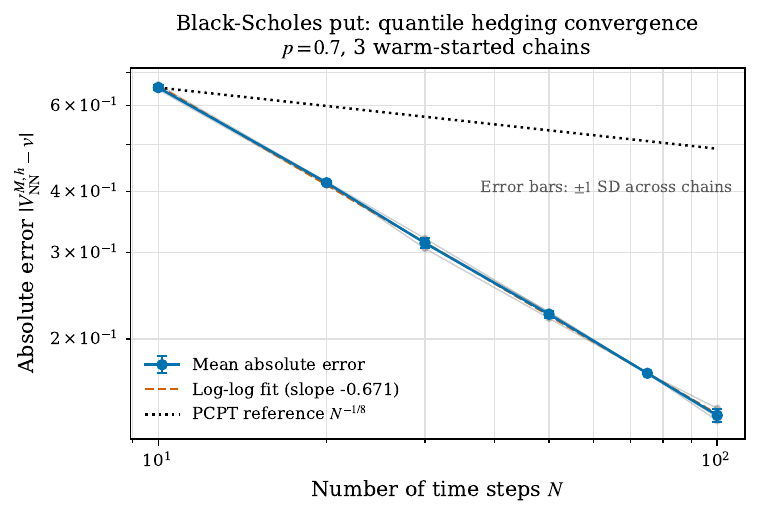}
        \caption{Black--Scholes case.}
    \end{subfigure}
    \begin{subfigure}{0.49\textwidth}
        \centering
        \includegraphics[width=\textwidth]{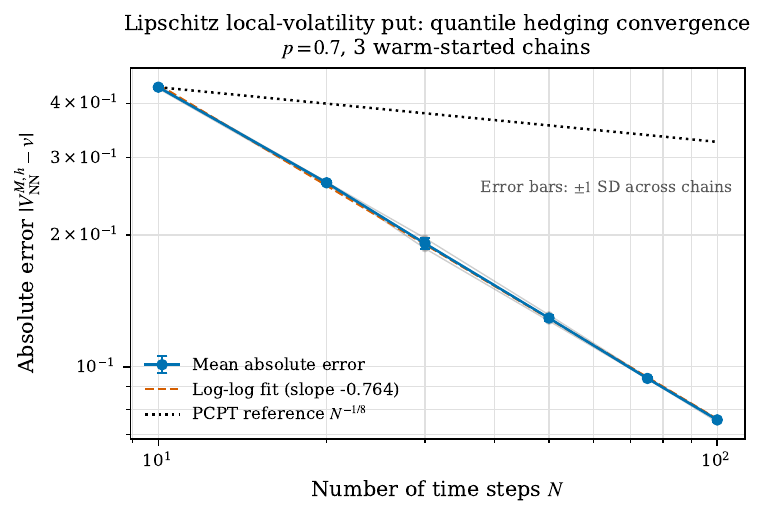}
        \caption{Local volatility case.}
    \end{subfigure}
    \caption{Mean absolute difference from the Monte Carlo reference at $(x,p)=(100,0.7)$ as a function of $N$ with $M=N^{1/8}$, $\eta_{\mathrm{stab}}=0.6$, $500$ training iterations at each value of $N$ and three warm-started chains. The error bars are one sample standard deviation across the three chains.}
    \label{fig:qh-convergence}
\end{figure}

For each model, the errors are measured against a single Monte Carlo reference computed as above with a separate sample and held fixed across all grids and chains. Figure~\ref{fig:qh-convergence} shows that the mean absolute difference from this reference decreases from $0.653$ to $0.140$ in Black--Scholes and from $0.435$ to $0.0756$ in the local-volatility model as $N$ increases from $10$ to $100$. A log--log fit to the mean errors over the six grids gives slopes of $-0.671$ and $-0.764$, respectively. The dotted line uses the limiting exponent $1/8$ of Corollary~\ref{cor:rates} as a visual reference. That corollary concerns the continuously stopped PCPT scheme; the fitted slopes are finite-grid observations for the trained neural scheme, not convergence rates.

The test estimates of the martingale defect $\mathbb E[P_N]-p$, averaged across the three chains, move from $-5.45\times10^{-3}$ at $N=10$ to $-1.63\times10^{-3}$ at $N=100$ in Black--Scholes, and from $-4.75\times10^{-3}$ to $-1.38\times10^{-3}$ in the local-volatility model. These endpoint comparisons do not imply monotonicity in $N$.

As a final experiment, we compare the learned control for the Black--Scholes put with an explicit optimal control of the continuous-time problem. In this setting, for fixed $1>p > p_0:=\bP(Z_Tg(X_T)=0)$, it is known (see \cite[Section 3.2.3.1]{benezet2019study}) that an optimal terminal state is given by
\[
P_T^*=\mathbf 1_{\{Z_Tg(X_T)\le q_p\}} = \mathbf 1_{\{W_T \ge c_p\}},
\qquad
\mathbb E[P_T^*]=p,
\]
where $q_p$ is the $p$-quantile of $Z_Tg(X_T)$ and $
c_p:=-\sqrt T\,\Phi^{-1}(p)$.

The corresponding optimal probability process is the conditional probability of this fixed terminal event:
\begin{align}\label{eq:bs-put-optimal-probability}
P_t^*
&=\mathbb E[P_T^*\mid\mathcal F_t]
 =\mathbb P(W_T\ge c_p\mid\mathcal F_t)
 =\mathbb P(W_T-W_t \ge c_p-W_t\mid\mathcal F_t)
 =\Phi\left(\frac{W_t-c_p}{\sqrt{T-t}}\right),
\qquad 0\le t<T,
\end{align}
as $W_T-W_t$ is independent of $\mathcal F_t$ and has distribution $\mathcal N(0,T-t)$.
To identify the optimal control, define $u(t,w):=\Phi((w-c_p)/\sqrt{T-t})$. A direct calculation gives
\[
u_w(t,w)=\frac{\varphi((w-c_p)/\sqrt{T-t})}{\sqrt{T-t}},
\]
so
\begin{align}\label{eq:bs-put-optimal-alpha-path}
dP_t^*&=\alpha_t^*\,dW_t, \mbox{ with }
\alpha_t^*
=u_w(t,W_t)=\frac{1}{\sqrt{T-t}}\,
  \varphi\left(\frac{W_t-c_p}{\sqrt{T-t}}\right)=\frac{1}{\sqrt{T-t}}\,
  \varphi\left(\Phi^{-1}(P^*_t)\right) = \alpha^\star(t,P^*_t),
\end{align}
extended by $0$ at $p=0$ and $p=1$.

\begin{figure}[htbp]
    \centering
    \includegraphics[width=\textwidth]{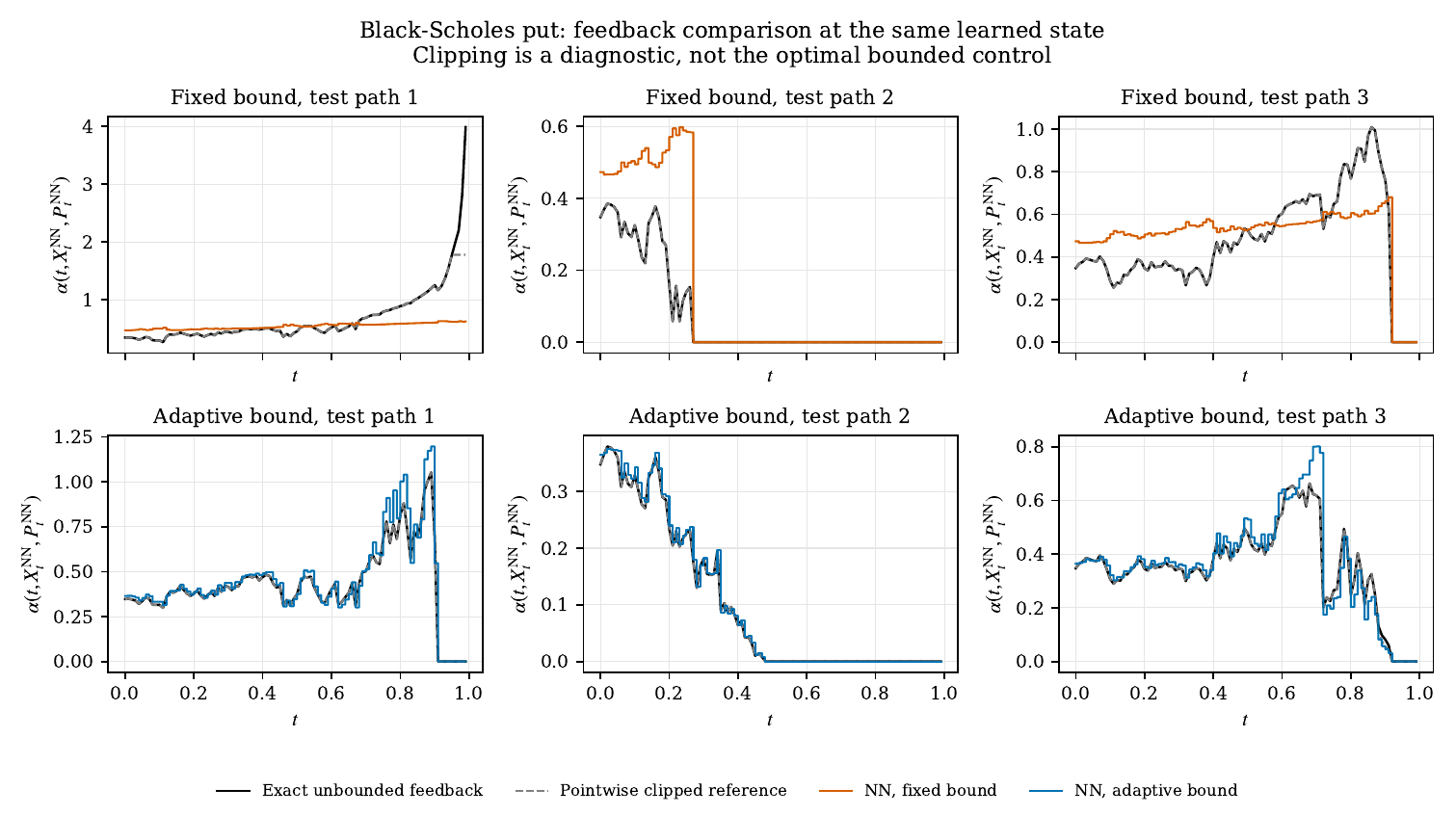}
    \caption{Black--Scholes put: comparison at the same learned state, $\alpha_k^{\mathrm{NN}}$ versus $\alpha^\star(t_k,P_k^{\mathrm{NN}})$, on the first three test paths. The dashed reference is pointwise clipped to the numerical bound and is not asserted to be optimal for the bounded problem.}
    \label{fig:bs-put-alpha-feedback}
\end{figure}

\begin{figure}[htbp]
    \centering
    \includegraphics[width=\textwidth]{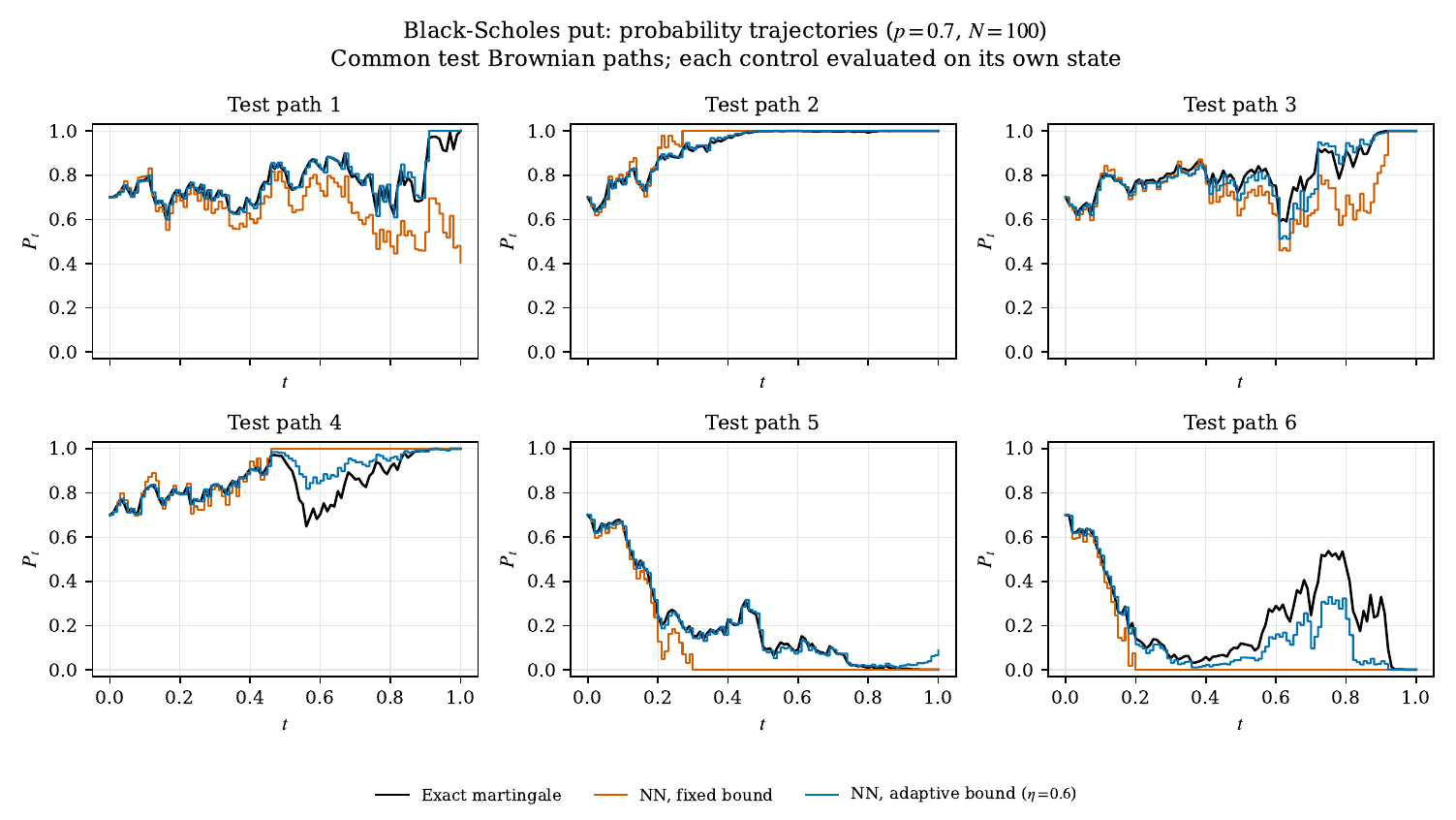}
    \caption{Black--Scholes put: learned probability processes and the exact martingale \eqref{eq:bs-put-optimal-probability} on the same six test paths.}
    \label{fig:bs-put-probability-paths}
\end{figure}

For the numerical comparison, we use $p=0.7$, $N=100$, $M=h^{-1/8}$, $\eta_{\mathrm{stab}}=0.6$, and $500$ training iterations. Here $p_0\simeq0.655422<p$ and $\alpha^\star(0,p)\simeq0.347693$. After selection of the neural controls using the validation sample, we replay the independent test Brownian paths.
In Figure~\ref{fig:bs-put-alpha-feedback}, we compare $\alpha_k^{\mathrm{NN}}$ and $\alpha^\star(t_k,P_k^{\mathrm{NN}})$ at the same learned state for each bound. Figure~\ref{fig:bs-put-probability-paths} compares the learned probability processes to $P^*$ along the same Brownian paths. Across all test paths and decision dates, the root mean squared feedback discrepancy is $0.259$ with the fixed bound and $0.113$ with the adaptive bound. The corresponding discrepancies for the probability processes are $0.126$ and $0.039$, respectively. These comparisons are with a particular continuous-time optimum; they do not establish convergence of the learned controls.

\clearpage
\appendix

\section{Proofs of technical lemmas}
\subsection{Proof of Lemma \ref{lem:props-XZQ}}\label{sec:proof-lem-props-XZQ}
    We fix $(t,x)\in[0,T]\times\bR^d$ and drop the superscripts when unambiguous.
    \begin{enumerate}
        \item
        For any $m\in\bR$, set
        \[
            c_m:=|m|\bar b+\frac{|m(m-1)|}{2}\bar\lambda^2,
            \qquad
            \mathcal M_s:=\cE\Big(\int_0^{\cdot}m\,\lambda(t+u,X_u)^\top dW_u\Big)_s.
        \]
        Since $|m\lambda|\le |m|\bar\lambda$, Novikov's condition implies that $\mathcal M$ is a martingale. Moreover, for every $s\in[0,\bar T]$,
        \begin{align*}
            Z_s^m
            &=\exp\!\Big(\int_0^s\!\big(mb(t+u,X_u)+\tfrac{m(m-1)}{2}|\lambda(t+u,X_u)|^2\big)du\Big)\mathcal M_s
            \le e^{c_m\bar T}\mathcal M_s.
        \end{align*}
        In addition,
        \begin{align*}
            \esp{\mathcal M_{\bar T}^2}
            &=\esp{\cE\Big(\int_0^{\cdot}2m\,\lambda(t+u,X_u)^\top dW_u\Big)_{\bar T}
            \exp\!\Big(m^2\int_0^{\bar T}|\lambda(t+u,X_u)|^2du\Big)}
            \le e^{m^2\bar\lambda^2\bar T},
        \end{align*}
        where we used again Novikov's condition, now with $2m\lambda$. Hence, by Doob's maximal inequality,
        \begin{align*}
            \esp{\sup_{s\in[0,\bar T]}Z_s^m}
            &\le e^{c_m\bar T}\esp{\sup_{s\in[0,\bar T]}\mathcal M_s}
            \le 2e^{c_m\bar T}\esp{\mathcal M_{\bar T}^2}^{1/2}
            \le 2\exp\!\Big(\big(c_m+\tfrac{m^2}{2}\bar\lambda^2\big)\bar T\Big).
        \end{align*}

        \item
        By definition of $X$ and using Jensen's inequality, Burkholder--Davis--Gundy inequality and the linear growth of $\mu$ and $\sigma$ from Assumption~\ref{h:coeffs}, we get, for any $s\le\bar T$,
        \begin{align*}
            \varphi(s) &:= \esp{\sup_{u\le s}|X_u|^m} \\&\le C_m\Big(|x|^m+\esp{\Big(\int_0^s|\mu(t+u,X_u)|du\Big)^m}+\esp{\sup_{u\le s}\Big|\int_0^u\sigma(t+v,X_v)dW_v\Big|^m}\Big) \\
            &\le C_m\left(|x|^m + s^{m-1} \esp{\int_0^s |\mu(t+u,X_u)|^m du} + \esp{\left(\int_0^s \|\sigma(t+u,X_u)\|^2 du\right)^{\frac m2}}\right) \\
            &\le C_{\bar T,m}\left(|x|^m + \esp{\int_0^s (1+|X_u|)^m du} \right) \\
            &\le C_{\bar T,m} \left(1 + |x|^m + \int_0^s \varphi(u) du \right),
        \end{align*}
        and Gronwall's inequality yields $\varphi(\bar T)\le C_{\bar T,m}(1+|x|)^m$.

        \item
        One has, from Jensen inequality and Burkholder--Davis--Gundy inequality, using the previous estimates and the linear growth of $\mu$ and $\sigma$ from Assumption~\ref{h:coeffs},
        \begin{align*}
            \esp{|X_{\bar T}-X_{\bar T-\varepsilon}|^m} &= \esp{\Big|\int_{\bar T-\varepsilon}^{\bar T}\mu(t+u,X_u)\,du + \int_{\bar T-\varepsilon}^{\bar T}\sigma(t+u,X_u)\,dW_u\Big|^m} \\
            &\le C_m\Big(\esp{\Big|\int_{\bar T-\varepsilon}^{\bar T}\mu(t+u,X_u)\,du\Big|^m} + \esp{\sup_{\bar T-\varepsilon \le s \le \bar T} \Big|\int_{\bar T-\varepsilon}^{s}\sigma(t+u,X_u)\,dW_u\Big|^m}\Big) \\
            &\le C_m\Big(\varepsilon^{m-1}\int_{\bar T-\varepsilon}^{\bar T}\esp{|\mu(t+u,X_u)|^m}du + \esp{\Big(\int_{\bar T-\varepsilon}^{\bar T}\|\sigma(t+u,X_u)\|^2 du\Big)^{m/2}}\Big) \\
            &\le C_m\Big(\varepsilon^{m-1}\int_{\bar T-\varepsilon}^{\bar T}\esp{(1+|X_u|)^m}du + \varepsilon^{m/2-1}\,\esp{\int_{\bar T-\varepsilon}^{\bar T}\|\sigma(t+u,X_u)\|^m du}\Big) \\
            &\le C_{m} \varepsilon^{m/2} (1+\sup_{u \le \bar T}\esp{|X_u|^m}) \le C_{\bar T,m}\,(1+|x|)^m\,\varepsilon^{m/2}.
        \end{align*}

        \medskip\noindent Similarly for $Z$, one computes, using the previous moment bounds and the boundedness of $b$ and $\lambda$,
        \begin{align*}
            &\esp{|Z_{\bar T}-Z_{\bar T-\varepsilon}|^m} \\ =& \esp{\Big|\int_{\bar T-\varepsilon}^{\bar T} Z_u\,b(t+u,X_u)\,du + \int_{\bar T-\varepsilon}^{\bar T} Z_u\,\lambda(t+u,X_u)^\top dW_u\Big|^m} \\
            \le& C_m\Big(\esp{\Big|\int_{\bar T-\varepsilon}^{\bar T} Z_u\,b(t+u,X_u)\,du\Big|^m} + \esp{\sup_{\bar T-\varepsilon \le s \le \bar T} \Big|\int_{\bar T-\varepsilon}^{s} Z_u\,\lambda(t+u,X_u)^\top dW_u\Big|^m}\Big) \\
            \le& C_m\Big(\varepsilon^{m-1}\int_{\bar T-\varepsilon}^{\bar T}\esp{Z_u^m\,|b(t+u,X_u)|^m}du + \esp{\Big(\int_{\bar T-\varepsilon}^{\bar T}Z_u^2\,|\lambda(t+u,X_u)|^2 du\Big)^{m/2}}\Big) \\
            \le& C_m\Big(\bar b^m\,\varepsilon^{m}\,\sup_{u\le\bar T}\esp{Z_u^m} + \bar\lambda^m\,\varepsilon^{m/2}\,\sup_{u\le\bar T}\esp{Z_u^m}\Big) \\
            \le& C_{\bar T,m} \varepsilon^{m/2}.
        \end{align*}

        \item By H\"older's inequality,
        \begin{align*}
            \esp{|Z_{T-t}\,g(X_{T-t})|^r}
            \le \esp{|Z_{T-t}|^{2r}}^{1/2}\,\esp{|g(X_{T-t})|^{2r}}^{1/2} \le C_{r}\,(1+|x|)^{r},
        \end{align*}
        which concludes the proof.
    \end{enumerate}

\subsection{Proof of Lemma \ref{lem:Lp-stability-mollified}}\label{sec:proof-lem-Lp-stability-mollified}

    We fix $(t,x) \in [0,T]\times\bR^d$ and set $\delta X_s:=X^{t,x}_s-\tilde X^{t,x}_s$, $\delta Z_s:=Z^{t,x}_s-\tilde Z^{t,x}_s$, $\delta\mu_s:=\mu(t+s,X_s)-\tilde\mu(t+s,\tilde X_s)$ and $\delta\sigma_s:=\sigma(t+s,X_s)-\tilde\sigma(t+s,\tilde X_s)$ for $s\in[0,T-t]$.

    First, by It\^o's formula and Young's inequality, for any $s\in[0,T-t]$,
    \begin{align*}
        \esp{|\delta X_s|^{2p}} =& \esp{\int_0^s 2p\,|\delta X_u|^{2p-2}\,\delta X_u^\top \delta \mu_u + p\,|\delta X_u|^{2p-2}\,\big\|\delta\sigma_u\big\|^2\, + 2p(p-1)\,|\delta X_u|^{2p-4}\,|\delta X_u^\top \delta\sigma_u|^2\,du} \\
        \le& C_p \esp{\int_0^s |\delta X_u|^{2p-1}|\delta\mu_u| + |\delta X_u|^{2p-2}\big\|\delta\sigma_u\big\|^{2}\,du} \\
        \le& C_p \esp{\int_0^s |\delta X_u|^{2p} + |\delta\mu_u|^{2p} + \big\|\delta\sigma_u\big\|^{2p}\,du}.
    \end{align*}
    Using the Lipschitz property of $\mu$, $\sigma$ in Assumption \ref{h:coeffs} and Lemma \ref{lem:mollified-coeffs-props}, we have
    \begin{align*}
        |\delta\mu_u| &= |\mu(t+u,X_u)-\tilde\mu(t+u,\tilde X_u)| \\
        &\le |\mu(t+u,X_u)-\mu(t+u,\tilde X_u)| + |\mu(t+u,\tilde X_u)-\tilde\mu(t+u,\tilde X_u)| \\
        &\le L|\delta X_u| + C_0\varepsilon,
    \end{align*}
    and similarly, $\|\delta\sigma_u\|\le L|\delta X_u| + C_0\varepsilon$. Therefore,
    \begin{align*}
        \esp{|\delta X_s|^{2p}} \le& C_p \esp{\int_0^s |\delta X_u|^{2p} + (L|\delta X_u| + C_0\varepsilon)^{2p} + (L|\delta X_u| + C_0\varepsilon)^{2p}\,du} \\
        \le& C_p \esp{\int_0^s |\delta X_u|^{2p}  + \varepsilon^{2p}\,du} \\
        =& C_p\int_0^s \esp{|\delta X_u|^{2p}}\,du + C_p\,T\,\varepsilon^{2p},
    \end{align*}
    and the bound for $\esp{|\delta X_s|^{2p}}$ follows from Gr\"onwall's inequality.

    Regarding $Z$, we similarly have by It\^o's formula and Young's inequality, for any $s\in[0,T-t]$,
    \begin{align*}
        \esp{|\delta Z_s|^{2p}} 
        \le& C_p \esp{\int_0^s |\delta Z_u|^{2p} + \big|b(t+u,X_u)Z_u - \tilde b(t+u,\tilde X_u)\tilde Z_u\big|^{2p} \right.\\&\qquad\left.+ \big\|Z_u\lambda(t+u,X_u)^\top - \tilde Z_u\tilde\lambda(t+u,\tilde X_u)^\top\big\|^{2p}\,du}.
    \end{align*}
    Using the Lipschitz and boundedness property of $b$, $\lambda$ in Assumption \ref{h:coeffs} and Lemma \ref{lem:mollified-coeffs-props}, we have
    \begin{align*}
        &\big|b(t+u,X_u)Z_u - \tilde b(t+u,\tilde X_u)\tilde Z_u\big| \\
        \le& \big|b(t+u,X_u)Z_u - b(t+u,X_u)\tilde Z_u\big| + \big|b(t+u,X_u)\tilde Z_u - \tilde b(t+u,X_u)\tilde Z_u\big| + \big| \tilde b(t+u,X_u)\tilde Z_u - \tilde b(t+u,\tilde X_u)\tilde Z_u\big| \\
        \le& \bar b |\delta Z_u| + C_0\varepsilon|\tilde Z_u| + L|\delta X_u||\tilde Z_u|,
    \end{align*}
    and similarly, $\big\|Z_u\lambda(t+u,X_u)^\top - \tilde Z_u\tilde\lambda(t+u,\tilde X_u)^\top\big\| \le \bar\lambda |\delta Z_u| + C_0\varepsilon|\tilde Z_u| + L|\delta X_u||\tilde Z_u|$. Therefore, using Cauchy-Schwarz inequality, the fact that $\sup_u \esp{|\tilde Z_u|^m} \le C_m < \infty$ for all $m \ge 1$, and the previous estimates, we get
    \begin{align*}
        \esp{|\delta Z_s|^{2p}} \le& C_p \esp{\int_0^s |\delta Z_u|^{2p} + \big(\bar b|\delta Z_u| + C_0\varepsilon|\tilde Z_u| + L|\delta X_u||\tilde Z_u|\big)^{2p}\,du} \\
        \le& C_p \esp{\int_0^s |\delta Z_u|^{2p} + |\tilde Z_u|^{2p} \varepsilon^{2p} + |\tilde Z_u|^{2p} |\delta X_u|^{2p}\,du} \\
        \le& C_p \int_0^s \esp{|\delta Z_u|^{2p}}\,du + C_p\int_0^s \esp{|\tilde Z_u|^{4p}}^{\frac12}\esp{|\delta X_u|^{4p}}^{\frac12} du + C_p\,T\,\varepsilon^{2p} \\
        \le& C_p \int_0^s \esp{|\delta Z_u|^{2p}}\,du + C_p\,T\,\varepsilon^{2p},
    \end{align*}
    and we conclude by Gr\"onwall's inequality.

\subsection{Proof of Lemma \ref{lem:bounds-JH}}\label{sec:proof-lem-bounds-JH}
    See e.g. \cite[Chapter V, Theorem 39]{Protter2004}, for the representation by $J$ and $H$ of the partial derivatives with respect to the initial condition.
    We detail the estimate on $\tilde J$. Set, for $s\in[0,T-t]$,
    \[
        A_s:=\partial_x \tilde\mu(t+s,\tilde X^{t,x}_s)\in\bR^{d\times d},\qquad
        C^{(k)}_s:=\partial_x \tilde\sigma^{(k)}(t+s,\tilde X^{t,x}_s)\in\bR^{d\times d},\qquad k\in\{1,\dots,q\},
    \]
    so that the first equation in \eqref{eq:sde deriv} rewrites as
    \[
        d\tilde J^{t,x}_s = A_s\,\tilde J^{t,x}_s\,ds + \sum_{k=1}^q C^{(k)}_s\,\tilde J^{t,x}_s\,d(W_s)^k,\qquad \tilde J^{t,x}_0=I_d.
    \]
    By Lemma \ref{lem:mollified-coeffs-props}, under Assumption \ref{h:coeffs}, the maps $x\mapsto \tilde\mu(t,x)$ and $x\mapsto \tilde\sigma(t,x)$ are globally Lipschitz uniformly in $t$, hence there exists $L\ge 0$ such that $\|A_s\|+\sum_{k=1}^q\|C^{(k)}_s\|\le L$ for all $s$, almost surely, uniformly in $(t,x)$. Using It\^o's formula, we get
        \begin{align*}
            \esp{\| \tilde J^{t,x}_{s}\|^2} &= \|I_d\|^2 + \int_0^{s} \esp{2\,\langle \tilde J^{t,x}_u, A_u\tilde J^{t,x}_u\rangle + \sum_{k=1}^q \|C^{(k)}_u\tilde J^{t,x}_u\|^2}\,du \\
            &\le d + C\int_0^{s} \esp{\|\tilde J^{t,x}_u\|^2}\,du,
        \end{align*}
    where we used $|\langle J,AJ\rangle|\le \|A\|\,\|J\|^2$ and $\|C^{(k)}J\|\le \|C^{(k)}\|\,\|J\|$.\\
    Gronwall's lemma gives $\esp{\|\tilde J^{t,x}_s\|^2}\le d e^{Cs}$ for all $s\in[0,T-t]$, hence in particular $\sup_{(t,x)}\esp{\|\tilde J^{t,x}_{T-t}\|^2}<\infty$. Finally, by Cauchy--Schwarz,
    \[
        \esp{\|\tilde J^{t,x}_{T-t}\|}\le \esp{\|\tilde J^{t,x}_{T-t}\|^2}^{\frac12}\le \sqrt{d}\,e^{\frac{C}{2}T}<\infty.
    \]
    We now detail the bound on $\tilde H$. For $s\in[0,T-t]$, define
    \[
        a_s:=|\tilde\lambda(t+s,\tilde X^{t,x}_s)|^2-\tilde b(t+s,\tilde X^{t,x}_s),\qquad
        B_s:=2\, \partial_x \tilde\lambda(t+s,\tilde X^{t,x}_s)^\top \tilde\lambda(t+s,\tilde X^{t,x}_s)-\partial_x \tilde b(t+s,\tilde X^{t,x}_s),
    \]
    so that \eqref{eq:sde deriv} can be rewritten as
    \begin{align*}
        d\tilde H^{t,x}_s &= a_s\,\tilde H^{t,x}_s\,ds + \tilde Q^{t,x}_s\,(\tilde J^{t,x}_s)^\top\,B_s\,ds
        - \Gamma_s\,dW_s, \text{ and} \\\Gamma_s&:=\tilde Q^{t,x}_s\,(\tilde J^{t,x}_s)^\top(\partial_x \tilde\lambda(t+s,\tilde X^{t,x}_s))^\top + \tilde H^{t,x}_s\,\tilde\lambda(t+s,\tilde X^{t,x}_s)^\top.
    \end{align*}
    By Lemma~\ref{lem:mollified-coeffs-props} and the boundedness of $(\tilde\lambda,\tilde b)$, we have $|a_s|\le C$ and $\|B_s\|+\|\partial_x\tilde\lambda(t+s,\tilde X^{t,x}_s)\|\le C$. Moreover, using $|\tilde\lambda|\le \bar\lambda$, we have the pointwise bound
    \[
        \|\Gamma_s\|^2
        \le 2\,|\tilde Q^{t,x}_s|^2\,\| (\tilde J^{t,x}_s)^\top(\partial_x \tilde\lambda(t+s,\tilde X^{t,x}_s))^\top\|^2 + 2\,|\tilde H^{t,x}_s|^2\,\|\tilde\lambda(t+s,\tilde X^{t,x}_s)\|^2
        \le C\big(|\tilde Q^{t,x}_s|^2\,\|\tilde J^{t,x}_s\|^2+|\tilde H^{t,x}_s|^2\big).
    \]
    Using It\^o's formula, the previous estimates together with Young's inequality, we get
    \begin{align*}
        \esp{|\tilde H^{t,x}_s|^2}
        &= \int_0^s \esp{2a_r|\tilde H^{t,x}_r|^2 + 2\,\tilde H^{t,x}_r\cdot(\tilde Q^{t,x}_r (\tilde J^{t,x}_r)^\top B_r) + \|\Gamma_r\|^2}\,dr \\
        &\le C\,\int_0^{s}\esp{|\tilde H^{t,x}_r|^2}\,dr
        + C\,\int_0^{s}\esp{\|\tilde J^{t,x}_r\|^2\,|\tilde Q^{t,x}_r|^2}\,dr, 
     \end{align*}
    Gronwall's lemma yields
    \[
        \esp{|\tilde H^{t,x}_{T-t}|^2}
        \le C\,\int_0^{T-t}\esp{\|\tilde J^{t,x}_r\|^2\,|\tilde Q^{t,x}_r|^2}\,dr.
    \]
    Finally, by Cauchy--Schwarz together with the moment bounds for $\tilde J$ and Lemma \ref{lem:props-XZQ}, the right-hand side is bounded by $C$ (since $\esp{\|\tilde J\|^4}$ and $\esp{|\tilde Q|^4}=\esp{\tilde Z^{-4}}$ are both bounded uniformly in $x$). Hence
    \begin{align*}
        \esp{|\tilde H^{t,x}_{T-t}|}\le \esp{|\tilde H^{t,x}_{T-t}|^2}^{\frac12}\le C.
    \end{align*}

\subsection{Proof of Proposition \ref{prop:subdpp-hatvMh}}\label{app:dpp proof}

If $t+h\ge T+\varepsilon^2$, the terminal condition
$\hat v^{M,h}(T+\varepsilon^2,x,p)=p\,g(x)$ and the definition of
$\hat v^{M,h}$ give equality in the stated inequality. Assume henceforth
that $t+h<T+\varepsilon^2$, and set $R:=T+\varepsilon^2-(t+h)$.

We use a countable-partition argument for the one-step DPP; see, e.g.,
\cite{Krylov1999,Jakobsen2019}. Fix $\kappa\in(0,1)$,
$\delta\in(0,1]$, $\eta>0$, and first-step controls
$(\alpha,\beta)\in\cA^{M,h}_p\times\cB^{\varepsilon,h}$.
Choose a countable Borel partition $(B_i)_{i\ge1}$ of
$\bR^d\times[0,1]$ into sets of diameter at most $\delta$, with
representatives $(x_i,p_i)\in B_i$. At each $(t+h,x_i,p_i)$, choose an
$\eta$-optimal pair of continuation controls
$(\alpha^i,\beta^i)\in\cA^{M,h}_{p_i}\times\cB^{\varepsilon,h}$.

For $p'\in[0,1]$, define
\[
 c_i(p'):=
 \begin{cases}
 p'/p_i,&p'<p_i,\\
 1,&p'=p_i,\\
 (1-p')/(1-p_i),&p'>p_i,
 \end{cases}
 \qquad \alpha^{i,p'}:=c_i(p')\alpha^i.
\]
Writing
$P^i:=P^{p_i,\alpha^i}$, we have, for all $u\ge0$,
\[
 P^{p',\alpha^{i,p'}}_u=
 \begin{cases}
 c_i(p')P^i_u,&p'<p_i,\\
 P^i_u,&p'=p_i,\\
 1-c_i(p')(1-P^i_u),&p'>p_i.
 \end{cases}
\]
These martingales remain in $[0,1]$. Since $0\le c_i(p')\le1$, the
rescaling reduces the amplitudes and, if $c_i(p')>0$, preserves the original stopping
times on each grid interval (if $c_i(p')=0$, we obtain the zero control, which also has the PCPT structure taking e.g. $\tau^k=kh$), hence $\alpha^{i,p'}\in\cA^{M,h}_{p'}$.
Moreover, the two martingales are ordered and have respective means
$p'$ and $p_i$, so
\[
 \esp{\left|P^{p',\alpha^{i,p'}}_R-P^i_R\right|}=|p'-p_i|.
\]

To compare continuation costs, put
\[
 H_i(x'):=\hat Z^{t+h,x',\beta^i}_R\,
 g(\hat X^{t+h,x',\beta^i}_R).
\]
Since the martingale difference is bounded by $1$, H\"older's inequality
and Lemma~\ref{lem:stability-shaken-mollified} give
\[
 \esp{H_i(x')\left|P^{p',\alpha^{i,p'}}_R-P^i_R\right|}
 \le \|H_i(x')\|_{L^{1/\kappa}}\,|p'-p_i|^{1-\kappa}
 \le C(1+|x'|)|p'-p_i|^{1-\kappa}.
\]
For $(x',p')\in B_i$, the same stability estimates at fixed controls,
using $0\le P^i_R\le1$ and $|x'-x_i|\le\delta\le1$, also yield
\[
 \esp{\left|H_i(x')-H_i(x_i)\right|P^i_R}
 \le C(1+|x'|)|x'-x_i|.
\]
Combining these bounds with the $\eta$-optimality at $(x_i,p_i)$ and
Lemma~\ref{lem:holder-p-hatvMh} yields
\[
 \esp{H_i(x')P^{p',\alpha^{i,p'}}_R}
 \le \hat v^{M,h}(t+h,x',p')
 +C(1+|x'|)(\delta+\delta^{1-\kappa})+\eta.
\]

Choose Brownian-functional versions of the continuation controls and
realise their copies after $h$ using $W^h_u:=W_{h+u}-W_h$, which is
independent of $\cF_h$. On the event
$E_i:=\{(\hat X^{t,x,\beta}_h,P^{p,\alpha}_h)\in B_i\}\in\cF_h$,
concatenate the first-step controls with the copies of
$(c_i(P^{p,\alpha}_h)\alpha^i,\beta^i)$.
The functions $c_i$ are Borel measurable, and $h$ is a grid point.
Thus this countable pasting preserves progressive measurability, the
PCPT structure, the amplitude bounds, and the state constraint.
The concatenated pair is therefore admissible.

Conditioning on $\cF_h$, the flow property of $\hat X$ and the
multiplicative flow property of $\hat Z$ give
\begin{align*}
 \hat v^{M,h}(t,x,p)
 &\le \esp{\hat Z^{t,x,\beta}_h\,
 \hat v^{M,h}(t+h,\hat X^{t,x,\beta}_h,P^{p,\alpha}_h)}\\
 &\quad+C(1+|x|)(\delta+\delta^{1-\kappa})+e^{\bar b h}\eta,
\end{align*}
where we used
$\esp{\hat Z^{t,x,\beta}_h(1+|\hat X^{t,x,\beta}_h|)}\le C(1+|x|)$
and $\esp{\hat Z^{t,x,\beta}_h}\le e^{\bar b h}$.
Letting $\eta\downarrow0$ and $\delta\downarrow0$, then taking the
infimum over the first-step controls, proves the result.

\printbibliography

\end{document}